\documentclass[12pt,preprint]{emulateapj}

\makeatother

\def\simgt{\lower.5ex\hbox{\gtsima}}

\newcommand{\fgas}{$M_{\rm gas} / [M_{\rm gas} + M_{\ast}]$}
\newcommand{\fg}{$f_{\rm gas}$}

\newcommand{\gdr}{$\delta_{ \rm GDR}$}

\newcommand{\s}{{\it Spitzer}}
\newcommand{\h}{{\it Herschel}}
\newcommand{\msol}{$\rm M_{\odot}$}
\newcommand{\md}{$M_{\rm dust}$}

\newcommand{\lsol}{L$_{\odot}$}
\newcommand{\lir}{$L_{\rm IR}$}
\newcommand{\ms}{$M_{\ast}$}
\newcommand{\lco}{$L^{\prime}_{\rm CO}$}

\newcommand{\mol}{$M_{\rm H_{\rm 2}}$}
\newcommand{\td}{$T_{\rm d}$}
\newcommand{\aco}{$\alpha_{\rm CO}$}

\newcommand{\mgas}{$M_{\rm gas}$}

\lefthead{Magdis et al..~}

\slugcomment{to appear in ApJ}
\shorttitle{Dust and Gas in Star Forming Galaxies to z = 2}
\begin{document}
 \title{The Evolving Interstellar Medium of Star Forming Galaxies since z=2 as Probed by Their Infrared Spectral Energy Distributions}
 \author{Georgios E. Magdis $\!$\altaffilmark{1},
 	      E. Daddi $\!$\altaffilmark{2},     
		 M. B\'ethermin $\!$\altaffilmark{2},
		 M. Sargent $\!$\altaffilmark{2},
		 D. Elbaz $\!$\altaffilmark{2},
		 M. Pannella $\!$\altaffilmark{2},
		 M. Dickinson $\!$\altaffilmark{3},
		 H. Dannerbauer $\!$\altaffilmark{4},
		 E. da Cunha $\!$\altaffilmark{5},
		 F. Walter $\!$\altaffilmark{5},
		 D. Rigopoulou $\!$\altaffilmark{1,6},
		 V. Charmandaris $\!$\altaffilmark{7,8,9},
		 H.S Hwang $\!$\altaffilmark{10}
		 J. Kartaltepe $\!$\altaffilmark{3} 
		 }

\altaffiltext{1}{Department of Physics, University of Oxford, Keble Road, Oxford OX1 3RH}
\altaffiltext{2}{CEA, Laboratoire AIM, Irfu/SAp, F-91191 Gif-sur-Yvette, France}
\altaffiltext{3}{NOAO, 950 N. Cherry Avenue, Tucson, AZ 85719, USA}
\altaffiltext{4}{Universit\"at Wien, Institut \"ur Astronophysik, T\"urkenschanzstrasse 17, 1180 Wien, Austria}
\altaffiltext{5}{Max-Planck-Institut f\"ur Astronomie, K\"onigstuhl 17, D-69117 Heidelberg, Germany}
\altaffiltext{6}{Space Science \& Technology Department, Rutherford Appleton Laboratory, Chilton, Didcot, Oxfordshire OX11 0QX}
\altaffiltext{7}{Department of Physics \& ICTP, University of Crete, GR-71003, Heraklion, Greece} 
\altaffiltext{8}{IESL/Foundation for Research \& Technology-Hellas, GR-71110, Heraklion, Greece}
\altaffiltext{9}{Chercheur Associ\'e, Observatoire de Paris, F-75014, Paris, France}
\altaffiltext{10}{Smithsonian Astrophysical Observatory, 60 Garden Street, Cambridge, MA 02138, USA}

\begin{abstract}
Using data from the mid-infrared to millimeter wavelengths for individual galaxies and for stacked ensembles at $0.5 < z < 2$, we derive robust estimates of dust masses (\md) for main sequence (MS) galaxies, which obey a tight correlation between star formation rate (SFR) and stellar mass ($M_{\ast}$), and for star-bursting galaxies that fall outside that relation. Exploiting the correlation of gas to dust mass with metallicity (\mgas/\md$-Z$), we use our measurements to constrain the gas content, CO-to-H$_{\rm 2}$ conversion factors (\aco) 
and star formation efficiencies (SFE) of these distant galaxies. Using large statistical samples, we confirm
 that \aco ~and SFE are an order of magnitude higher and lower, respectively, in MS galaxies at high redshifts  compared to the values of local galaxies with equivalently high infrared luminosities (\lir ~$> 10^{12}$ \lsol). For galaxies within the MS, we show  that the variations of specific star formation rates (sSFR=SFR/$M_{\ast}$) are driven by varying gas fractions. For relatively massive galaxies like those in our samples, we show that the hardness of the radiation 
field, $\langle U \rangle$, which is proportional to the dust mass weighted luminosity (\lir/\md), and the primary parameter defining the shape of the IR-spectral energy distribution (SED), is equivalent  to SFE/Z. For MS galaxies with stellar mass $\log(M_{\ast}/M_\odot) \ge 9.7$ we measure this quantity, $\langle U \rangle$, showing that it
 does not depend significantly on either the stellar mass or the sSFR. This is explained as a simple consequence of the existing correlations between SFR$-M_{\ast}$,  $M_{\ast}-Z$  and  \mgas$-$SFR. Instead, we show that $\langle U \rangle$ (or equally  \lir/\md) does evolve, with MS galaxies having harder radiation fields and thus warmer temperatures as redshift increases from $z = 0$ to 2, a trend which can also be understood based on the redshift evolution of the $M_{\ast}-Z$ and SFR$-M_{\ast}$ relations. These results motivate the construction of a universal set of SED templates for MS galaxies, that are independent of their sSFR or $M_{\ast}$, but which vary as a function of redshift with only one parameter, $\langle U \rangle$.
\end{abstract}


\section{INTRODUCTION}
Deep and wide multi-wavelength extragalactic surveys have greatly enhanced our understanding of galaxy evolution. It has now been well established that the star formation rates (SFRs) in galaxies were on average higher in the past, with galaxies emitting the bulk of their bolometric energy in the infrared (IR) progressively dominating the star formation density of the Universe that peaked  at $z \geq 1$ (e.g., Le Borgne et al.\ 2009).

A recent major step forward in characterizing the nature of star formation in distant galaxies has been 
the discovery that the majority of star-forming galaxies at every redshift define a narrow locus in the 
stellar mass ($M_{\ast}) ~-$ star formation rate (SFR) plane. This correlation has been observed in the local Universe (Brinchmann et al.\ 2004; Peng et al.\ 2010), as well as at intermediate redshifts $0.5 < z < 3$ (Noeske et al.\ 
2007; Elbaz et al.\ 2007; Daddi et al.\ 2007; Pannella et al.\ 2009; Rodighiero et al.\ 2010a, Karim et al.\ 
2011, Magdis et al.\ 2010), and beyond (e.g., Daddi et al.\ 2009, Stark et al.\ 2010), with a normalization factor that increases rapidly with look back time, mirroring the increase in the star formation activity of the early galaxies. 

The fact that this correlation between SFR and $M_{\ast}$ appears to be present at all epochs has been used to define a characteristic specific star formation rate ($\mathrm{sSFR} \equiv \mathrm{SFR}/M_{\ast}$) at each redshift and stellar mass, and from its evolution with cosmic time, a main sequence (MS) mode of star formation that is followed by the majority of star-forming galaxies. This main sequence also serves as a tool to distinguish starburst (SB) from normal{\footnote{Throughout the paper we will use the terms ``normal'' and ``main sequence'' galaxy interchangeably}  galaxies at any redshift, simply by measuring the excess sSFR of the galaxy with respect to the SFR--$M_{\ast}$ correlation at that redshift. Such outliers are known to exist at any redshift. In the local Universe, a class of galaxies that are usually identified as starbursts are the (Ultra) Luminous Infrared Galaxies (ULIRGs, \lir ~$> 10^{12}$ \lsol, Sanders \& Mirabel 1996, Elbaz et al.\ 2007). However, in the distant Universe it has proven to be harder than initially thought to associate pure starbursts with a specific class of high$-z$ galaxies, although at least a fraction of Submillimeter Galaxies (SMGs) seem to exhibit elevated specific star formation rates (Tacconi et al.\ 2006; 2008; Daddi et al.\ 2007; 2009, Takagi et al.\ 2008).

The remarkable uniformity of MS galaxies, as indicated by the SFR-\ms ~correlation, and their contrasting nature to that of starburst galaxies, has recently been further manifested with respect to two more sets of observable parameters.  
Using direct measurements of the total infrared luminosity (\lir=$L_{8-1000 \mu m}$) ~of distant galaxies that have now become possible with the \h ~Space Observatory (Pilbratt et al.\ 2010), Elbaz et al.\ (2011) showed that galaxies that follow the SFR-\ms ~correlation are also part of a secondary,
infrared main sequence, defined by a universal total-to-mid-IR luminosity ratio, IR8 $\equiv$ \lir/$L_{\rm 8}$, where $L_{\rm 8}$ is the rest-frame 8$\,\mu$m luminosity. 
The majority of MS sequence galaxies, at all luminosities and redshifts, are found to obey 
this linear correlation between \lir ~and $L_{\rm 8}$, suggesting that they share a common
IR spectral energy distribution (SED), with a mid-to-far-IR shape that has evolved little with cosmic time. Interestingly, starburst galaxies, which systematically fall above the SFR-\ms ~main sequence, are also outliers to the \lir-$L_{\rm 8}$ relation, exhibiting elevated IR8 values 
with respect to normal galaxies.

The third quantity that highlights the distinct nature of MS and 
starbursts encompasses the main driver of star formation activity: the molecular gas mass (\mol), i.e., the raw material out of which galaxies form stars.  Recent studies, by Daddi et al.\ 
(2010) and Genzel et al.\ (2010), found that normal galaxies at any redshift have 
lower star formation efficiencies ($\mathrm{SFE} \equiv \mathrm{SFR}$/\mol), compared to star-bursting systems that exhibit an accelerated mode of star formation activity, probably triggered by a major merger. These studies all suggest that there are two regimes of star formation: i) a long--lasting mode, followed by the majority of galaxies at every redshift that also form  a tight MS in the SFR$-$\ms ~and in the \lir$-L_{\rm 8}$ planes, and ii) a short--lived starburst mode for galaxies that can become strong outliers from both relations. However, this picture heavily relies on \mol ~estimates which are still a matter of significant debate due to the poorly determined conversion factor to derive molecular gas mass from CO luminosities (\aco = \mol/\lco), which is known to vary as a function of metallicity and, perhaps, of the intensity of the radiation field (e.g., Leroy et al.\ 2011, Magdis et al.\ 2011b).

Although the separation between starbursts and MS galaxies is already apparent in the observed \lir-\lco ~plane,
 the distinct nature of the star formation activity of the two populations becomes truly evident when different \aco ~factors are used for the 
 two classes of galaxies. This differentiation of the \aco ~value has clearly been seen in the nearby universe, with local ULIRGs (starbursts) having, a value of \aco\ that is a factor of $\sim$ 6 smaller than that for
local spiral galaxies\footnote{We note that Papadopoulos et al.\ (2012) presented evidence that  higher \aco ~values are possible in local ULIRGs} (e.g., Downes \& Solomon 1998), and seems to be true at high redshift too. Using a dynamical and a dust-to-gas mass ratio analysis of star-forming disks at $z \sim 1.5$, Daddi et al.\ (2010a) and Magdis et al.\ (2011b) argued for a CO conversion factor\footnote{The units of \aco, \msol pc$^{-2}$ (K km s$^{-1}$)$^{-1}$, are omitted from the text for brevity. Also, \aco ~estimates in this work account for the presence of Helium coexisting with the molecular hydrogen} \aco~$\sim$ 4.0 for this kind of objects, similar to that of the Milky Way and local spirals. On the other hand, several studies place an upper limit 
 of \aco ~$\sim$ 0.8 for several SMGs (e.g., Tacconi et al.\ 2008, Carilli et al.\ 2010, Magdis et al.\ 2011b, Hodge et al.\ 2012). Furthermore, recent numerical simulations indicate a clear distinction between the \aco ~values for disks and mergers at all redshifts, although with a considerable scatter (e.g., Narayanan et al.\ 2011, Feldmann et al.\ 2012). These findings challenge the common approach of ``blindly'' applying a (local) ULIRG-like \aco~$=0.8$ value to derive the \mol ~of high$-z$ ULIRGs. Instead, they highlight the necessity of determining \aco ~in larger samples of high$-z$ galaxies and of investigating how it varies between normal galaxies and mergers/starburst, but also among galaxies on the main sequence.
 
To this end, in Magdis et al.\ (2011b), we applied, for the first time at high redshift, a method for measuring \aco ~that is commonly used in the local universe. The method relies on measuring the total dust mass of a galaxy (\md) and assuming that it is proportional to \mgas ~(e.g., Leroy et al.\ 2011). Taking advantage of the detailed characterization of the peak of the SED enabled by \h ~data, as well as of the Rayleigh-Jeans tail from ground based mm observations, we acquired robust \md ~estimates for a normal disk at $z \sim1.5$ and a star-bursting SMG at $z = 4.05$. The dust mass estimates were 
 subsequently used to infer an  \aco ~value of $\sim$ 4.0 for the disk and an upper limit of $\sim$ 1 for the starburst galaxy, in agreement with previous dynamical estimates (Daddi et al 2010a, Hodge et al. 2012). This study also offered a first hint of a close link between the SED shape, as traced by the dust-
 ghted luminosity (\lir/\md) and \aco ~and SFE, with MS galaxies exhibiting lower \lir/\md ~values compared to SB galaxies.
 
Having demonstrated the applicability of the method at high redshift, here we extend this exercise to a  larger sample of main sequence disks at $z \sim 1.5$ and $z \sim 0.5$ that have extensive photometry from rest-frame UV to mm wavelengths, including state of the art \h ~data 
from the \h ~Great Observatories Origins Deep Survey (GOODS-\h, PI D.\ Elbaz) as well as low-transition CO observations and mm continuum interferometric measurements. We use this to investigate 
the average \aco ~value for high$-z$ MS galaxies, finding consistent values with what is predicted by the local \aco ~versus Z trend and what is inferred by dynamical analyses (Daddi et al.\ 2010b).

Pushing the method further, we here analyse large statistical samples of galaxies at $z \sim 1$ and $z \sim 2$  to derive their average dust and gas mass content. While the lack of CO measurements for such large samples
prevent us from learning anything on their conversion factor, the recovered information is invaluable to study the distribution of gas in high redshift galaxies.
 This allows us to investigate possible variations of gas fractions and SFE within the MS, providing insights about the origin of the ``thickness'' of the SFR-$M_{\ast}$ relation that has been shown to reflect the variation of the physical properties of MS galaxies (e.g., Salmi et al.\ 2012, Elbaz et al.\ 2011). Furthermore, the available data allow us to investigate the SED shape of MS galaxies as function of redshift and also as a function of their their offset from the main sequence. Put together, in this study we attempt to characterize the gas, dust and SED properties of main sequence galaxies throughout cosmic time and therefore
provide a better understanding of  the nature of the sources that dominate the star formation density at all redshifts (e.g., Rodighiero et al.\ 2011, Sargent et al.\ 2012).

This paper is organized as follows. In Section 2, we  describe our sample and the multi-wavelength observations. In Section 3 we present a detailed analysis of the far-IR properties. In Section 4 we discuss the method to derive \aco ~and present 
\aco ~and SFE estimates for MS and starburst galaxies. In Section 5, we explore the 
variations of SFE within the MS, and discuss two possible scenarios to explain the observed dispersion of the SFR-$M_{\ast}$ correlation. In Section 6 we investigate these two scenarios through stacking, while in Section 7 we explore the shape of the SED of MS galaxies as a function of cosmic time and build template SEDs of MS galaxies in various redshift bins. Finally, in Section 8, we provide a discussion motivated  by the results of this study, while a summary of the latter is presented in Section 9.  Throughout the paper we adopt $\Omega_{\rm m} = 0.3, ~H_{\rm 0} = 71$ km s$^{-1}$ Mpc$^{-1}$, $\Omega_{\rm \Lambda}$ = 0.7 and a Chabrier IMF (Chabrier et al.\ 2003).

\section{OBSERVATIONS AND SAMPLE SELECTION}
The aim of this study is to investigate the far-IR and gas 
properties of high$-z$, normal, main sequence galaxies. Instead of using a 
large, heterogeneous sample, we decided to focus on a small but well defined set of sources selected to meet the following criteria: 1) available spectroscopic redshifts; 2) rich rest-frame UV to mid-IR photometry; 3) no excess in the specific star formation rate, i.e., sSFR/sSFR$_{\rm MS}$ $<$ 3 (where sSFR$_{\rm MS}$ is the MS trend, as detailed below)  4) available [1-0] or [2-1] low-transition CO observations, to enable CO 
luminosity estimates, without the caveat of the uncertainties introduced by the excitation corrections affecting higher CO transition lines;  5) \h ~detections that offer a detailed sampling of the far-IR part of the SED; and, if possible, 6) mm continuum data that provide a proper 
characterization of the Rayleigh-Jeans tail, which is crucial for robust dust mass estimates. 

To derive a characteristic sSFR$_{\rm MS}$ at a given redshift and a given stellar mass, we define a main sequence, SFR$_{\rm MS}(z,M_{\ast})$, that varies with stellar mass with a slope of 0.81 (e.g., Rodrighero et al. 2011), and evolves with time  as $(1+z)^{2.95}$ (e.g., Elbaz et al. 2011, Pannella et al. 2009). In what follows we describe the \h ~observations used in this study, as well as the sample of galaxies considered here.


\subsection{\h ~ Data}
We use deep 100 and 160$\,\mu$m PACS and 250, 350, and 500$\,\mu$m SPIRE observations from the GOODS-\h\ program. Details about the observations are given in Elbaz et al.\ (2011). \h ~fluxes are derived from point-spread function (PSF) fitting using {\it galfit} (Peng et al.\ 2002). A very extensive set of priors, including all galaxies detected in the ultra-deep Spitzer Multiband Imaging Photometer (MIPS) 24$\,\mu$m imaging,  is used for source extraction and photometry at 100, 160 and 250$\,\mu$m, which effectively allow us to obtain robust flux estimates for relatively isolated sources, even beyond formal confusion limits at 250$\,\mu$m. For 350 and 500$\,\mu$m, this approach does not allow accurate measurements due to the increasingly large PSFs. Hence, we use a reduced set of priors based primarily on Very Large Array (VLA) radio detections, resulting in flux uncertainties consistent with the confusion noise at the IR wavelengths. Our measurements are in good agreement with the alternative catalogs used in Elbaz et al. (2011). The advantage of using {\it galfit} for PSF fitting is in its detailed treatment of the covariance matrix to estimate error bars in the flux measurements, which is crucial to eventually derive reliable estimate of flux errors for the case of blended/neighbouring sources. The effective flux errors in each band can vary substantially with position, depending on the local density of prior sources over areas comparable to the PSF.
A detailed description of the flux measurements and Monte Carlo (MC) derivations of the uncertainties will be presented elsewhere (E. Daddi et al., in preparation). We correct the PACS photometry for a 
small flux bias introduced (due to source filtering for background subtraction) during data reduction 
(see H-GOODS public data release\footnote{http://hedam.oamp.fr/GOODS-Herschel/index.php}; Popesso et al.\ in preparation).
\begin{table*}
{\footnotesize
\caption{PACS, SPIRE and 1.3mm photometry for MS galaxies in this study}             
\label{tab:4}      
\centering                          
\begin{tabular}{l c c c c c c c c c c c c}        
\hline\hline                 
Source &RA$^{1}$&DEC& $S_{100}$ & $S_{160}$ & $S_{250}$ & $S_{350}$ & $S_{500}$ & $S_{1.3mm}$ \\
& J2000&J2000&$mJy$ &$mJy$&$mJy$& $mJy$ &$mJy$&$mJy$\\
\hline
ID-8049&188.9751587& 62.1787071 & 10.7$\pm$0.7&20.1$\pm$1.5&23.0$\pm$1.5&23.7$\pm$5.2&12.5$\pm$4.5&-\\ 
ID-5819&189.1774139& 62.1594429 & 12.2$\pm$0.7&19.5$\pm$1.3&13.6$\pm$1.8&5.7$\pm$6.5&3.2$\pm$7.1 &-\\ 
ID-7691& 188.9502106& 62.1763153& 14.3$\pm$0.7&19.0$\pm$1.3 &  13.6$\pm$1.5& 7.6$\pm$6.5&  4.6$\pm$7.2 &   -\\
BzK-4171& 189.1106110& 62.1431656& 3.4$\pm$0.4 &10.3$\pm$1.1&15.1$\pm$3.1&7.5$\pm$5.6&7.0$\pm$5.8&0.04$\pm$0.14\\
BzK-12591& 189.4224091& 62.2142181&10.2$\pm$ 0.7 &19.3$\pm$1.4& 25.6$\pm$2.0& 19.7$\pm$4.4 &9.2$\pm$5.6& -\\
BzK-25536&189.3679962& 62.3152504 &0.8$\pm$0.7 &3.0$\pm$0.9&5.2$\pm$1.2&0.2$\pm$6.1&5.6$\pm$5.6&-\\
BzK-21000& 189.2942505& 62.3762665& 9.1 $\pm$0.5& 17.0$\pm$1.4& 25.3$\pm$4.4& 20.1$\pm$4.7& 11.6$\pm$7.5& 0.83$\pm$0.36\\
BzK-17999&189.4658661& 62.2556038&4.9$\pm$0.5&12.5$\pm$1.1&15.3$\pm$1.3&11.1$\pm$5.4&5.9$\pm$4.8&0.30$\pm$0.10\\
BzK-16000& 189.1253967& 62.2410736& 1.8$\pm$0.5&4.4$\pm$0.7&10.2$\pm$4.4&9.5$\pm$5.2&3.3$\pm$5.9&0.53$\pm$0.13\\
\hline   
\end{tabular}\\
Notes:\\
1: coordinates are from VLA 1.4 GHz continuum emission (Morrison et al. 2010). The VLA 1.4 GHz map has a resolution of 1.8'', and the typical position
accuracy for our sources is 0.20''}
\end{table*}

\subsection{A Sample of Main Sequence Galaxies at $z \sim 1.5$ and $z \sim 0.5$}
Daddi et al.\ (2010) presented PdBI CO[2-1] emission line detections of six star-forming galaxies at $z \sim 1.5$, originally selected by 
using the ``star-forming BzK'' color criterion (Daddi et al.\ 2004b). All 
six sources (BzK-21000, BzK-17999, BzK-4171, BzK-16000, BzK-12591 and BzK-22536) have spectroscopic redshifts (now confirmed by multiple CO detections; for optical redshifts see 
Stern et al.\ in preparation, Cowie et al.\ 2004) and robust PACS and/or SPIRE detections. Four of these sources also benefit 
from 1.3\ mm continuum data  derived as a by-product from the CO[5-4] emission line observations (Dannerbauer et al.\ 2012 in prep). For BzK-21000, continuum detections and uper limits are also obtained at 1.1, 2.2, and 3.3\ mm (Daddi et al.\ 2010a; Dannerbauer et al.\ 2009).  In addition to IRAC and MIPS 24$\,\mu$m data, the sources are seen in the 16$\,\mu$m InfraRed Spectrograph peak-up image (Teplitz et al.\ 2011). Although we will revisit (and confirm) the star formation rates of these sources based on the \h ~data, the existing UV, mid-IR and radio SFR estimates, along with stellar mass measurements derived by fitting the 
Bruzual \& Charlot (2003) model SEDs to their rest-frame UV to near-IR photometry (Daddi et 2010a), consistently place them in the SFR$-M_\ast$ main sequence at $z \sim1.5$ (Daddi et al.\ 2010). Finally, the UV rest-frame morphologies, the double-peaked CO profiles, the large spatial extent of their CO reservoirs, and the low gas excitation of the sources, all provide strong evidence that they are large, clumpy, rotating disks (Daddi et al.\ 2010a).

We also consider three $z \sim 0.5$ normal disks with S\'ersic index $n < 1.5$ and spectroscopic redshifts, for which we have CO[2-1] emission line detections (Daddi et al.\ 2010b). Similar to the $z \sim1.5$ sample, 
the sources are detected in the PACS and/or SPIRE bands and are part of the SFR$-M_{\ast}$ main sequence based on their 24$\, \mu$m--derived IR luminosities and SFRs, which are known to be robust in this redshift range (Elbaz et al.\ 2010, 2011). The \h ~and millimetre photometry of the main sequence galaxies considered here are summarised in Table 1.

\subsection{A Sample of High$-z$ SMGs}
As a comparison sample to our MS galaxies, we also 
include in our analysis a small sample of SMGs: GN20, SMMJ2135-0102 and  HERMES J105751.1+573027. The selection of the targets was driven by the need for available multi-wavelength photometry, including \h ~and mm continuum observations, as well as CO[1-0] emission line detections. 

GN20 was already studied in Magdis et al.\ (2011). It is one of the best-studied SMGs to date, the most luminous and also one of the most distant ($z = 4.05$, Pope et al.\ 2006, Daddi et al.\ 2009) in the GOODS-N field. \h ~photometry and the far-IR/mm properties of the source have already been presented in detail in Magdis et al.\ (2011b). In brief, the source is detected in all \h ~bands (apart from 100$\,\mu$m) and in the AzTEC 1.1 mm map (Perera et al.\ 2008) while continuum emission is also measured at 2.2, 3.3, and 6.6\,mm (Carilli et al.\ 2011) as well as at 1.4 GHz with the VLA (Morrison et al.\ 2010). Furthermore, Carilli et al.\ (2010) reported the detection of the CO[1-0] and CO[2-1] lines with the VLA, and CO[6-5] and CO[5-4] lines with the Plateau de Bure Interferometer (PdBI) and the Combined Array for Research in Millimeter Astronomy (CARMA), respectively.  A compilation of the photometric data is given in Table 1 of Magdis et al.\ (2011b).

SMMJ2135-0102 (SMM-J2135 hereafter) is a highly magnified SMG at $z = 2.325$, with an amplification factor of $\mu$ = 32.4 $\pm$ 4.5, serendipitously discovered by Swinbank et al.\ (2010) behind the  cluster MACS J2135-01. The source has exquisite photometric coverage from rest-frame UV to radio 
 including SPIRE broadband data (Swinbank et al.\ 2010 Table 1 and  Ivison et al.\ 2010 Table 1). Swinbank et al.\ (2010) also report the detection of  CO[1-0] and CO[3-2] emission lines using the Zpectrometer on the Green Bank Telescope and  PdBI observations, while Ivison et al.\ (2010), using the SPIRE Fourier Transform Spectrometer (FTS), presented the detection of the [CII]158$\,\mu$m cooling line.

Finally, HERMES J105751.1+573027 (HSLW-01 hereafter), is a \h/SPIRE-selected galaxy at z=2.957, multiply-lensed (magnification factor $\mu=10.9 \pm 0.7$) by a foreground group of galaxies. The source was discovered in Science Demonstration Phase \h/SPIRE observations of the Lockman-SWIRE field as part of the Herschel Multi-tiered Extragalactic Survey (HerMES; S. Oliver et al.\ 2012). The optical to mm photometry of the source is presented in Table 1 of Conley at al. (2011), while Riechers et al.\ (2011)  reports the detection of  CO[5-4], CO[3-2], and CO[1-0] emission using the PdBI and CARMA and the Green Bank Telescope. 

We note that due to the high magnification factor for SMM-J2135 and HSLW-01, these sources might 
not be representative of the bulk population of SMGs, traditionally selected as $S_{\rm 850} > 5$ mJy. Also, a large number of non-lensed SMGs with CO[3-2] observations is available in the literature. However,  we choose not to consider them in this study, as the uncertain gas excitation 
results in dubious CO[1-0] estimates (e.g., Ivison et al.\ 2010, Riechers et al. 2011), which are essential for investigating the gas properties of the sources. 

\subsection{A Statistical Sample of $z \sim 1$ and $z \sim 2$ MS Galaxies}
In addition to our samples of individually-detected MS galaxies at $z \sim 1.5$ and $z \sim 0.5$, we also perform a stacking analysis to derive the average SED of MS galaxies at $z \sim 1$ and $z \sim 2$. Specifically, we use the $z \sim 1$ GOODS samples of Salmi et al.\ (2012; see also Daddi et al. 2007 and Pannella et al.\ in prep). In order to remove bulge-dominated galaxies with low sSFR, 
we discard objects with a S\'ersic index $n > 1.5$ (Salmi et al.\ 2012), as measured from the GOODS ACS images (Giavalisco et al.\ 2004), and consider only 24$\mu$m detected galaxies from the GOODS ultra-deep \s ~imaging. To remove starbursts we proceed as follows: i) from $M_{\ast}$ and redshift, we compute sSFR$_{\rm MS}$, i.e., the fiducial mean sSFR expected for MS galaxies with a given mass and redshift, derived as described at the start of \S 2 ii) from \h ~data we derive \lir ~and subsequently SFR estimates, and exclude galaxies with measured sSFR/sSFR$_{\rm MS}$ $>$ 3. 
Since not all starburst systems are expected to be detected by \h ~we also omit sources with sSFRs/sSFR$_{\rm MS}$ $>$ 3 using the UV-based, corrected for extinction, SFR estimates by Daddi et al.\  (2007) and the \s ~estimates of Salmi et al.\ (2012). The $z \sim 2$ sources are drawn from the BzK--selected sample of Daddi et al.\ (2007; now extended to deeper K-band depths by Pannella et al.\ in prep.). Since this color-selection technique naturally excludes quiescent galaxies, we only exclude starbursts identified using the same method as for the $z \sim 1$ sample. The total number of $z \sim 1$ and $z \sim 2$ galaxies considered in the stacking analysis is 1569 and 3618, respectively. The average SEDs of these two samples are then measured  by stacking, using various standard techniques  depending on the wavelength:

\begin{itemize}
\item 16$\,\mu$m: We use the Teplitz et al.\ (2011) maps. The stacking is performed using the IAS stacking library (B\'ethermin et al.\ 2010a) and PSF-fitting photometry.
\item 24 and 70$\,\mu$m: We use the 24$\,\mu$m and 70$\,\mu$m (Frayer et al.\ 2006b) maps of GOODS, the IAS stacking library and aperture photometry with parameters similar to that in B\'ethermin et al.\ (2010a).
\item 100 and 160$\,\mu$m: We use the GOODS-\h\  PACS data (Elbaz et al.\ 2011), the IAS stacking library and PSF-fitting. We apply the appropriate flux correction for faint, non masked, sources to the PACS stacks (Popesso et al.\ in preparation).
\item 250, 350 and 500\,$\mu$m: We use the GOODS-\h ~SPIRE (Elbaz et al.\ 2011) data and the mean of pixels centred on sources as in B\'ethermin et al.\ (2012a). The bias due to clustering is estimated to be about 20\% at 500\,$\mu$m and thus negligible compared with the statistical uncertainties. This bias is smaller at shorter wavelengths where the beam is narrower (B\'ethermin et al.\ 2012a).
\item 870\,$\mu$m and 1100\,$\mu$m: We used the Wei{\ss} et al. (2009) LABOCA map of eCDFS and the Perera et al. (2008) AzTEC map of GOODS-N. Contrary to SPIRE, (sub-)mm data are noise-limited, it is thus optimal to beam-smooth the map before stacking (Vieira et al. in prep.). We thus apply the Marsden et al. (2009) method to perform our stacking, which takes the mean of pixels centered on sources in the beam-smoothed map.
\end{itemize}

At all wavelengths, we used a bootstrap technique to estimate the uncertainties (Jauzac et al.\ 2011), and the mean fluxes measured in the two fields are combined quadratically to produce a mean SED at $z\sim 1$ and $z\sim 2$. For the $z \sim 2$ normal galaxies we also construct average SEDs in three stellar mass bins in the range of $9.7 < \log(M_{\ast}/M_{\odot}) <11.2$ with a bin size of $\Delta \log(M_{\ast}/M_{\odot}) = 0.5$, 
as well as in four bins of sSFR/sSFR$_{\rm MS}$ over the range $-0.4 < \log(\mathrm{sSFR/sSFR_{MS}}) < 0.4$, with a bin size of $\Delta \log(\mathrm{sSFR/sSFR_{MS}}) = 0.2$.

\section{Derivation of Far-IR Properties}
Several key physical properties of distant galaxies, such as infrared luminosities (\lir), dust temperatures (\td) and dust masses (\md),  can be estimated by fitting their mid-to-far-IR SEDs with various models and templates 
calibrated in the local Universe. However,  the lack of sufficient data for a proper characterization of  the SED of distant galaxies has often limited this kind of analysis  
to models suffering from (necessarily) over-simplified assumptions and large generalizations.  
The \s, \h, and millimeter data available for the galaxies in our high$-z$ sample 
provide thorough photometric sampling of their SEDs, allowing the use of more realistic models that 
previously have been applied mainly in the analysis of nearby galaxies.
Here we consider both the physically-motivated Draine \& Li (2007) (DL07 hereafter) models, as well as the more simplistic, but widely used, modified black body model (MBB).

\subsection{The Draine \& Li 2007 Model}  
We  employ the dust models of DL07, which constitute an update of those developed by Weingartner \& Draine (2001) and Li \& Draine (2001), and which were successfully applied to the integrated photometry of the Spitzer Nearby Galaxy Survey (SINGS) galaxies (Draine et al.\ 2007). These models describe the interstellar dust as a mixture of carbonaceous  and amorphous silicate grains, whose size distributions are chosen to mimic the observed extinction law in the Milky Way (MW), the Large Magellanic Cloud (LMC), and the Small Magellanic Cloud (SMC) bar region. The properties of these grains are parametrized by the PAH index, $q_{\rm PAH}$, defined as the fraction of the dust mass in the form of PAH grains. The majority of the dust is supposed to be located in the diffuse ISM, heated by a radiation field with a constant intensity $U_{min}$. A smaller fraction $\gamma$ of the dust is exposed to starlight with intensities ranging from $U_{min}$ to $U_{max}$, representing the dust enclosed in photo-dissociation regions (PDRs).  Although this PDR component contains only a small fraction of the total dust mass, in some galaxies it contributes a substantial fraction of the total power radiated by the dust. Then, according to 
DL07, the amount of dust $d$\md ~exposed
to radiation intensities between $U$ and $U + dU$ can be expressed as a 
combination of a $\delta$-function and a power law:\\
\begin{equation}
\begin{centering}
{\rm \frac{dM_{\rm dust}}{dU} = (1-\gamma) M_{\rm dust} \delta(U-U_{\rm min})+  \gamma M_{\rm dust} {\alpha-1 \over U_{\rm min}^{1-\alpha} - U_{\rm max}^{1-\alpha}} U^{-\alpha}}
\end{centering}
\end{equation}
\noindent with ($U_{min} \le U_{max}, \alpha \ne 1$), \md ~the total dust mass, $\gamma$ the fraction  of the dust mass that is associated with the power-law part of the starlight intensity distribution, and  $U_{min} , U_{max} , \alpha$  characterizing the distribution of starlight intensities in the high-intensity regions. 
\begin{figure*}
\centering
\includegraphics[scale=0.6]{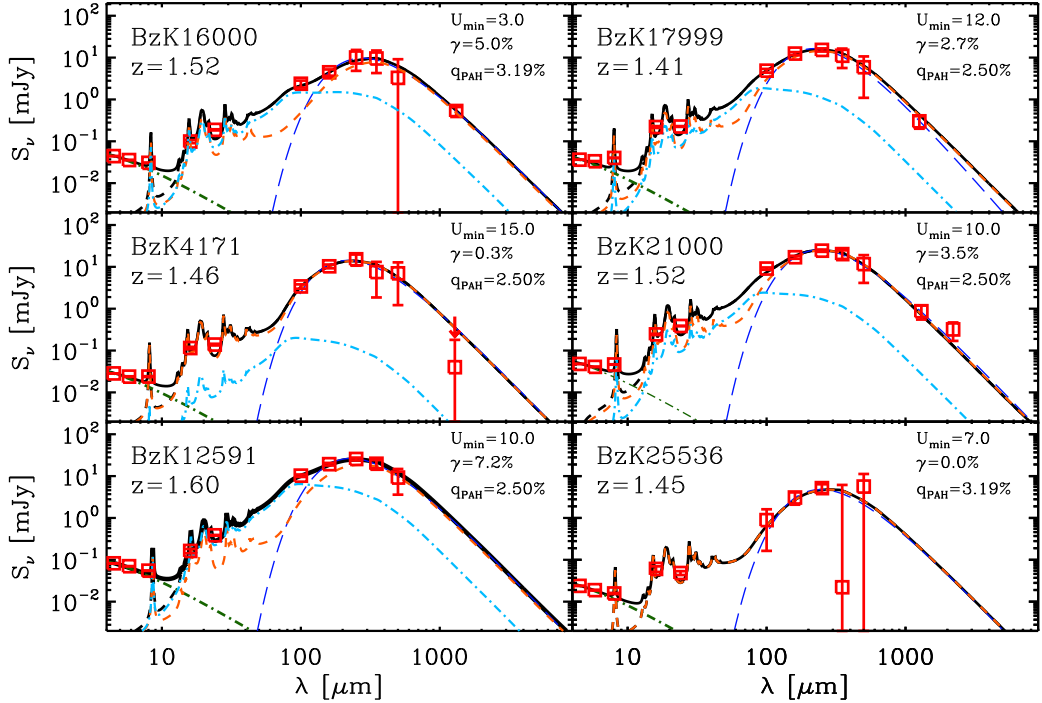}
\caption{Observed SEDs of $z\sim1.5$ BzK galaxies, overlaid with the best fit Draine \& Li (2007) models (black) and the best fit single-temperature modified blackbody (blue). The black dashed line is the DL07 model without the stellar component that is depicted with a green dotted-dashed line. Orange dashed and cyan dot-dashed lines show separate contributions of starlight and emission from dust heated by $U=U_{min}$ (diffuse ISM component) and dust heated by $U_{min} < U < U_{max}$ (``PDR'' component) respectively. The fitted parameters from the best fit Draine \& Li (2007) model fits are listed within each panel.}
\label{fig:sed1} %
\end{figure*}

\begin{figure*}
\centering
\includegraphics[scale=0.45]{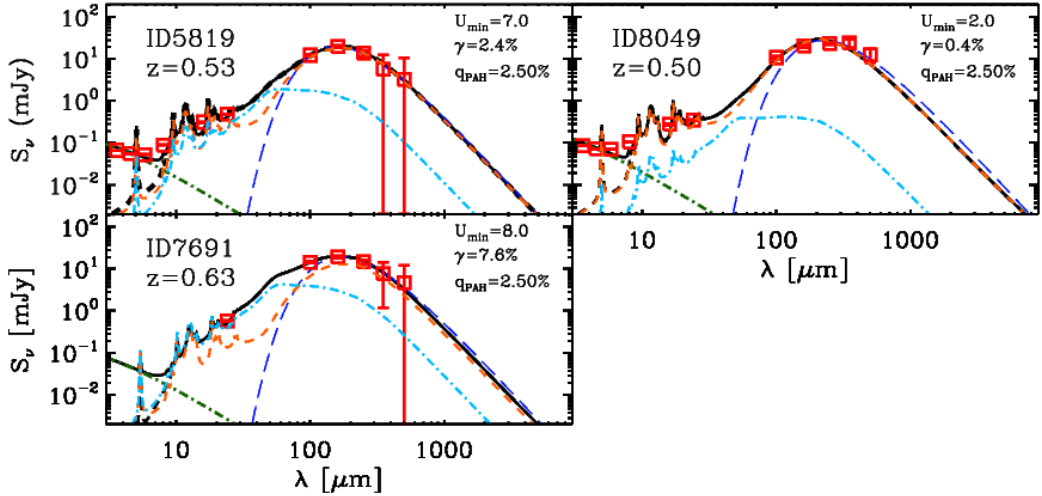}
\caption{Same as Figure 1, but for the $z\sim0.5$ galaxies in our sample.}
\label{fig:sed2} %
\end{figure*}
\begin{figure*}
\centering
\includegraphics[scale=0.45]{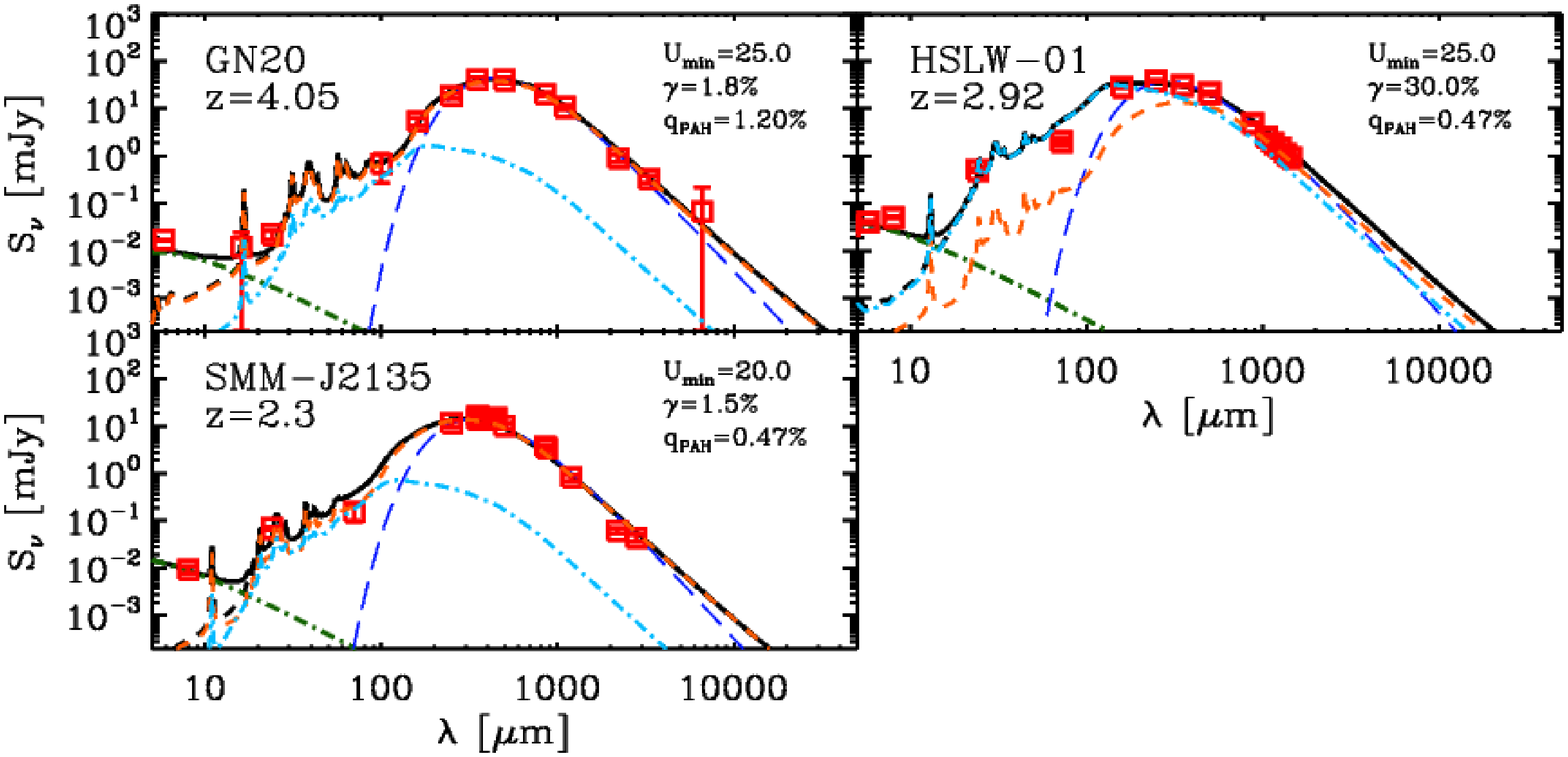}
\caption{Same as Figure 1, but for the SMGs in our sample.}
\label{fig:sed3} %
\end{figure*}

Following DL07, the spectrum of a galaxy can be described by a linear combination of one stellar component approximated by  a  blackbody with color temperature $T_*$ = 5000K, and two dust components, one arising from dust in the diffuse ISM, heated by a minimum radiation field $U_{min}$ (``diffuse ISM'' component), and 
one from dust heated by a power-law distribution of starlight, associated with the intense  photodissociation regions (``PDR'' component). Then, the model emission spectrum of a galaxy at distance $D$ is:
\begin{eqnarray}
{\rm f_{\nu}^{\rm model} = \Omega_* B_\nu(T_*) + {M_{\rm dust} \over 4 \pi D^2} \left[ (1-\gamma) p_\nu^{(0)}+ \gamma p_\nu \right]},
\end{eqnarray}
\noindent where $\Omega_*$ is the solid angle subtended by stellar photospheres, $p_\nu^{(0)}= p_\nu^{(0)}(q_{\rm PAH}, U_{\rm min})$ and $ p_\nu =p_\nu(q_{\rm PAH}$ $,U_{\rm min},U_{\rm max},\alpha)$ are the emitted power per unit frequency per unit dust mass for dust heated by a single starlight intensity $U_{\rm min}$, and dust heated by a power-law distribution of starlight intensities $dM/dU \propto U^{-\alpha}$ extending from $U_{\rm min}$ to $U_{\rm max}$. 

In principle, the dust models in their most general form
are dictated by seven free parameters, ($\Omega_{*},  q_{\rm PAH}, U_{\rm min}, U_{\rm max}, \alpha, \gamma$ and $M_{\rm d}$). However, Draine et al.\ (2007) showed that the overall fit is insensitive to the adopted dust model (MW, LMC and SMC) and the precise values of $\alpha$ and  $U_{max}$. In fact they showed that fixed values of  $\alpha=2$ and $U_{max}=10^6$ successfully described the SEDs of galaxies with 
a wide range of properties. They also favor the choice of MW dust models for which a set of models with $q_{PAH}$ ranging from 0.4\% to 4.6\% is available. Furthermore, since small $U_{min}$ values correspond to dust temperatures below $\sim15$ K that cannot be constrained by far-IR photometry alone, in the absence of rest-frame submm data,  they suggest using 0.7 $\le U_{min} \le$ 25. While this lower cutoff 
for $U_{min}$ prevents the fit from converging to erroneously large amounts of cold dust heated by weak starlight ($U_{min}$ $<$ 0.7), the price to pay is a possible underestimate of the total dust mass if large amounts of cold dust are indeed present. However, Draine et al.\ (2007)  concluded that omitting submm data from the fit increases the scatter of the derived masses up to 50\% but does not introduce a systematic bias in the derived total dust masses.

Under these assumptions, we fit the mid-IR to mm data points of each galaxy in our sample, searching for the best fit model by $\chi^2$ minimization and parametrizing  the goodness of fit by the value of the  reduced $\chi^2$, $\chi^2_\nu \equiv \chi^2 / df $ (where $df$ is the number of degrees of freedom). The best fit model yields a total dust mass (\md), $U_{min}$, $\gamma$ and $q_{\rm PAH}$ while to derive \lir\footnote[1]{$L_{\rm dust}$ quoted below is similar to $L_{\rm IR}$, but integrated from 0 to $\infty$} ~estimates we integrate the emerging SEDs from 8- to 1000$\mu$m:
\begin{equation}
  L_{\rm IR} = \int_{8\mu m}^{1000\mu m}  L_\nu (\lambda) \times \frac{c}{\lambda^{2}}~\rm d\lambda.
  \label{eq:LIR}
\end{equation}

A by-product of the best fit model is also the 
dust weighted mean starlight intensity scale factor, $\langle U \rangle$, defined as:
\begin{equation}
  \langle U\rangle = 
  \left\{
  \begin{array}{ll}
    \displaystyle
\frac{L_{\rm dust}}{P_{0} M_{dust}}  ~~~~~~or\\
\displaystyle
\left[(1-\gamma)U_{min}+\frac{\gamma \ln(U_{max}/U_{min})}{U_{min}^{-1}-U_{max}^{-1}}\right], ~~~for ~\alpha=2 
  \end{array}
  \right.
  \label{eq:avU}
\end{equation}
\noindent where $P_{0}$ is the power absorbed per unit dust mass in a radiation field with $U=1$. Note that essentially $\langle U \rangle$ is proportional to \lir/\md, and as we will discuss later for the definition of \lir ~adopted here, i.e. $L_{\rm 8-1000\,\mu m}$, our data suggest that  $P_{0}$ $\approx$125. Uncertainties in \lir ~and \md\ are quantified using Monte Carlo simulations. 
To summarise, for each galaxy a Gaussian random number generator was used to create 1000 artificial flux sets from the original fluxes and measurement errors. These new data sets were then fitted in the same way, and the standard deviation in the new parameters was taken to represent the uncertainty in the parameters found from the real data set. Best fit values along with 
their corresponding uncertainties are listed in Table 2 for all sources in our sample. To check for possible 
contamination of the submm continuum broad band photometry from C+ (158$\,\mu$m) emission (e.g., Smail et al.\ 2011), we repeated the fit excluding the affected bands (i.e., 350$\,\mu$m and 500$\,\mu$m for $z \sim $0.5 and $z \sim1.5-2.0$ respectively without noticing any effect in the derived parameters.The best fit models along with the observed SEDs of the $z \sim1.5$,  $z \sim 0.5$ galaxies and SMGs are shown in Figures \ref{fig:sed1}, \ref{fig:sed2} and \ref{fig:sed3} respectively.

\begin{table*}
{\footnotesize
\caption{Physical properties of individually detected sources, based on Draine \& Li 2007 Models }             
\label{tab:4}      
\centering                          
\begin{tabular}{l c c c c c c c c c c c c}        
\hline\hline                 
Source & $z_{\rm spec}$&$\chi^2_{\nu}$ & $\log L_{\rm IR}$ & $\log M_{\rm d}$ (DL07) & $U_{\rm min}$ & $\gamma$ & $q_{PAH}$& $\langle U \rangle$& $T_{\rm d}$ $^{a}$ & $\beta$ $^{a}$\\

&  && $L_\odot$ & M$_\odot$ &  &\%&\%&&K&\\
\hline
ID-8049 &0.507& 2.1 & 11.22$\pm$0.02 & 8.81$\pm$0.13 & 2.0 & 0.4 &2.50&2.0& 22$\pm$2 & 1.5  $^{b}$\\ 
ID-5819 &0.530& 1.78 & 11.26$\pm$0.01 & 8.22$\pm$0.10 & 7.0 & 2.4 &2.50&8.8& 31$\pm$1 & 1.5 $^{b}$\\
ID-7691 &0.637& 0.65 & 11.54$\pm$0.03 & 8.28$\pm$0.10 & 8.0 & 7.6 &2.50&14.5& 31$\pm$2 & 1.5 $^{b}$\\
BzK-4171 &1.465& 1.56 & 11.98$\pm$0.04 & 8.70$\pm$0.09 & 15.0 & 0.3 &2.50&15.4& 37$\pm$2 & 1.5 $^{b}$  \\
BzK-12591 &1.600& 1.04 & 12.44$\pm$0.02 & 9.09$\pm$0.09 & 10.0 & 7.2 &2.50&17.6& 37$\pm$1 & 1.5 $^{b}$\\
BzK-25536 &1.459& 1.94 & 11.46$\pm$0.06 & 8.52$\pm$0.26 & 7.0 & 0.0 &3.19&7.0& 33$\pm$3 & 1.5 $^{b}$\\
BzK-21000 &1.523& 1.12 & 12.32$\pm$0.01 & 9.07$\pm$0.06 & 10.0 & 3.5 &2.50&13.7& 35$\pm$2 & 1.4$\pm$0.2\\
BzK-16000 &1.522& 1.71 & 11.87$\pm$0.03 & 9.11$\pm$0.07 & 3.0 & 5.0 & 3.19 & 4.7& 30$\pm$1 & 1.5$\pm$0.2\\
BzK-17999 &1.414& 0.86 & 12.06$\pm$0.02 & 8.78$\pm$0.07 & 12.0 & 2.7 &2.50&15.3& 33$\pm$1 & 1.9$\pm$0.2  \\ 
GN20 &4.055& 1.44 & 13.25$\pm$0.06 & 9.67$\pm$0.06 & 25.0 & 1.8 &1.20&29.6& 33$\pm$2 & 2.1$\pm$0.2 \\ 
SMM-J2135 $^{c}$ &2.325& 3.41 & 12.30$\pm$0.06 & 8.85$\pm$0.06 & 20.0 & 1.5 &0.47&21.9& 30$\pm$1 & 2.0$\pm$0.2 \\
HSLW-01 $^{c}$ &2.957& 6.61 & 13.13$\pm$0.06 & 8.93$\pm$0.08 & 25.0 & 30.0 &0.47&134.0& 39$\pm$2 & 2.1$\pm$0.2\\
Stack-z1 $^{d}$&0.98& 3.91 & 11.21$\pm$0.09 & 8.14$\pm$0.13 & 8.0 & 2.0 &3.90&9.7& 32$\pm$2 & 1.5$^{b}$\\
Stack-z2 $^{d}$&1.97& 2.01 & 11.55$\pm$0.06 & 8.27$\pm$0.11 & 12.0 & 2.5 &3.19&15.1& 38$\pm$2 & 1.5$^{b}$\\
\hline                                   
\end{tabular}\\
Notes:\\
a: Derived based on single temperature MBB\\
b: $\beta$ fixed to this value in the MBB fit\\
c: Values corrected for magnification\\
d: Best fit values accounting for  the redshift distribution of the sample\\ }
\end{table*}

\subsection{The Importance of Millimeter Data}
While \h ~data accurately probe the peak of the SED in the far-IR emission of distant galaxies, rest-frame submm observations ($\lambda_{\rm rest} \ge 250\,\mu$m) are necessary to sample the Rayleigh-Jeans tail. Since the available far-IR photometry for five of our sources is 
restricted to \h ~observations, it is important to explore possible biases or systematics introduced by the lack of mm continuum data. Given the ever increasing number of distant galaxies with  
\h ~photometry, this investigation will also provide guidance for similar studies in the future. To assess the significance of adding mm photometry  in the derivation of the far-IR properties of the galaxies, and particularly their dust mass, we repeat the fitting procedure, this time excluding any data at wavelengths longer than rest-frame 200$\,\mu$m, for the four $z \sim 1.5$ BzK and the three SMGs with available mm continuum observations. 

A comparison between the derived dust masses with and without 
the use of mm data is shown in Figure \ref{fig:mmnomm} (left). We find no evidence for a strong systematic bias, with an average ratio between the two estimates $\langle M^{mm}_{\rm dust}/M^{nomm}_{\rm dust} \rangle = 0.80 \pm 0.2$. Interestingly, the sources with the largest discrepancies are the SMGs, although the sample is too small to clearly demonstrate the existence of a systematic effect. However, the addition of mm data has a noticeable impact on the uncertainties of the derived \md ~estimates, which are reduced, on average, by a factor of $\sim 2$. Studies in the local universe reach similar conclusions, with the addition that in the absence of rest-frame sub-mm data,  dust masses tend to be underestimated for metal-poor galaxies. This could serve as an indirect indication that our sample mainly consists of metal-rich sources, something that we will also argue in Section 4.

The derived $\gamma$ and $U_{\rm min}$ are also in broad agreement between the two cases.  
However, we notice  a weak systematic bias towards higher $\langle U \rangle$ values, i.e., stronger mean radiation fields, when mm data are considered in the fit, reflecting the fact that $\langle U \rangle \propto$ \lir/\md. 
We conclude that such trends in $\langle U \rangle$ and \md\ suggest that mm data can place better constraints on the diffuse ISM emission, as well as on the  relative contribution of a PDR component to the total radiation field, and consequently on the \md ~of individual high$-z$ galaxies.  Finally, we note that the detailed sampling of the peak of the far-IR emission provided by \h ~data can result in robust  \lir ~estimates without the need of mm data.
\begin{figure*}
\centering
\includegraphics[scale=0.36]{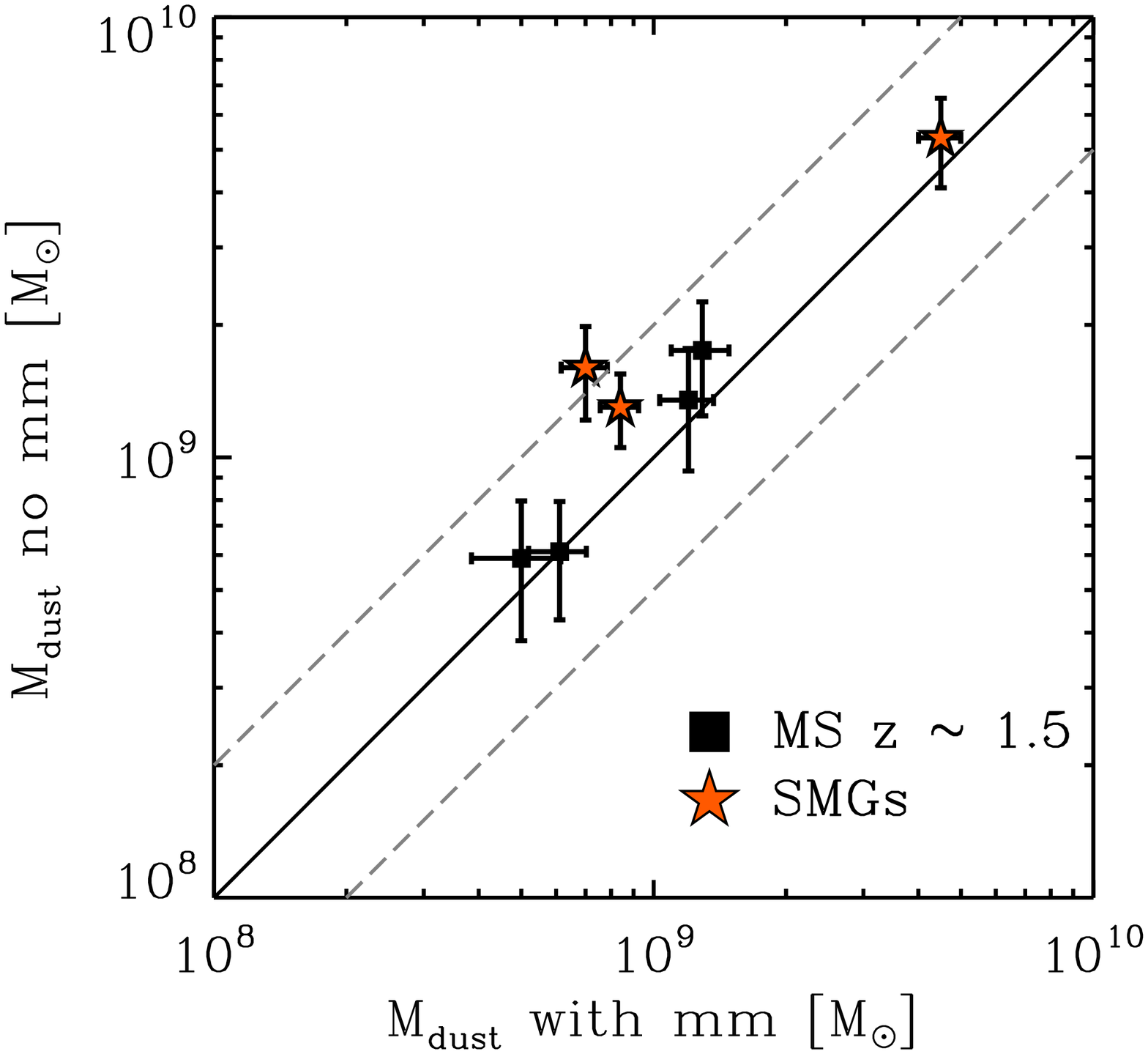}
\includegraphics[scale=0.36]{{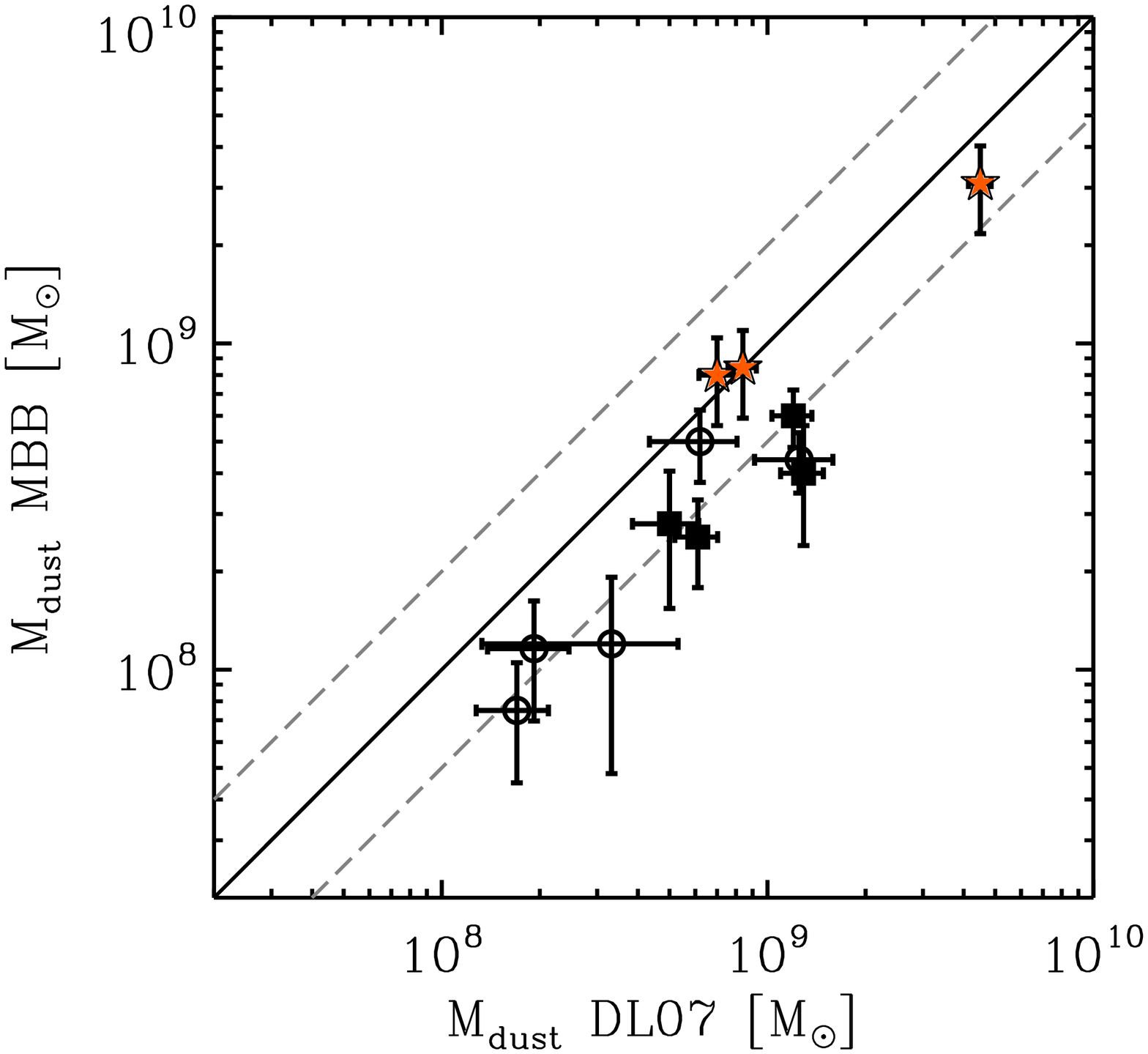}}
\caption{{\bf Left:} The significance of the addition of mm continuum data for the derivation of dust masses. A comparison between the DL07 \md ~estimates inferred with and without the addition of mm data to the fit. Although, the two estimates seem to be overall in good agreement, we note that in absence of mm continuum data we tend to overestimate the \md ~in SMGs. {\bf Right:} Comparison between dust masses derived based on a single temperature 
modified black body (MBB) and those derive by the Draine \& Li (2007) models. Filled squares and open circles indicate sources with and without mm data. For the former $\beta$ was treated as a free parameter in the MBB fit while for the latter it was fixed to $\beta = 1.5$. Orange stars represent the SMGs considered in this study. In both panels the black solid line corresponds to unity  and the two grey dashed lines its offset by a factor of 2 and 0.5. }
\label{fig:mmnomm} %
\end{figure*}

\subsection{Comparison With Modified Black Body Fits}
Another method for deriving estimates for the dust masses and other far-IR properties such as dust temperatures and dust emissivity indices ($\beta$), is to fit the far-IR to submm SED of the galaxies with a single-temperature modified blackbody, expressed as:
\begin{equation}
  f_{\rm \nu} \propto \frac{\nu^{3+\beta}}{e^{\frac{h\nu}{kT_{d}}}-1}
\end{equation}
\noindent where \td ~is the effective dust temperature and $\beta$ is the effective dust emissivity index.
Then, from the best fit model, one can estimate  \md ~from the relation:  
\begin{equation}
 M_{d} = \frac{S_{\nu}D^{2}_{L}}{(1+z)\kappa_{\rm rest}B_{\nu}(\lambda_{rest},T_{d})}, ~~with~~  \kappa_{\rm rest}=\kappa_{\rm 0}(\frac{\lambda_{o}}{\lambda_{rest}})^{\beta}
\end{equation} 
\noindent where $S_{\rm \nu}$ is the observed flux density, $D_{\rm L}$ is the luminosity distance, and
 $\kappa_{rest}$ is the rest-frame dust mass absorption coefficient at the observed wavelength. 
While this is a rather simplistic approach, mainly adopted due to the lack of sufficient sampling of the SED
of distant galaxies, it has been one of the most widely-used methods in the literature. Therefore, an analysis based on MBB-models not only provides estimates of the effective dust temperature and dust emissivity of the galaxies in our sample, but also a valuable comparison between dust masses inferred with the MBB and DL07 methods.

We fit the standard form of a modified black body, considering observed data points with $\lambda_{rest} > 60 \,\mu m$, to avoid emission from very small grains that dominate at shorter wavelengths. For the cases where mm 
data are available we let $\beta$ vary as a free parameter, while for the rest we have assumed a fixed value of $\beta = 1.5$, typical of disk-like, main sequence galaxies (Magdis et al.\ 2011b, Elbaz et al.\ 2011). 
From the best fit model, we then estimate the total \md ~through equation 6. For consistency with the DL07 models we adopt a value of $\kappa_{250\mu m}$= 5.1 cm$^{2}$ g$^{-1}$ (Li \& Draine 2001). To obtain the best fit models and the corresponding  uncertainties of the parameters, we followed  the same procedure as for the DL07 models. The derived parameters are summarized in Table 2 and the best fit models are shown in Figures 1, 2 and 3. 

A comparison between dust masses derived by DL07 and MBB models in shown in Figure \ref{fig:mmnomm} (right). We see that a modified black body appears to infer dust masses that are lower than those derived based on DL07 models on average by a factor of $\sim 2$ ($\langle M^{DL07}_{\rm dust}/M^{MBB}_{\rm dust}\rangle = 1.96 \pm 0.5$). A similar trend was recently reported by Dale et al.\ (2012), who found that the discrepancy between the two dust mass estimates is smaller for sources with warmer $S_{70}/S_{160}$ colors. Interestingly, when we convolve the best fit DL07 rest-frame SEDs of our galaxies with the 70- and 160$\,\mu$m PACS filters, we find that the sources with the warmest $S_{70}/S_{160}$ are indeed those with the best agreement in the dust masses derived from the two methods. The reason behind 
this discrepancy has been addressed by Dunne et al.\ (2000). The grains of a given size and material are exposed to different intensities of the interstellar radiation field  and thus attain different equilibrium temperatures which will contribute differently to the SED. The single-temperature models cannot account for this range of \td ~in the ISM and attempt to simultaneously fit both the Wien side of the grey body, which is dominated by warm dust, as well as the Rayleigh-Jean tail, sensitive to colder dust emission. The net effect is that the temperatures are driven towards higher values, consequently resulting in lower dust masses. While restricting the fit to $\lambda \le 100\,\mu$m does not provide an adequate solution  to the problem (Dale et al.\ 2012), a two-temperature black body fit is more realistic and returns dust masses that are larger by a factor of $\sim 2$ compared to those derived based on a single \td ~MBB (Dunne \& Eales 2001), in line with those inferred by the DL07 technique.

Despite the large uncertainties associated with the MBB analysis, it is  worth noting that for all SMGs in our sample we measure an effective dust emissivity index $\beta \sim 2.0$, in agreement with the findings of Magnelli et al.\ (2012). On the other hand, the far-IR to mm SEDs of all BzK galaxies  are best described by  $\beta \sim 1.5$, similar to that found by Elbaz et al.\ (2011) for MS galaxies.  However, we stress that the effective $\beta$ does not necessarily reflect the intrinsic $\beta$ of the dust grains due to the degeneracy between $\beta$ and the temperature distribution of dust grains in the ISM. The $\beta$-\td ~degeneracy also prevents  a meaningful comparison between the derived \td ~for  the galaxies in our sample.

\section{ The Gas Content of Main Sequence and Star-bursting Galaxies}
Several studies in the local Universe have revealed that  the total gas-to-dust mass ratio (\gdr) 
 is correlated with the gas-phase oxygen abundance, in the sense that more metal-rich galaxies 
 tend to exhibit lower \gdr ~values (e.g., Mu{\~n}oz-Mateos et al.\ 2009, Leroy et al.\ 2011 and references within). In addition to crucial information regarding the amount of metals trapped in the dust, this 
 correlation can serve as a valuable tool for deriving indirect estimates of the CO to molecular gas mass (\mol) ~conversion factor. In particular, if we know the metallicity, the dust mass and the mass of the atomic gas mass ($M_{\rm HI}$) of a galaxy, then we can estimate the \gdr ~through the \gdr$-Z$ relation, and \mol ~through the following relation:
\begin{equation}
\delta_{ \rm GDR} M_{\rm dust} \equiv M_{\rm gas} = M_{\rm H_{2}} + M_{\rm HI}
\end{equation}
\noindent Subsequently, if \lco ~is measured then we can estimate \aco, since \mol= \aco ~$\times$ \lco. This method has successfully been applied  in the local Universe (e.g., Leroy et al.\ 2011) and has recently been extended to high$-z$ galaxies by the pilot study of Magdis et al.\ (2011b). Even though restricted to only two galaxies, which are also part of this study (GN20 and BzK21000), Magdis et al.\ (2011b) showed that this approach provides \aco ~estimates that are consistent with other independent measurements and dynamical arguments.  
Although it has been a common practice to adopt a value of \aco~= 0.8, typical of local ULIRGs, for high$-z$ ULIRGs, 
our earlier study suggested that such low values apply only to the subset of genuinely star-bursting systems at high redshift, while high$-z$ main sequence galaxies, even those with $L_{\rm IR} > 10^{12} L_\odot$ (i.e., those classified as ULIRGs) have larger \aco ~values, similar to those of local spirals.  Here, we wish to extend this experiment, this time using a substantially larger sample, and to investigate possible correlations between \aco ~and starburst versus main sequence indicators. 
\begin{figure*}
\centering
\includegraphics[scale=0.28]{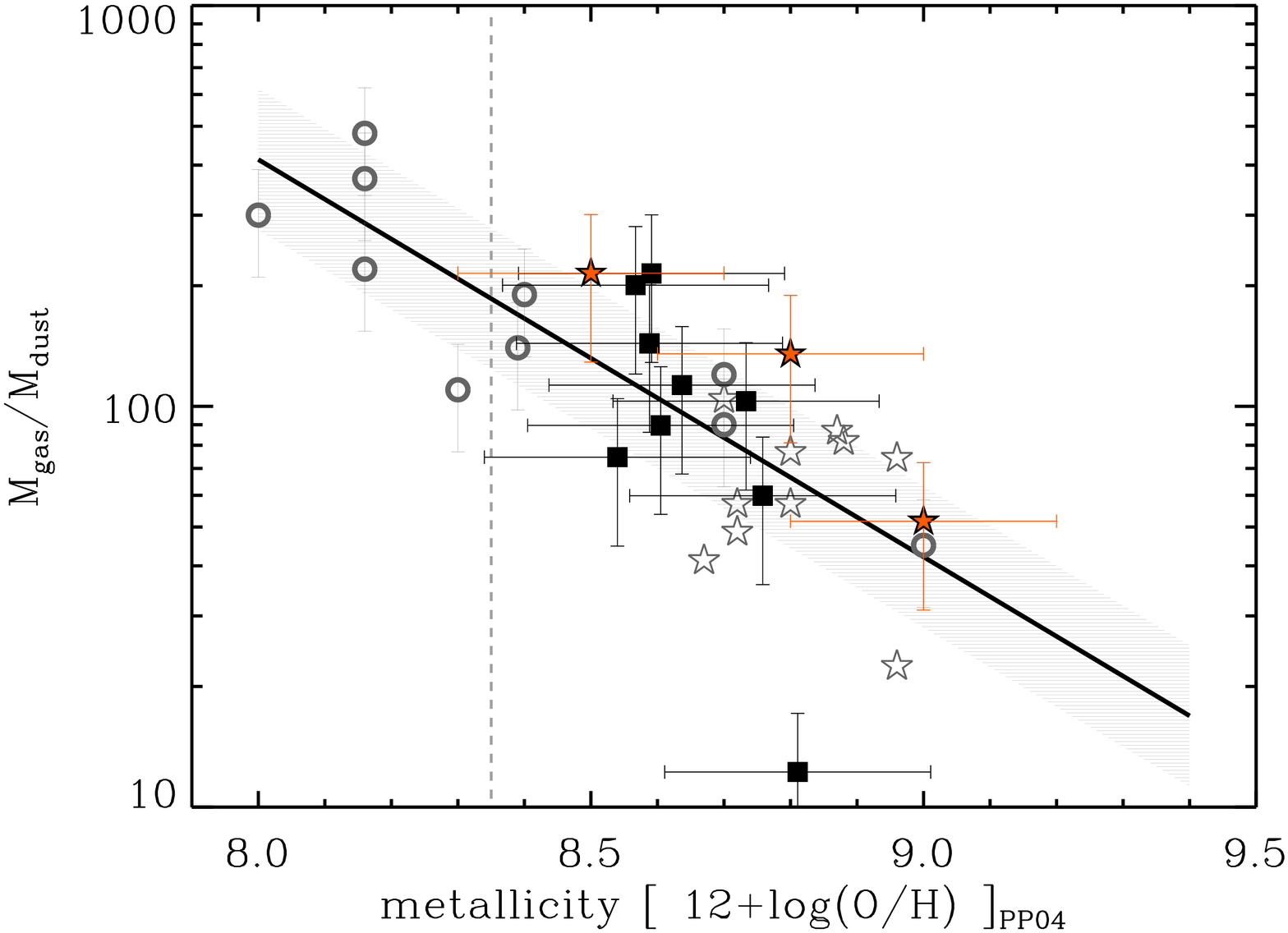}
\includegraphics[scale=0.28]{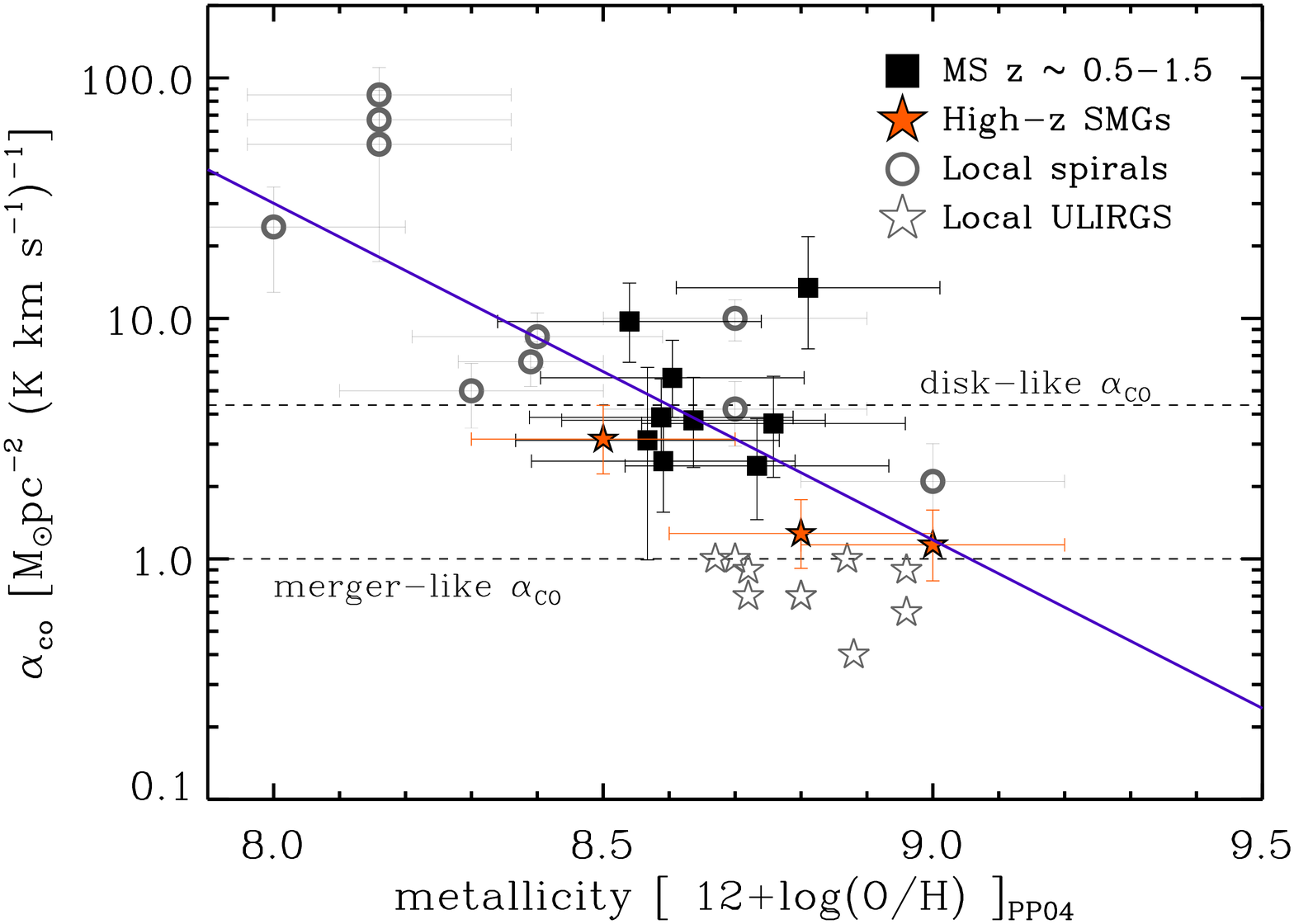}
\caption{ {\bf Left:} \mgas/\md ~vs.\ metallicity for a sample of galaxies in the Local Group by Leroy et al.\ (2011) (grey circles) and local ULIRGs by Downes \& Solomon (1998) (grey stars). The solid black line is the best linear regression fit to Leroy's sample and the grey shadowed area depicts the dispersion of the correlation. Filled black squares and orange stars indicate the position of $z \sim 0.5-1.5$ MS galaxies and high$-z$ SMGs respectively, based on the \aco ~vs.\ metallicity relation derived in the right panel (see eq. 8). The vertical, dashed line indicates the limiting metallicity for which the presented relation is used in this study.
{\bf Right:} Inferred \aco ~values, based on the \gdr ~method, against metallicity for high$-z$ MS galaxies (black squares), SMGs (orange stars). Similar to the left panel, we include the sample of galaxies in the Local Group by Leroy et al.\ (2011) (grey circles), as well as the local ULIRGs by Downes \& Solomon (1998) (grey stars).  All metallicities have been calculated on PP04 scale. The blue solid line represents the best regression fit to the data
(excluding local ULIRGs), with a slope of -1.39.}
\label{fig:aco_met} %
\end{figure*}

\subsection{Estimating Metallicities}
A key ingredient of this method is the metallicity of the galaxies, for which we have to rely on indirect measurements. We first derive stellar mass estimates by fitting the Bruzual \& Charlot  (2003) model SEDs to their rest-frame UV to near-IR spectrum. Then for the $z \sim1.5$ BzK galaxies and SMMJ2135-0102 we use the $M_{\ast}-Z$ relation at $z \sim 2$ from Erb et al.\ (2006), while for the $z \sim 0.5$ sources we adopt the fundamental metallicity relation (FMR) of Mannucci et al.\ (2010) that relates the SFR and the stellar mass to metallicity. To check for possible systematics, we also derive metallicity estimates for the $z \sim 1.5$ sources with the FMR relation and find that the two methods provide very similar results. For GN20 and HLSW-1 the situation is more complicated, as the FMR formula is only applicable up to $z \sim 2.5$, and the Erb et al.\ (2006) relation saturates above $M_{\ast} \sim 10^{11}$\msol. For these two sources we adopt the line of reasoning of Magdis et al.\ (2011b). Namely, in addition to the metallicity estimates based on the FMR relation, we also consider the extreme case where the huge SFR of the two sources ($\sim 1500$~\msol~yr$^{-1}$) is due to a  final burst of star formation triggered by a major merger that will eventually transform the galaxy into a massive elliptical. Once star formation ceases, the mass and metallicity of the resulting galaxy will not change further, and one might therefore apply the mass--metallicity relation of present-day elliptical galaxies (e.g., Calura et al.\ 2009). Combining the metallicity estimates based on this scenario with those derived based on the FMR relation, we estimate that the metallicities could fall in the ranges $12 + \log[O/H] =$ 8.8 to 9.2 for GN20, and 8.6 to 9.0 for HLSW-1. For the analysis we will adopt $12 + \log[O/H] = 9.0 \pm 0.2$ for GN20 and  $12 + \log[O/H] = 8.8 \pm 0.2$ for HLSW-1. We also adopt a typical uncertainty of 0.2 for the rest of the sources.  The assumed metallicities, all calibrated at the Pettini \& Pagel (2004) (PP04) scale, along with the stellar masses, are given in Table 3.

\subsection{Derivation of the CO to \mol ~Conversion Factor}
To derive \aco ~estimates for our sample, we first derive a \gdr$-Z$ relation using the local sample presented by Leroy et al.\ (2011), after converting all metallicities to the PP04 scale (Figure  \ref{fig:aco_met} left). The data yield a 
tight correlation between the two quantities\footnote{Adopting the relation quoted by Leroy et al.\ (2011), i.e., $\log$~\gdr$ = (9.4 \pm 1.1) - (0.85 \pm 0.13) \times (12+\log[O/H])$, has virtually no impact on the results presented here, with a mean difference between the derived \aco ~estimates of a factor of $\sim1.15$} described as: 
$\log$~\gdr$ = (10.54 \pm  1.0) - (0.99 \pm 0.12) \times (12+\log(O/H))$,
with a scatter of 0.15~dex. Having obtained the \md ~and metallicity estimates, we then use the \gdr$-Z$ relation to derive \mgas ~and  subsequently estimate \aco ~from the equation \mgas ~= \aco $\times$ \lco. For the last step we have  assumed that at high$-z$, \mol~ $\gg ~M_{\rm HI}$ or equivalently that \mgas~$\approx$ \mol, as supported by both observational evidence (e.g., Daddi et al.\ 2010; Tacconi et al.\ 2010,  Geach et al.\ 2011) and theoretical arguments (Blitz \& Rosolowsky 2006; Bigiel et al.\ 2008, Obreschkow et al.\ 2009). All values and their corresponding uncertainties, which take into account both the dispersion of the \gdr$-Z$ relation and the uncertainties in $Z$ and \md, are summarized in Table 3. 
In Figure \ref{fig:aco_met} (right), we plot the derived \aco ~values as a function of metallicity, along with the estimates of Leroy et al.\ (2011) for a sample of local normal galaxies as well as for a sample of local ULIRGs from Downes \& 
Solomon (1998), for which we were able to compute their metallicities on the PP04 scale. Fitting high$-z$ MS galaxies along with the local sample of Leroy et al.\ (2011), yields: \\
\begin{equation}
\log (\alpha_{\rm CO} ) = -(1.39\pm0.3)\times(12+\log[O/H])_{\rm PP04} +12.8{\pm2.2}
\end{equation}
with \aco ~decreasing for higher metallicities, in agreement with previous studies (e.g., Wilson\ 1995, Israel et al.\ 1997, Schruba et al.\ 2012, and references within). It is evident that main sequence galaxies at any redshift have higher \aco ~values than those of local star-bursting ULIRGs and more similar to those of local normal galaxies. In particular, for high$-z$ MS galaxies, we find a mean \aco ~of 5.5$\pm$0.4, very close to the MW value of 4.6 (e.g., Strong
\& Mattox 1996, Dame et al.\ 2001). Interestingly, SMM-J2135 falls close to the locus of main sequence galaxies with  \aco ~$\sim 3$, while for the remaining SMGs we find \aco~$\sim 1$, close to the average  value of local ULIRGs (\aco~$\sim 0.8$, e.g., Solomon et al.\ 1997, Tacconi et al. 2008). Finally, simply  as a consistency check, we derive and plot in Figure \ref{fig:aco_met} (left) the \mgas/\md ~values of our targets, using \aco ~estimates as obtained from equation 8.
 
\subsection{Limitations}
Before trying to interpret our findings regarding the \aco ~values, it is important to 
discuss possible caveats and limitations of the method.
A key ingredient for the derivation of \aco\ is  the \md ~estimate. These estimates rely heavily on the adopted model of the grain-size distribution. Dust masses based on 
different grain size compositions and opacities in the literature 
can vary even by a factor of $ \sim 3$, indicating that the absolute values should be treated with caution (e.g., Galliano et al.\ 2011).  However, the relative values of dust masses derived based on the same assumed dust model should be correct and provide a meaningful comparison, as long as the dust has similar properties in all galaxies. Therefore, any trends arising from the derived \md ~estimates are essentially insensitive to the assumed dust model.

One important assumption of our technique is that the  
\gdr$-Z$ relation defined by local galaxies does not evolve substantially with cosmic time. These quantities are very intimately related, 
as dust is ultimately made of metals. 
For example, solar metallicity of $Z_{\odot}$ = 0.02 means that 2\% of the gas is in metals. 
The local \gdr $- Z$ relation suggests an average trend  where 1\% of the gas into dust. This means that about 50\% of the metals are locked into dust. This is already quite an efficient process of metal condensations into dust, and it is hard to think how this could be even more
efficient at high-$z$ (where, e.g., less time is available for the evolution of low-mass stars into the AGB phase), with a much higher proportion of metals locked into the dust phase, a situation that would lead to an overestimate of
\mgas with our technique.
Similarly, chemical evolution studies suggest a small  evolution of  dust-to-metal ratio, i.e., by a maximum factor of $\sim$ 2 from 2 Gyr after the Big Bang up to the present time (e.g.,  Edmunds 2001, Inoue 2003, Calura et al.\ 2008). Furthermore, if we 
plot the ratio of the direct observables, \lco/\md, as a function of metallicity, both local and high$-z$ galaxies follow the same trend, suggesting that the local \gdr$-Z$ relation is valid at high$-z$ (Figure  \ref{fig:lcoz}). Finally, while the Leroy et al.\ (2011) sample yields a slope of $-1.0$ in the \gdr$-Z$ relation, Mu{\~n}oz-Mateos et al (2009), find a rather steeper slope of -2.45. However, adopting the Mu{\~n}oz-Mateos relation (after rescaling it to PP04 metallicity), returns very similar \aco ~values, as the two relations intercept at metallicities very close to the average metallicity  of our sample ($12 + \log[O/H] \approx 8.6$). We note though that for GN20 and HLSW (the sources with the highest Z estimates), the  Mu{\~n}oz-Mateos relation would reduce \aco ~values by a factor of $\sim 3$.

\begin{figure}[!t]
\centering
\includegraphics[scale=0.37]{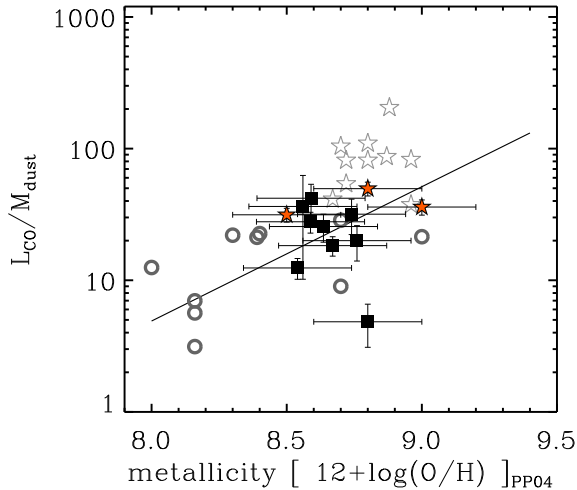}
\caption{\lco/\md ~against metallicity for the same sample of galaxies as in Figure \ref{fig:aco_met}.
This plot serves as a sanity check for the assumption that the \mgas/\md ~vs.\ 
metallicity relation does not evolve considerably with cosmic time.}
\label{fig:lcoz} %
\end{figure}
\begin{figure*}
\centering
\includegraphics[scale=0.39]{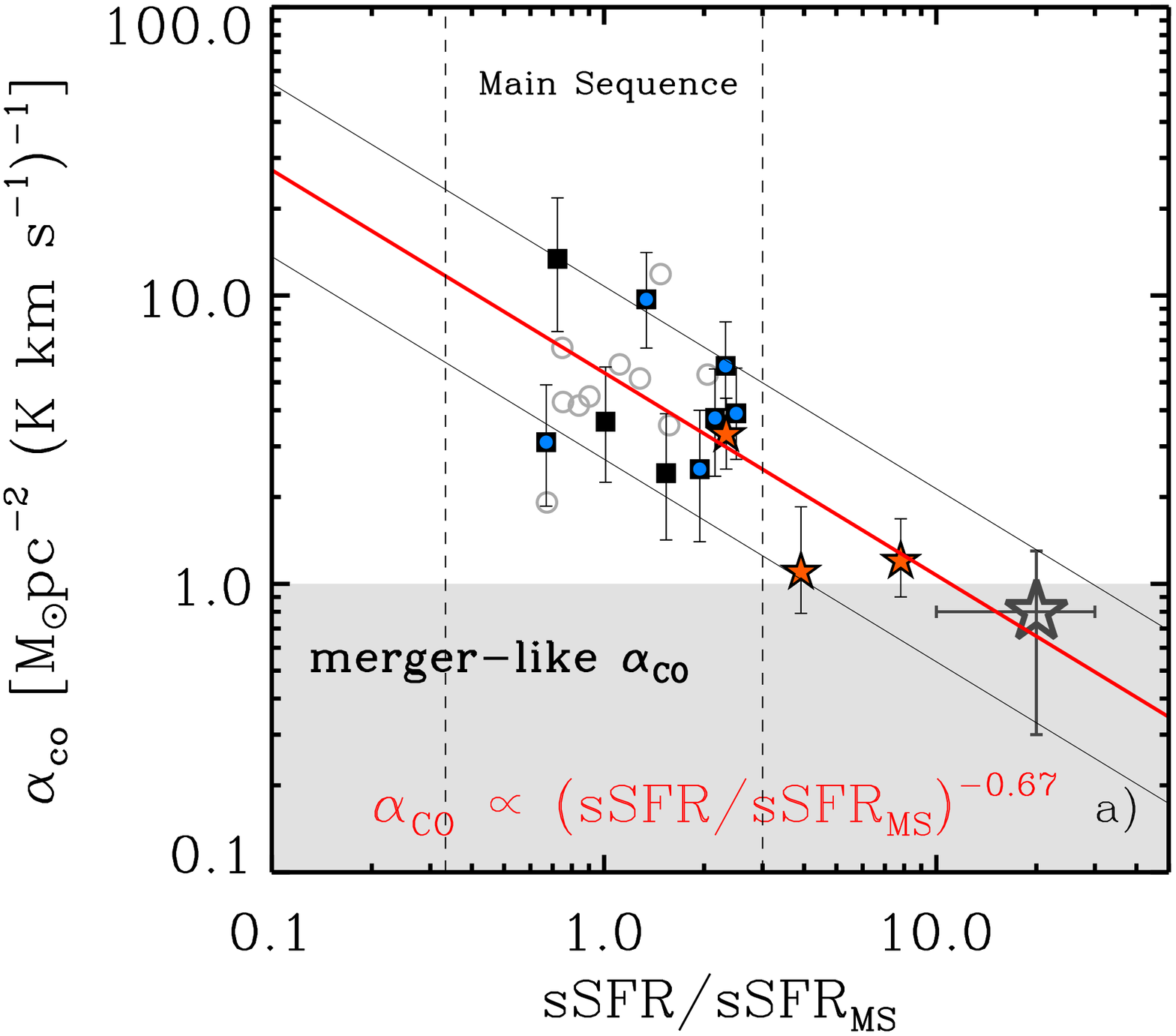}
\includegraphics[scale=0.39]{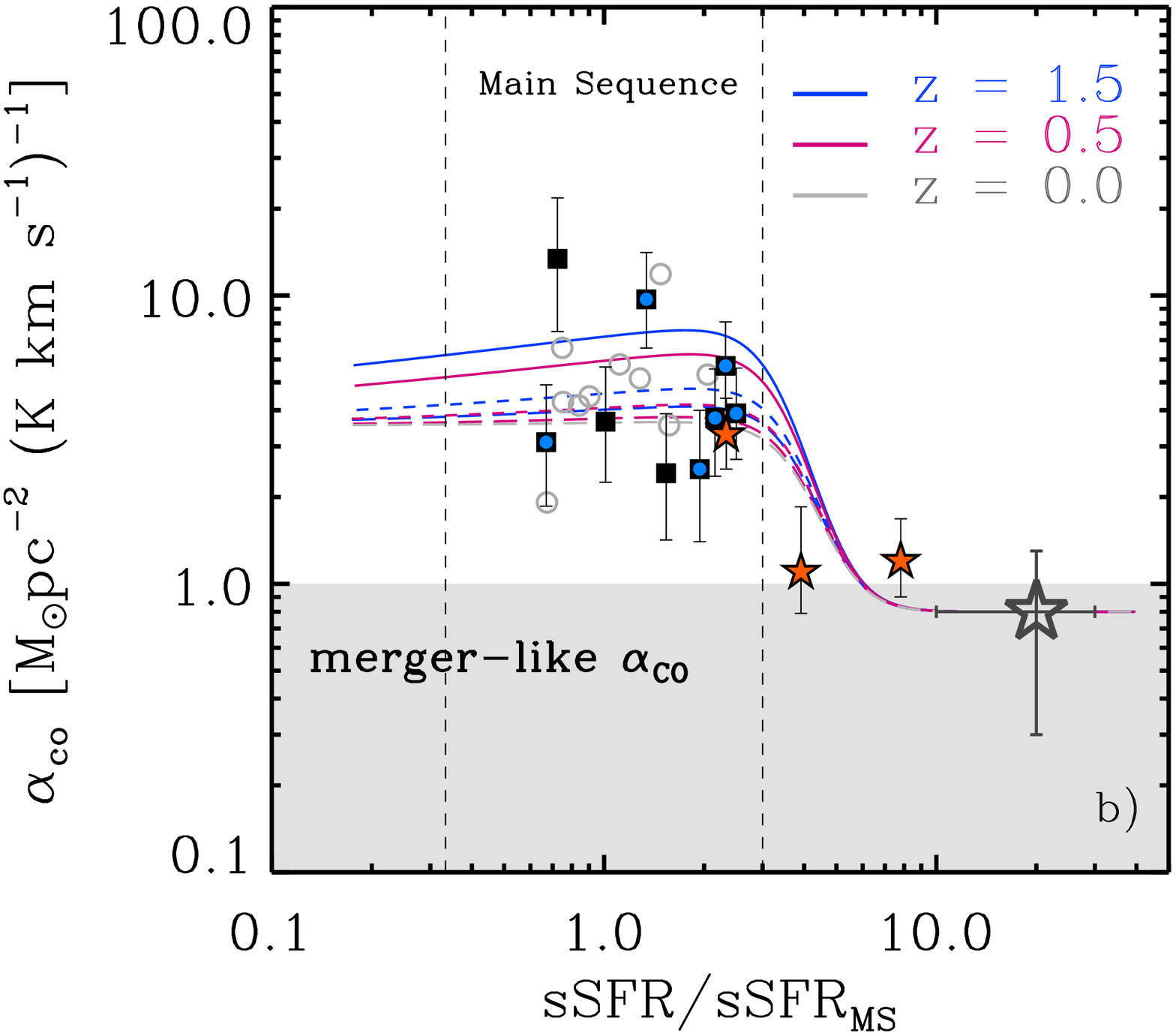}
\caption{{\bf a)} Inferred \aco ~values against the offset from the main sequence. a) Inferred \aco ~values for the 
MS galaxies (black squares) and SMGs (orange stars) considered in this study against their 
offset from the main sequence, parametrized as sSFR/sSFR$_{\rm MS}$. Empty grey circles represent local disks from Leroy et al.\ (2008) and Schruba et al.\ (2012), while the empty grey star the locus of local ULIRGs. The data (only high$-z$ galaxies) are best fit with a slope of -0.67 (red solid line) with a scatter of 0.3~dex and a Spearman's rank correlation coefficient of $\rho = -0.51$. {\bf b)} Same as the left panel but for the case of a metallicity-dependent \aco. Tracks 
show the variation predicted by Sargent et al.\ (2012b, in prep)  based on (i) the relative importance of the 
starburst and main-sequence mode of star formation as a function of sSFR/sSFR$_{\rm MS}$ and (ii) a ``metallicity-
dependent''  \aco, where the metallicity of main-sequence galaxies is inferred based on the stellar mass$-$star 
formation$-$metallicity relation of Mannucci et al.\ (2010). Blue (purple), solid, dashed and long-dashed lines 
correspond to $z =1.5$   ($z = 0.5$) for $M_{\ast}$ = 1.0 $\times$ 10$^{10}$ \msol, 5.0 $\times$ 10$^{10}$ 
\msol, and 1.0 $\times$ 10$^{11}$ \msol. The grey long-dashed line corresponds to $z = 0.0$ and $M_{\ast}$ = 1.0 
$\times$ 10$^{11}$ \msol.}
\label{fig:ssfr} %
\end{figure*}

\subsection{\aco ~as a Function of Specific Star Formation Rate}
With robust \lir ~measurements in hand, we are in a position to derive accurate estimates of the SFR and 
sSFR for galaxies in our sample. We first convert the derived $L_{\rm IR}$ to star formation rates by using the Kennicutt (1998) relation, scaled for a Chabrier IMF for consistency with our stellar mass estimates. We then infer the specific star formation rate of each source and compare it to the characteristic sSFR$_{\rm MS}$ of the main sequence galaxies at the corresponding redshift of the source, using the methodology described in \S2 up to $z=2.5$.
For the two SMGs at $z \sim 3$ and $z \sim 4$, we use the SFR-$M_{\ast}$ relations of Magdis et al.\ (2010a) and Daddi et al. (2009) respectively. In order to conservatively identify MS galaxies, we classify starbursts as galaxies with an excess sSFR relative to that of the MS galaxies by at least a factor of 3, i.e., with  sSFR/sSFR$_{\rm MS}$ $>$ 3. We note however that the sSFR/sSFR$_{\rm MS}$ indicator is only a statistical measure of the star formation mode of the galaxies, and there is no rigid limit  that separates main sequence from starburst galaxies. 

Our analysis suggests that high$-z$ MS galaxies, associated with a secular star formation mode,  exhibit higher \aco ~values at any redshift compared to those of merger-driven local ULIRGs. Since the latter are known to be strong outliers from the local SFR-M$_{\ast}$ relation, we expect a dependence of the \aco ~value with sSFR. However, the functional dependence between the two quantities is not straightforward. In the local universe, there is evidence for a bimodality in the \aco ~value, linked with the two known star formation modes: normal disks (for $Z \sim Z_{\odot}$) are associated with \aco ~$\sim$ 4.4 and merger driven ULIRGs with \aco~$\sim$ 0.8. 
As we will discuss later, if the \aco ~of normal galaxies depends only on metallicity, then we expect only a weak increase of \aco ~as a function of sSFR/sSFR$_{\rm MS}$ within the MS. Consequently, in order to incorporate the observed steep decrease of \aco ~in starburst systems, \aco ~and the relative offset from the main sequence (i.e., sSFR/sSFR$_{\rm MS}$) will be related with a step-function. On the other hand, 
if \aco\ also depends on other parameters than metallicity, such as the compactness and/or the clumpiness of the ISM of normal galaxies, and if the ISM conditions of MS galaxies do vary significantly, then \aco ~could smoothly decrease as we depart from the MS, and move towards the starburst regime (e.g., Narayanan et al.\ 2012).

The scenario of a  continuous variation of \aco ~with sSFR/sSFR$_{\rm MS}$ is depicted in  Figure \ref{fig:ssfr}a. In addition to our galaxies we also include  a sample of local normal galaxies for which we have robust $M_{\ast}$ and SFR estimates from Leroy et al.\ (2008) and \aco ~measurements from Schruba et al.\ (2012). A linear regression fit yields the following relation:
\begin{equation}
\alpha_{\rm CO} = (5.8\pm 2.0) \times [{\rm sSFR/sSFR_{MS}}]^{(-0.67 \pm 0.25)}
\end{equation}
with a scatter of 0.3~dex. We stress that, given the lack of a sufficient number of starbursts in our sample, this relation is poorly constrained in the starburst regime. Nevertheless, the locus of local ULIRGs, seems to be consistent with the emerging trend. Note that since we lack a sample of local ULIRGs for which both \aco ~and sSFR are accurately determined, the position of local ULIRGs in Figure \ref{fig:ssfr}  indicates the average \aco ~and  sSFR  of the population, as derived by Downes \& Solomon (1998) and Da Cunha et al. (2010b), respectively. To further check the robustness of the fit we also  perturbed the original the data within the errors and repeated the fit for 1000 realisations. The  mean slope derived with this method is very close to the one describing the original data. A Spearman's rank correlation test yields a correlation coefficient of $\rho = -0.51$, with a $p-$value of 0.03, suggesting a moderately significant correlation between sSFR/sSFR$_{\rm MS}$ and  \aco.  However, repeating the Spearman's test, this time excluding star-bursting systems, does not yield a statistically  significant correlation, with $\rho = -0.15$ and a $p-$value of 0.3, indicating a very small dependence, if any, between  \aco ~and sSFR within the MS. 

It is therefore unclear whether \aco\ varies {\it inside} the MS sequence, or if instead there is a bifurcation between the mean values appropriate for MS versus SB galaxies. If the \aco\ conversion factor primarily depends on metallicity, one would expect little or no variations with sSFR at constant mass within the MS. Detailed computations for this scenario are presented in  Sargent et al.\ (2012b, in prep.), in which 
 burst-like activity is assumed to be characterised by a constant, low conversion factor, chosen here to be $\alpha^{\rm SB}_{\rm CO}$ = 0.8. To predict the variations of $\alpha_{\rm CO}$ as a function of sSFR/sSFR$_{\rm MS}$ according to this scenario, Sargent et al.\ (2012) compute the relative importance of main-sequence and starburst activity at a given position within the stellar mass vs.\ SFR plane using the decomposition of the sSFR-distribution at fixed stellar mass derived in Sargent et al.\ (2012a; based on the data of Rodighiero et al.\ 2011). Metallicities, which form the basis for computing $\alpha_{\rm CO}$ for an ISM experiencing ``normal'' star-formation activity, vary smoothly as a function of stellar mass and star formation rate according to the calibration of the fundamental metallicity plane given in Mannucci et al.\ (2010; see also Lara-Lopez et al.\ 2010). To compute the corresponding value of the conversion factor, Sargent et al.\ (2012b in prep.) assume a variation of $\alpha_{\rm CO} \propto Z^{-1}$ as they find that this relation between metallicity and $\alpha_{\rm CO}$ reproduces the faint-end slope of the $z = 0$ CO-LF of Keres et al.\ (2003) best, under the set of assumptions just described.

The colored tracks in  Figure \ref{fig:ssfr}b delineate the variations of $\alpha_{\rm CO}$ with sSFR/$\langle {\rm sSFR}\rangle_{\rm MS}$ expected for galaxies in three bins of total stellar mass.
Three regions can be clearly distinguished: (i) the {\it main-sequence locus}, characterised by a gradual increase of $\alpha_{\rm CO}$ with sSFR that is caused by the slight rise in metallicity predicted by the FMR; (ii) the {\it starburst region} at high sSFR/$\langle {\rm sSFR}\rangle_{\rm MS} \gg 4$, where the conversion factor, by construction, assumes a mass- and redshift-independent, constant value; and (iii) a narrow {\it transition region} (spanning roughly sSFR/$\langle {\rm sSFR}\rangle_{\rm MS} \in [2,4]$) where $\alpha_{\rm CO}$ drops from an approximately Milky Way-like conversion factor to the starburst value. Note that, in contrast to the previous case,  \aco ~varies only slightly among MS galaxies. This weak dependence of \aco ~on  sSFR/sSFR$_{\rm MS}$ introduces a step as we move from MS to starburst systems. In the emerging picture, the majority of the star-forming population is dominated by either the main-sequence or starburst mode, and only a small fraction consists of ``composite'' star-forming galaxies hosting both normal and starburst activity. For composite sources, the $\alpha_{\rm CO}$ should be interpreted as a mass-weighted, average conversion factor that reflects the relative amount of the molecular gas reservoir that fuels star-forming sites experiencing burst-like and secular star-formation events, respectively. Note also that variations with redshift (indicated by different colored lines in Figure 7b)
are small for the limited range of stellar mass covered by our sample, and would likely be indistinguishable within the natural scatter about the median trends plotted in the Figure.

Both scenarios appear to be consistent with the data, leaving open the question 
of whether the transition from normal to starburst galaxies is followed by a smooth or a step-like 
variation of \aco. However, both scenarios agree in that \aco ~should not vary much {\it within} the 
main sequence. Indeed, even for the case of continuous variation, the decrease of \aco ~becomes statistically significant only for strongly star-bursting systems, as equation 9 indicates that \aco ~would vary at most by a factor of $\sim 2$ within the full range of the MS, well within the observed scatter of the relation. 
Therefore, the average value \aco~$\sim 5$, as derived in this study, can be regarded as representative for the whole population of MS galaxies at any redshift,
with stellar masses $\ge10^{10}M_\odot$. We remind the reader that because sSFR$_{\rm MS}$ increases with redshift as (1+z)$^{2.95}$, at least out to $z \approx 2.5$, galaxies with (U)LIRG-like luminosities can enter the MS at higher redshifts or at large stellar masses (e.g., Sargent et al.\ 2012). 
We stress that our analysis suggests that the \aco ~of these systems, i.e., high$-z$ main sequence ULIRGs, is on average a factor of $\sim$ 5 larger than that of local ULIRGs.

\subsection{Comparison with Theoretical Predictions}
The determination of the \aco ~value has also been the focus of several theoretical studies. 
In particular Narayanan et al.\ (2011), investigated the dependence of \aco ~on the galactic environment 
in numerical simulations of disk galaxies and galaxy mergers, and reported a relationship 
between \aco ~and the CO surface brightness of a galaxy.  Here, we compare our observationally constrained  \aco ~values with those predicted by their theoretical approach.  
According to Narayanan et al.\ (2011) :
\begin{equation}
{\rm \alpha_{\rm CO} = \frac{6.75 \times 10^{20} \langle W_{\rm CO} \rangle ^{-0.32}}{6.3 \times 10^{19} \times (Z/Z_{\odot})^{0.65}}} 
\end{equation}
where $\langle W_{\rm CO} \rangle$, is the CO surface brightness of the galaxy in K Km s$^{-1}$.
According to Solomon \& Vanden Bout (2005)
\begin{equation}
{\rm L^{\prime}_{\rm CO} = 23.5 ~\times ~\Omega_{B*S} ~D^{2}_{\rm L}  ~I^{'}_{\rm CO} ~(1 + z)^{-3}  }
\end{equation}
where \lco ~is measured in K km s$^{-1}$ pc$^{2}$, $\Omega_{\rm B*S}$  is the solid angle of the source convolved with the telescope beam measured in arcsec$^{2}$, and I$^{'}_{\rm CO}$ is the observed integrated line intensity in K km s$^{-1}$. In the case of an infinitely good resolution, as assumed in Narayanan's simulations, this becomes the solid angle subtended by the source. 
The CO line luminosity can also be expressed for a source of any size in terms of the total line flux: 
\begin{equation}
{\rm L^{'}_{\rm CO} = 3.25 \times 10^{7} ~S_{\rm CO} ~\Delta v ~\nu ^{-2}_{obs} ~D^{2}_{\rm L} ~(1 + z)^{-3} }
\end{equation}
\noindent where $S_{\rm CO}\Delta$v is the velocity integrated flux in Jy km s$^{-1}$. From equations 11 and 12, and for the case of an infinitely good resolution, we derive :
\begin{equation}
{\rm W_{\rm CO} = 1.38\times 10^{6}~ \frac{4 \times~\nu_{obs}^{-2}}{\pi (1+z)~\theta^{-2}} ~ \times  S_{\rm CO}\Delta v }
\end{equation}
where $\nu_{\rm obs}$ is the observed frequency and $\theta$ is the size of the source in arcsec\footnote{We introduce a $(1+z)$ term here, not
present in the original Narayanan et al.\ formula, which is required in order to make $\alpha_{\rm CO}$ independent on redshift for fixed physical properties of the galaxies. Without that term the systematic difference between predicted and measured $\alpha_{\rm CO}$ values would grow to a factor of 2}. For our sources we use, where possible, the  
  CO sizes, and UV sizes when CO observations are too noisy to reliably measure the source extent.  Using equations 10 and 13, we then get an estimate of \aco ~for each source in our sample according to the prescription of Narayanan et al.\ (2011), and in Figure \ref{fig:nara} we compare them to the values derived based on our method. Although there seems to be a small systematic offset (on average by a  factor of 
  $\sim$ 1.4) towards lower values for the theoretical approach, the overall agreement of the \aco ~values is very good when one considers the large uncertainties and assumptions of the two independent methods. We stress that the derived $\langle W_{\rm CO} \rangle$ and therefore the estimated \aco ~values are very sensitive to the choice of the size indicator.
\begin{figure}
\centering
\includegraphics[scale=0.37]{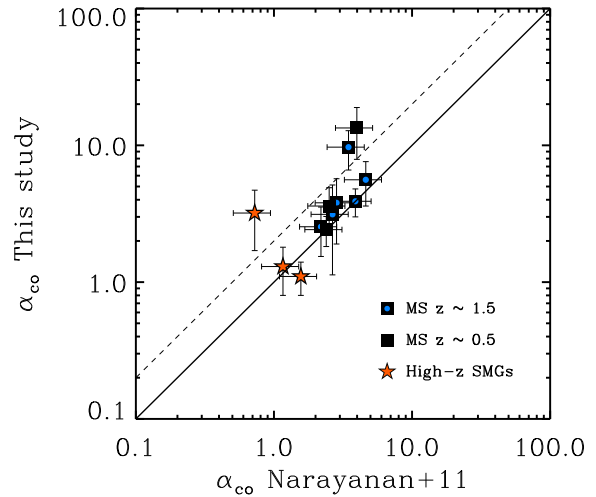}
\caption{Comparison between the \gdr ~inferred \aco ~values for the galaxies in  this study  and those predicted by the theoretical approach of Narayanan et al.\ (2011). MS galaxies at $z \sim 0.5$ and 
and $z \sim 1.5$ are shown with  black squares and black squares with blue circles respectively, while 
SMGs are depicted as orange stars. The solid line is the 1-to-1 relation and the dashed line its offset in y, by a factor of 2.}
\label{fig:nara} %
\end{figure}
\begin{figure*}
\centering
\includegraphics[scale=0.39]{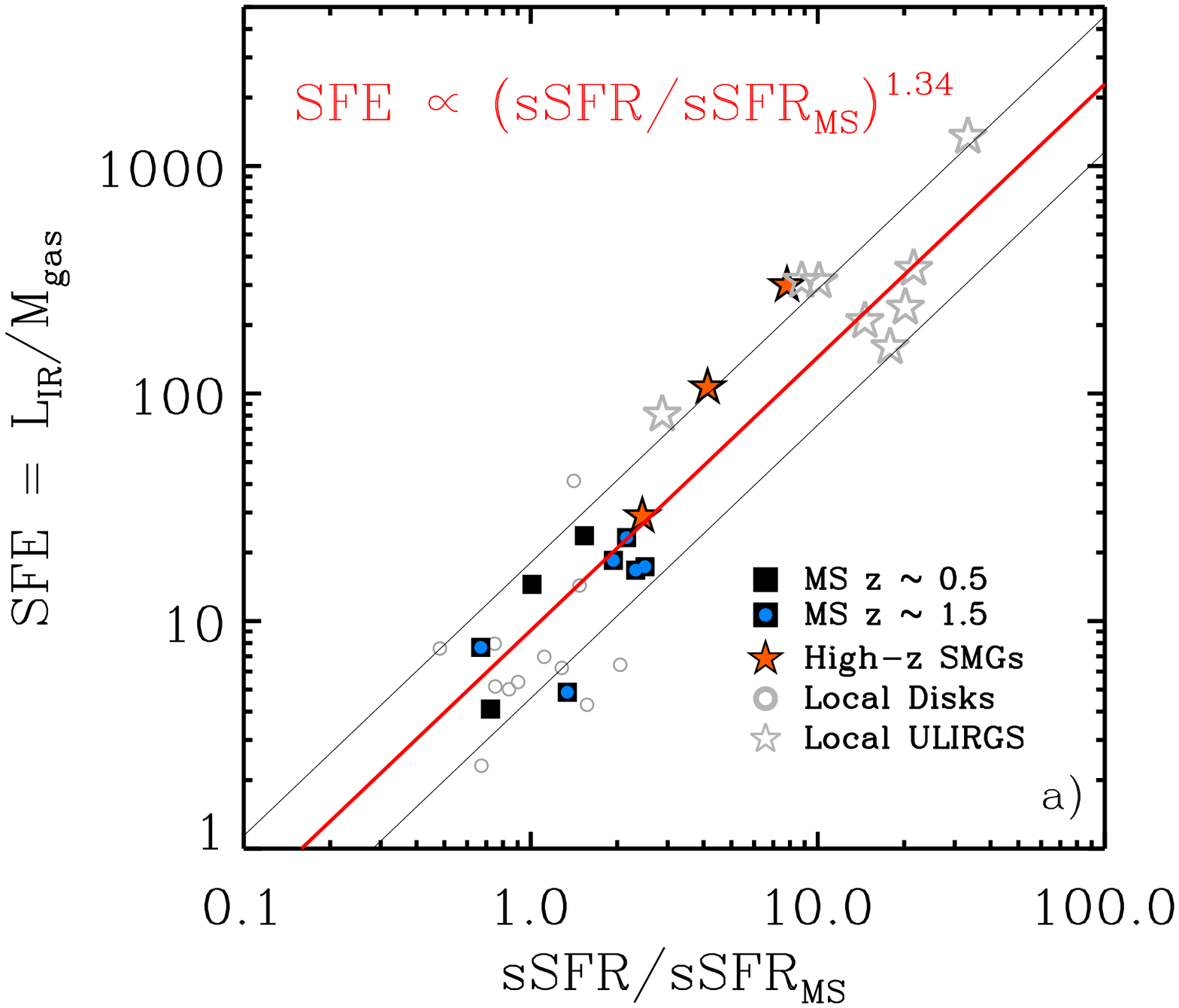}
\includegraphics[scale=0.39]{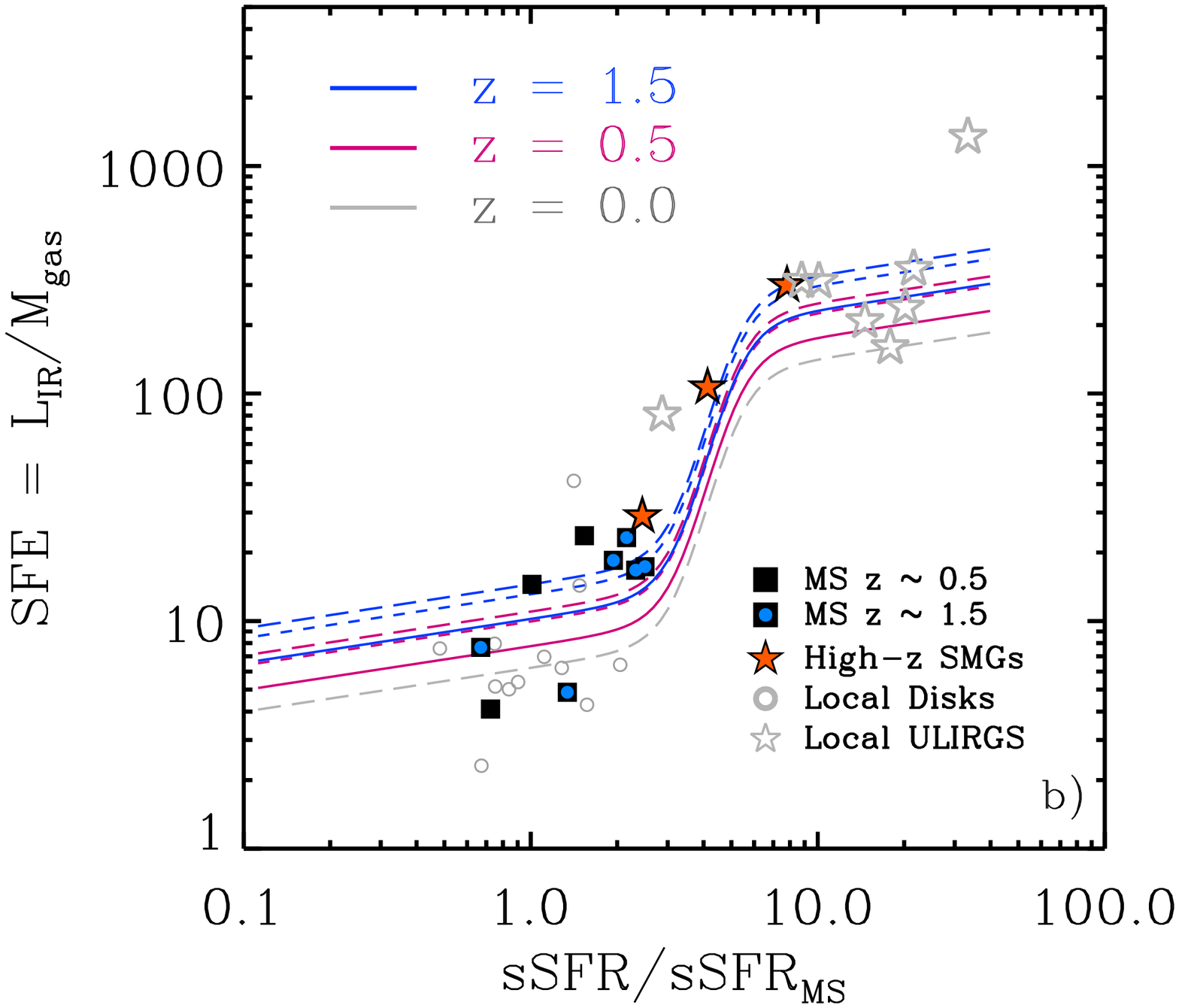}
\caption{{\bf a)} Star formation efficiency, SFE=\lir/\mgas, against the offset from 
the main sequence. Black squares and orange stars are the MS galaxies and the SMGs considered in this study. Empty grey circles are local normal galaxies (Leroy et al.\ 2008 and Schruba et al.\ 2012) and grey stars depict local ULIRGs  (Zaurin et al.\ 2009 and Solomon et al.\ 1997).  For local ULIRGs we have assumed (\aco=0.8). The red solid line depicts the best fit to the data with a slope of 1.34, while a Spearman's rank test indicates a tight 
 correlation between the plotted quantities with $\rho= 0.71$. {\bf b)} Same as the left panel but for the case of a weak increase of SFE with sSFR/sSFR$_{\rm MS}$ within the main sequence. Tracks show the dependence of SFE on sSFR/sSFR$_{\rm MS}$ as expected for a slightly sub-linear relation between \mgas ~and SFR  and the relative importance of the starburst and main-sequence mode of star formation as a function of sSFR/sSFR$_{\rm MS}$. Blue (purple), solid, dashed and long-dashed lines correspond to $z = 1.5$  
 ($z = 0.5$) for $M_{\ast}$ = 1.0 $\times$ 10$^{10}$ \msol, 5.0 $\times$ 10$^{10}$ \msol, and 1.0 $\times$ 10$^{11}$ \msol. Grey long-dashed line corresponds to $z = 0$ and $M_{\ast}$ = 1.0 $\times$ 10$^{11}$ \msol.}
\label{fig:sfe} %
\end{figure*}

\subsection{Implications for the  Star Formation Activity of High$-z$ MS Galaxies}
While the sSFR probes the star formation mode of a galaxy only indirectly and from a statistical point of view, accurate measurements of its star formation efficiency, SFE = SFR/\mgas, can provide more direct and reliable answers.
With the robust \aco ~estimates derived in this study, we can convert the CO measurements into \mgas 
~and subsequently, infer the  star formation efficiency of galaxies in our sample. In what follows we use SFE = SFR/\mgas ~ and SFE = \lir/\mgas ~interchangeably, since SFR is linearly related to \lir ~through SFR $=$ \lir  ~$\times 10^{-10}$  (i.e., through the Kennicutt 1998 relation, calibrated for a Chabrier IMF).

We have so far established that even though there  appears to be only small variation of \aco ~among MS galaxies, the \aco ~of high$-z$ MS galaxies is similar to that of local disks with lower \lir, and is approximately $\sim$  5 times larger than the average \aco ~value observed for local ULIRGs. The direct implication of this finding is that for a given \lco,
one could expect a large variation in the amount of molecular gas for galaxies between MS and starburst galaxies. This also suggests that for a given \lir ~(or equally, SFR), galaxies in the main sequence have lower star formation efficiencies, as compared to star-bursting systems. Indeed, using the inferred \aco ~values we derive SFE estimates for each source in our sample and plot them against sSFR/sSFR$_{\rm MS}$ in Figure \ref{fig:sfe}a. We supplement our sample with a compilation of local disks (Leroy et al.\ 2008),  as well as with a sample of local ULIRGs (Solomon et al.\ 1997, Rodr{\'{\i}}guez Zaur{\'{\i}}n et al., 2010).

Although the actual dependence of SFE on sSFR/sSFR$_{\rm MS}$ within the MS and the transition from MS to the starburst regime 
will be discussed in detail in the next section, here we can derive some crucial results about the SFE 
of the two populations (MS and starbursts galaxies) by employing a simple, empirical relation between SFE and sSFR/sSFR$_{\rm MS}$. Indeed, a linear regression fit and a Spearman's test to our high$-z$ data (including starbursts), yields a strong and statistically significant correlation ($\rho$ = 0.71, $p$-value = 0.0012)
between star formation efficiency and sSFR/sSFR$_{\rm MS}$, with:
\begin{equation}
 {\rm SFE =  (8.1\pm1.2) \times [{\rm sSFR/sSFR_{MS}}]^{1.34\pm0.13}}
\end{equation}
\noindent suggesting  substantially higher star formation efficiencies for galaxies with enhanced sSFR (Figure \ref{fig:sfe}a). For MS galaxies we find an average $\langle SFE \rangle = (14 \pm 2)$~\lsol/\msol, corresponding to an average gas consumption timescale ($\tau_{\rm gas}=$ \mgas $/ SFR$) of $\sim$ 0.7 Gyrs, indicative of long-lasting star formation activity. This is in direct contrast to the short-lived, merger-driven starburst episodes observed in local ULIRGs, with an average $\langle SFE \rangle = 200$ \lsol/\msol, $\sim 12$ times higher than that of MS galaxies. Since the majority of galaxies at any redshift are MS galaxies (e.g., Elbaz et al.\ 2011) and starbursts galaxies seem to play a minor role in the star formation density throughout the cosmic time (e.g., Rodrighero et al.\ 2011, Sargent et al.\ 2012), our results come to support the recently emerging picture of a secular, long lasting star formation as the dominant mode of star formation in the history of the Universe. 
\begin{figure}[!h]
\centering
\includegraphics[scale=0.37]{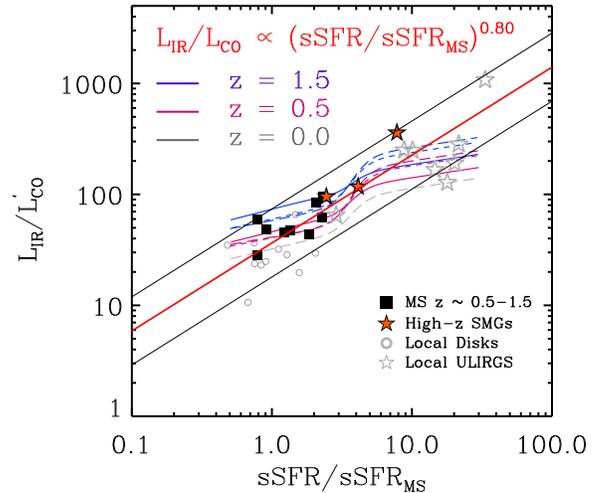}
\caption{Ratio of direct observables, \lir/\lco ~against the offset from the main sequence 
for the same set of objects as in Figure 9. The red solid line is the best fit to the data, with a slope of 0.80. Coloured tracks corresponds 
to the scenario by Sargent et al.\ 2012b, following the convention of Figure 9.}
\label{fig:lirlco} %
\end{figure}
\begin{table*}
{\footnotesize
\caption{Stellar mass, metallicity and gas properties}             
\label{tab:4}      
\centering                          
\begin{tabular}{l c c c c c c c c c c c c}        
\hline\hline                 
Source & $\log M_{\ast}$ & $Z_{\rm PP04}$ $^{a}$ & $\log L_{\rm CO}$ & $\alpha_{\rm CO}$ \\

& $M_\odot$ && $L_\odot$K$^{-1}$km$^{-1}$~s~pc$^{-2}$ & $M_\odot$K$^{-1}$km$^{-1}$~s~pc$^{-2}$\\
\hline
ID-8049 & 10.97 & 8.80 &9.47&13.4$\pm$5.5\\ 
ID-5819 & 10.81 & 8.76&9.53&3.7$\pm$1.3\\
ID-7691 & 10.83 & 8.74&9.78&2.4$\pm$1.0\\
BzK-4171 & 10.60 & 8.59&10.32&2.5$\pm$1.1\\
BzK-12591 & 11.04 & 8.63&10.50&3.8$\pm$2.0\\
BzK-25536 & 10.51 & 8.56&10.07&3.1$\pm$2.1\\
BzK-17999 & 10.59 & 8.58&10.23&3.9$\pm$1.2\\
BzK-16000 & 10.63 & 8.53&10.32&9.7$\pm$3.1\\
BzK-21000 & 10.89 & 8.60&10.34&5.7$\pm$2.0\\
GN20&11.36&9.00&11.21&1.0$\pm$0.3\\
SMM-J2135&10.34&8.50&10.30&3.2$\pm$1.0\\
HSLW-01&10.77&8.80&10.30&1.3$\pm$0.4\\
\hline                                   
\end{tabular}\\
Notes:\\
a: Derived using the $M_{\ast} - Z$ relation of Erb et al.\ (2006), Mannucci et al.\ 2010 or theoretical arguments (GN20 and HLSW-01). We assume a typical uncertainty of $\pm$0.2 for all cases.}
\end{table*}

To ensure that our result is not an artifact of the  assumed \aco ~values,  in Figure \ref{fig:lirlco}  we also plot the ratio of the direct observables, \lir/\lco ~versus the offset from the main sequence for the same set of objects, omitting any assumptions for \aco. 
We find the two values to strongly correlate with $\rho=0.79$ and 
a $p-$value of 3.4 $\times 10^{-6}$ and a functional relationship of:
\begin{equation}
 {\rm L_{\rm IR}/L^{\prime}_{\rm CO} =  34.6(\pm5.0) \times [{\rm sSFR/sSFR_{MS}}]^{0.80\pm0.1}}
\end{equation}
\noindent with  MS galaxies exhibiting lower \lir/\lco ~values by a factor of $\sim 4$. We note, however, that the absence of a well defined sample of high$-z$ starbursts is striking. SMGs, which are frequently regarded as prototypical high$-z$ starbursts, have recently turned out to be more likely a mixed ensemble of objects, with a significant fraction of them exhibiting star formation rates and star formation efficiencies typical of MS galaxies (Ivison et al.\ 2011, Rodighiero et al.\ 2011). Indeed, while we have only three SMGs in this study, we seem to face the same situation. Of the three SMGs considered here, only HSLW-01 appears to be a strong starburst, with sSFR/sSFR$_{\rm MS}$ $\sim$ 10 and SFE $\sim$ 300 \lsol/\msol. GN20 is only marginally outside the MS regime, with sSFR/sSFR$_{\rm MS} \sim 4$, although with high SFE~$\sim 120$ \lsol/\msol, while SMM-J2135 behaves like  a MS galaxy with 
sSFR/sSFR$_{\rm MS}$ $\sim$ 2.5 and SFE $\sim$ 30 \lsol/\msol, much lower than the star formation efficiency observed for local ULIRGs ($>$ 100 \lsol/\msol). Clearly a larger, objectively selected sample of high$-z$ starbursts is essential to improve this investigation.

\section{VARIATIONS OF SFE WITHIN THE MAIN SEQUENCE}
In the previous section, we demonstrated that the bulk of MS galaxies exhibit higher \aco ~ values and have lower star formation efficiencies than  starburst galaxies. Here we will attempt to investigate possible variations of SFE within the MS, taking into account that the thickness, i.e., the spread  of the SFR-$M_{\ast}$ correlation as traced by normally star-forming galaxies  at any redshift, is not just an artifact produced by random noise, but a manifestation of the variation of the 
physical properties of the main sequence galaxies, such as color and clumpiness (e.g., Salmi et al.\ 2012; see also Elbaz et al.\ 2011). Since the actual quantity that drives a galaxy above or below the main sequence is yet unknown, we will consider two limiting scenarios, where the relative position of a galaxy with respect to the MS is driven by i) variations in the gas fraction (\fg ~= \fgas, or equally in the gas to stellar mass ratio, \mgas/$M_{\ast}$) ~while the star formation efficiency remains roughly constant within the main sequence, or ii) variations in the star formation efficiency of MS galaxies, while \fg ~remains constant, in order to explain the observed dispersion in the SFR$-M_{\ast}$ plane.  Indeed, in practice there are two ways that a MS galaxy could have higher SFR: either it has 
more raw material (\mgas) to produce stars, or for the same amount of \mgas ~it 
is more efficient at converting that gas into stars. The two scenarios have direct implications on the star formation law (SF-law) in the galaxies. The first case implies the existence of a {\it global} \lir~--~\mgas ~relation that would apply to all MS galaxies, irrespective of their sSFR, like the one presented by Daddi et al.\ (2010b) and Genzel et al.\ (2010). 
On the other hand, the second scenario would imply {\it variations} of the star formation law, with parallel \mgas~--~\lir ~relations with constant slope, but with a normalisation factor that strongly increases with offset from the MS. While a combination of the two possibilities might also be plausible, we will consider the two limiting cases for simplicity.

\subsection{Limiting Cases}
\subsubsection{Scenario I: {\it a Global SF-Law for MS Galaxies; \fg ~as a Key Parameter}}
In this scenario, the physical parameter that drives the sSFR of a main sequence galaxy is the 
gas fraction, while the star formation efficiency remains roughly constant. In this case we have an SF law in the form: 
\begin{equation}
{\rm \log M_{\rm gas}= \xi_{\rm gas} \times \log SFR + C_{\rm 1}}, 
\label{eqX}
\end{equation}
where $\xi_{\rm gas}$ = 1 would imply \mgas/SFR = constant. However, there is observational evidence that $\xi_{\rm gas}$ $\approx$ 0.8, i.e.,  slightly lower than unity (e.g., Daddi et al.\ 2010, Genzel et al.\ 2010, Sargent et al.\ 2012 in prep.), suggesting a very mild dependence of SFE from SFR (or from sSFR/sSFR$_{\rm MS}$, for fixed stellar mass,)  with a slope of $\approx 0.2$. 
On the other hand, dividing equation 16 by stellar mass indicates that \mgas/$M_{\ast}$ would vary as a function of sSFR (or equally as a function of SFR at fixed stellar mass) with:
\begin{equation}
{\rm \frac{M_{\rm gas}}{M_{\ast}} \propto [\frac{sSFR}{sSFR_{MS}}]^{0.8}}
\end{equation}
\noindent In Figure 9b  we explore the variations of SFE with normalised sSFR for three representative bins of mass and redshift, as predicted by  Sargent et al.\ (2012 in prep.), based on the framework described above. The two factors that contribute to the evolutionary tracks are (i) the observation  that at fixed SFR (or \lir) starbursts display more than an order of magnitude higher SFE than that of ``normal'' galaxies (e.g., Daddi et al.\ 2010b,  Genzel et al.\ 2010, this study), and (ii) the fact that the dependence of SFR on \mgas ~is slightly supra-linear, such that SFE increases with \lir ~even when the main-sequence galaxies and starbursts are considered individually (i.e., eq. 16). Point (ii) is reflected in the weak increase of SFE throughout the main-sequence and starburst regime (sSFR/$\langle {\rm sSFR}\rangle_{\rm MS} < 2$ and sSFR/$\langle {\rm sSFR}\rangle_{\rm MS} > 4$, respectively), while the rapid rise in SFE in the transition region is the manifestation of point (i) above. Note that the ``jump'' in SFE, from the normal to the starburst galaxies is not forced by the assumption of a bimodal star formation activity but naturally results from the weak increase of SFE with sSFR/sSFR$_{\rm MS}$ within the MS. Similar tracks are presented in  Figure \ref{fig:lirlco}, considering \lir/\lco ~instead of \lir/\mgas. 
\subsubsection{Scenario II: {\it a Varying SF-Law for MS Galaxies; SFE as a Key Parameter}}
This scenario assumes that \fg ~for MS galaxies  remains  constant, and that the physical parameter that dictates the position of the galaxy on the SFR $-$ $M{_\ast}$ plane is the star formation efficiency. Since, \fg ~= const, then expressing SFE as a function of sSFR/sSFR$_{\rm MS}$, i.e., at fixed stellar mass, we have: 
\begin{equation}
{\rm \log SFE = \log(SFR) +C_{\rm 2}} 
\end{equation}
\noindent suggesting a linear dependence of SFE on sSFR/sSFR$_{\rm MS}$, in reasonable agreement with the slope derived by a linear fit ($\xi_{\rm SFE} = 1.34 \pm 0.13$, eq. 14). Although a similar slope is found when the fit is redone for MS galaxies alone, excluding the starbursts, a Spearman's test suggests that the correlation is statistically insignificant ($\rho$ = 0.58, $p-$value = 0.13), leaving open the question of whether the trend is valid {\it within} the MS, or if it is  driven solely by the starbursts. A similar picture emerges when, instead of \mgas, we consider the directly observable quantity \lco ~(Figure \ref{fig:lirlco}).  Again, in this case, excluding the starburst data results 
in a statistically insignificant correlation. If indeed the star formation efficiency of MS galaxies follows such a steep increase as a  function of sSFR/sSFR$_{\rm MS}$, aside the existence of parallel \mgas ~- \lir ~SF-laws, it would also imply a smooth, continuous transition from MS to starbursts galaxies.

Both scenarios considered here appear to be consistent with our 
data, and we cannot formally distinguish between them based on these individual objects. 
However, as we will see in the next section, the shape of the SED, both for individually 
elected sources as well as for populations averaged via stacking, can provide more definite 
answers.

\section{DISSECTING THE ``THICKNESS'' OF THE MAIN SEQUENCE}
So far, we have seen that the high-quality SEDs offered by \h ~and mm continuum data can be used to reveal a strong dependence of the \aco ~and SFE of a galaxy on its deviation from the main sequence (eq.9 and eq.14), highlighting significant differences between high$-z$ MS galaxies and local ULIRGs, even though the former may have (U)LIRG-like luminosities and SFRs. The well-sampled SEDs considered in this study can provide the tools to further explore the implications of the two scenarios presented in the previous section.

We have argued that the relation \mgas/\md ~$\propto Z^{-1}$, observed in the local universe, 
holds at high$-z$. A direct consequence of this assumption is that the ratio between the infrared luminosity and the dust mass of a galaxy at any redshift would be:
\begin{equation}
{\rm \frac{L_{\rm IR}}{M_{\rm dust}} \propto \frac{L_{\rm IR}}{M_{\rm gas} \times Z} = \frac{SFE}{Z}}
\end{equation}
\noindent with \lir ~$\propto$ SFR for dusty galaxies where the extinction is large.\footnote{In detail, \lir~$\propto ({\rm SFR_{tot} - SFR_{UV}})$, with SFR$_{\rm UV}$ becoming significant in the case of low extinction, hence typically at very low galaxy masses and metallicities or at high redshifts.} Based on this relation, we can make some 
predictions for the different trends between \lir/\md ~and $M_{\ast}$ and sSFR/sSFR$_{\rm MS}$, for the two scenarios described above. Afterwards, we will attempt to constrain the observational trend by stacking large samples of
mass-selected high-z galaxies.
\begin{figure*}
\centering
\includegraphics[scale=0.37]{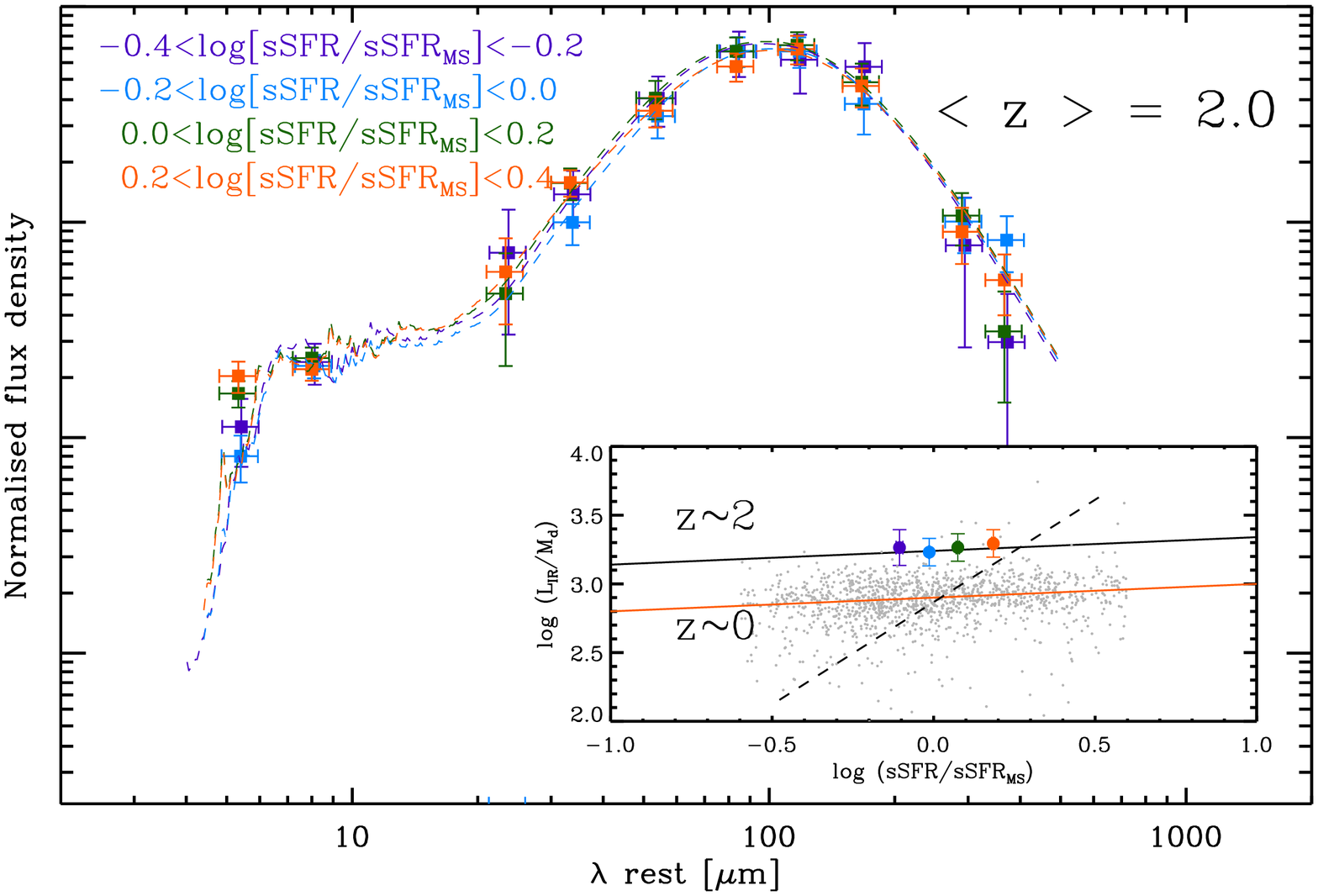}
\includegraphics[scale=0.37]{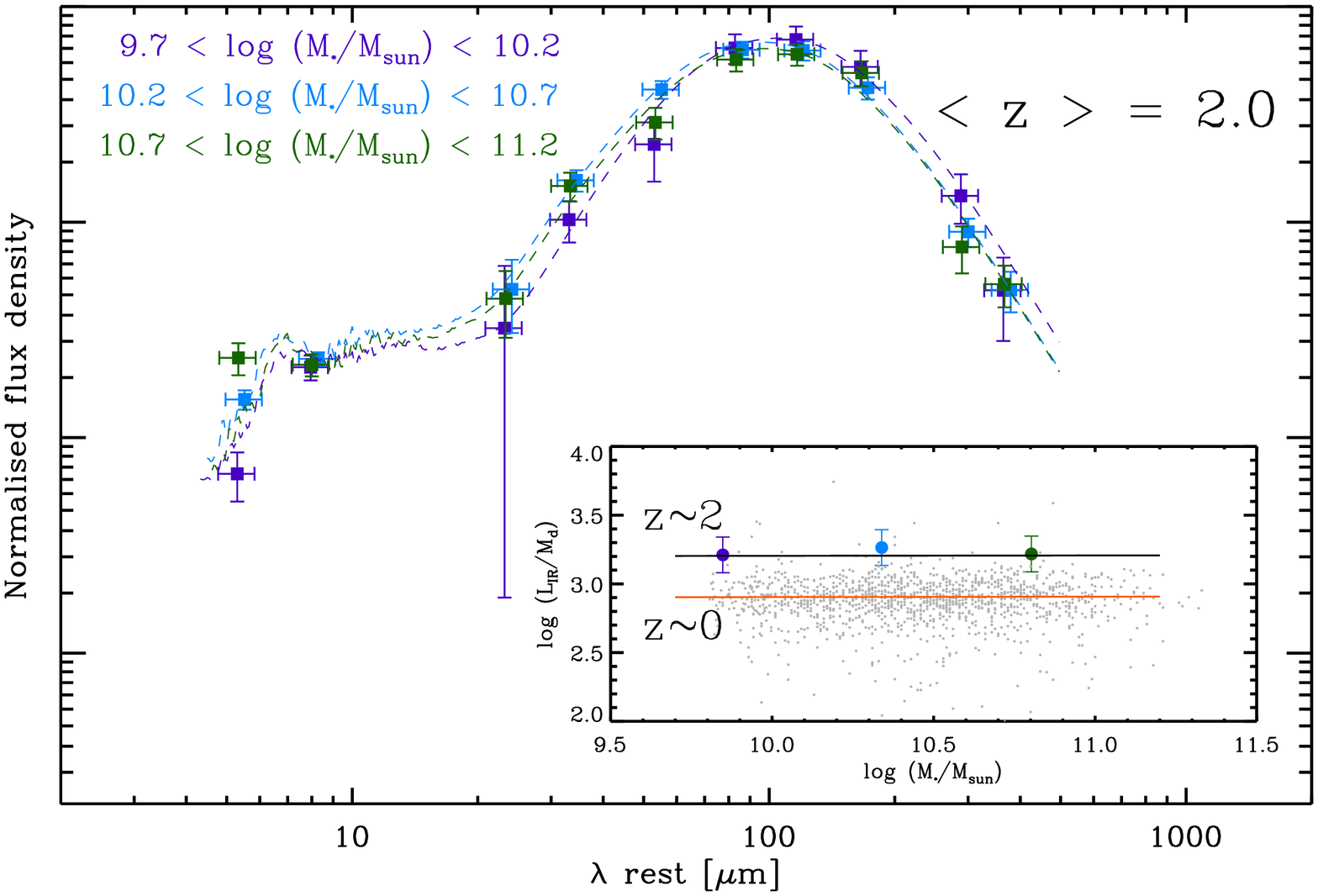}
\caption{ {\bf Top:} Observed average SEDs of $z \sim 2.0$ main sequence galaxies in four sSFR bins, derived by stacking and overlaid with the best fit Draine \& Li (2007) models. Symbols are color-coded based on the four sSFR bins. The inset panel shows the derived \lir/\md ~as a function of the offset from the main sequence for the $z \sim 2$ sample (colored squares) as well as for a set of normal  galaxies in the local universe by  da Cunha et al.\ 2010a (grey dots). For the $z \sim  2$ sample, the sSFR/sSFR$_{\rm MS}$ is derived based on \h ~measurements. The red solid line is the best fit to the local data, with a slope of 0.1. The black solid line shows the linear relation with a 0.5~dex offset. The black dashed line shows the expected trend for the case of a steep and continuous increase of SFE with sSFR/sSFR$_{\rm MS}$. {\bf Bottom:} Observed average SEDs of $z \sim 2.0$ main sequence galaxies in three stellar mass bins,  derived by stacking and overlaid with the best fit Draine \& Li (2007) models. Symbols are color-coded based on the three stellar mass  bins. The inset panel shows derived \lir/\md ~as a function of the stellar mass for the $z \sim 2$ sample (colored circles) and for the same set of  $z \sim 0$ normal galaxies as in the top panel. In both panels the SEDs of the stacked samples are normalised to \lir ~= 1 \lsol}
\label{fig:stack} %
\end{figure*}
\subsection{Trends in \lir/\md ~ for SFE $\approx$ constant}
For the case where \fg ~varies within the MS (scenario I, \S 5.1.1), we have Eq.~\ref{eqX} still holding.
Also, using the $z \sim 2$ stellar mass -- metallicity relation of Erb et al.\ (2006) 
 we get:
\begin{equation}
{\rm \log Z= \xi_{\rm z} \times \log M_{\ast} + C_{\rm 3}}
\end{equation}

\noindent with $\xi_{\rm z}$ $\sim$ 0.15
for a stellar mass range of $10 \leq \log (M_{\ast} /M_{\odot}) \leq 11$.  Finally, from the SFR$-M_{\ast}$ relation, we have:
\begin{equation}
{\rm \log SFR = \xi_{\rm ms} \times \log M_{\ast} + C_{\rm 4}}  
\end{equation}

\noindent with $\xi_{\rm ms} \approx 0.80$. Combining the above relations yields:

\begin{equation}
{\rm \log\frac{SFE}{Z} = (\xi_{\rm ms}-\xi_{\rm gas}*\xi_{\rm ms}-\xi_{\rm z}) \times \log M_{\ast}+C_{\rm 5}}.
\end{equation}
\noindent Substituting the values quoted above for the various $\xi$ coefficients very nearly cancels out any dependence on $\log {\rm M_{\ast}}$, giving:
\begin{equation}
{\rm \log\frac{L_{\rm IR}}{M_{\rm dust}} \propto \log\frac{SFE}{Z} \approx C_{\rm 5}}
\label{eqX2}
\end{equation}
\noindent i.e., a negligible dependence of the dust-mass-weighted luminosity \lir/\md ~with $M_{\ast}$. 
Expressing SFE as a function of sSFR, i.e., as a function of SFR for fixed stellar mass (and therefore metallicity) yields: 
\begin{equation}
{\rm \log\frac{L_{\rm IR}}{M_{\rm dust}} = \log SFE = (1-\xi_{\rm gas}) \times \log SFR +C_{\rm 1}},
\end{equation}
\noindent suggesting a mild dependence of \lir/\md ~with sSFR, with a slope of $\sim$ 0.2. 

\subsection{Trends in \lir/\md ~ for \fg ~$\approx$ constant}
In the case where \fg\ is constant among MS galaxies (scenario II, \S\ 5.1.2), where the thickness of the main sequence is the result of a strongly varying SFE (or equally, varying SF-law), we have: 
\begin{equation}
{\rm \log\frac{L_{\rm IR}}{M_{\rm dust}} = \log SFE + C_{\rm 6} = \log SFR + C_{\rm 7}}
\end{equation}
\noindent since ${\rm M_{gas}}$ is constant at a fixed stellar mass and \fg.  This suggests a strong dependence of \lir/\md 
~on sSFR/sSFR$_{\rm MS}$. Note that \lir/\md ~is not expected to vary with $M_{\ast}$, as was also the case for the previous 
scenario (Eq.~\ref{eqX2}).

Summarizing the analytical predictions, the two scenarios result in two distinct behaviors of \lir/\md ~and \fg ~as a function of sSFR/sSFR$_{\rm MS}$. In the case of a weak increase of SFE within the MS (SFE $\approx$ const, or equally, of a global SF-law for MS galaxies), we expect a variation of \fg ~(eq.\ 17) while \lir/\md ~remains roughly constant  within the MS (eq.\ 25). The opposite trends are expected for the case where \fg ~ remains constant (or equally, of a varying SF-law), i.e., a steep increase of SFE within the MS and a linear increase of  \lir/\md ~ with sSFR/sSFR$_{\rm MS}$ (eq.\ 26). We note that both scenarios predict only weak dependence of  \lir/\md ~with $M_{\ast}$.  Furthermore, the physical meaning of \lir/\md ~is the luminosity emitted per unit of dust mass, or the strength of the mean radiation field heating the dust, and could serve as a rough proxy of the effective dust temperature. Indeed, since, \lir ~$\propto$ $\sigma$\td$^{4}$,  two sources with the same \lir/\md\ should have similar \td. Similarly, for a given \md,  higher \lir ~indicates higher \td.  Therefore, the above analysis also suggests a strong variation of the SED shape of the galaxies within the MS for the scenario where \fg~$\approx$ constant, while only very little, if any, change in the shape of the SED for the case where SFE $\approx$ constant.

\subsection{\lir/\md ~and Gas Fractions in MS Galaxies}
Using the stacked samples of $z \sim 2$ MS galaxies in three stellar mass and four sSFR bins, 
described in section 2, we can investigate what  trends, if any, are present between \lir/\md $-$ $M_{\ast}$ and \lir/\md $-$ sSFR/sSFR$_{\rm MS}$. To derive the far-IR properties of the stacked samples we followed the same SED fitting procedure used for the individually detected sources. However, this time we take into account the redshift distribution of the stacked sources, to account for artificial SED broadening in the far-IR and smearing of the spectral features in the mid-IR regime. Namely, we construct the whole set of  DL07 models at various redshifts and create an average SED at the median redshift of the sources considered in the stack by co-adding the model SEDs at each redshift, weighted by the redshift distribution of stacked sources. The two methods return far-IR luminosities that are in good agreement, although not accounting for the redshift distribution of the stacked sources results in higher  \lir/\md ~values. This is  expected as in this case the artificial broadening drives the fit, erroneously, to higher $\gamma$ values, i.e. higher contribution of the PDR component, that result in lower \md ~ for a given \lir. We also validate the reliability of the inferred dust masses of the stacked samples against a possible artificial broadening of the SEDs from the dispersion of the shape of the SED of the sources included in the stacking. Namely, based on the value of $\langle U \rangle$ derived from the best fit to the stacked data, we generated 1000 artificial SEDs adopting a scatter of 0.2~dex in $\langle U \rangle$  and  assigned to each of them a redshift based on the redshift distribution of the original sample. Then we produced an average SED that we fit in the same manner as the original data. We find that the derived $\langle U \rangle$ value for the simulated data is in very good agreement with the one obtained for the real data. The best fit models for the various sSFR and stellar mass bins are shown in Figure \ref{fig:stack}, along with the stacked photometric points while the inferred parameters are presented in Table 4. We note that for every sSFR bin, the derived \lir ~implies an average SFR, and subsequently an average sSFR, that is very close to the initial corrected for extinction, UV-based  sSFR estimate. This is also the first time that \h ~data provide evidence, that the thickness of the main sequence at $z = 2$ is real and not a due to scatter arising from uncertainties in the derivation of SFR.
\begin{figure*}
\centering
\includegraphics[scale=0.39]{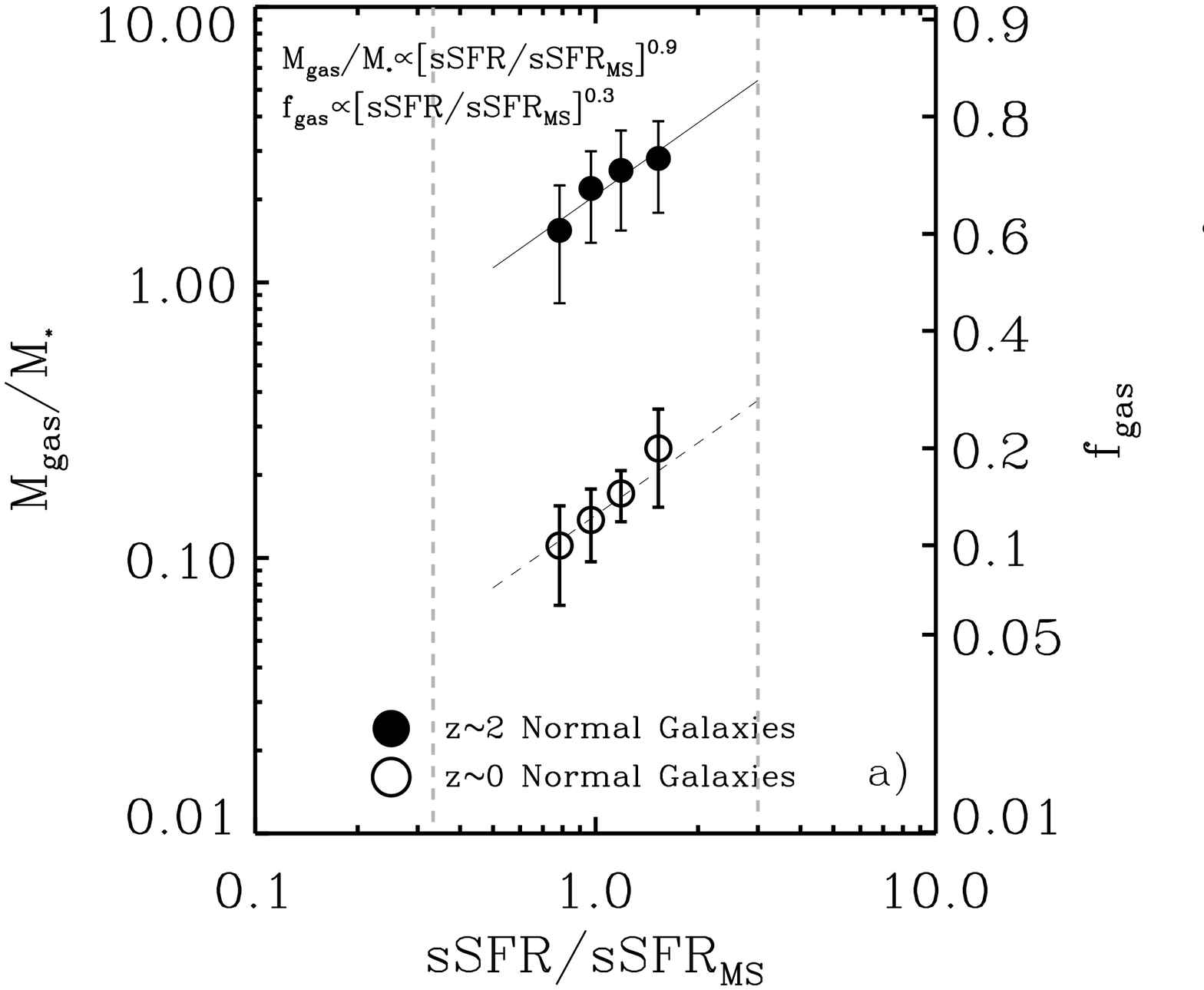}
\includegraphics[scale=0.39]{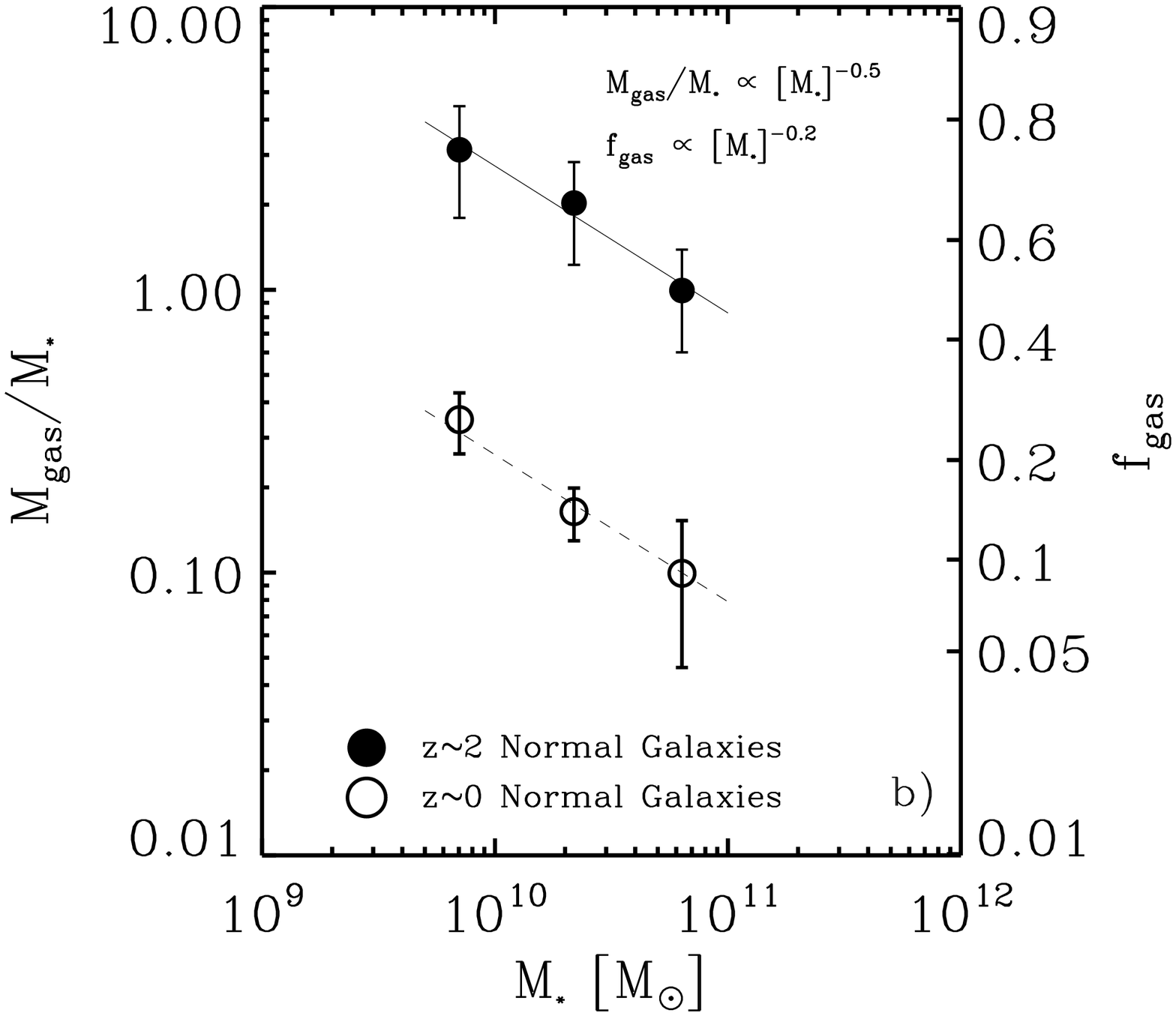}
\caption{ {\bf a)} Gas to stellar mass ratios (and gas fractions) as a function of deviation from the main sequence for the stacked sample 
of $z \sim 2.0 $ (filled circles) and the mean values of $z \sim 0 $ normal galaxies. Both samples are consistent with \mgas/$M_{\ast}$ $\propto$ [sSFR/sSFR$_{\rm MS}$]$^{0.9}$  suggesting a strong increase of  \mgas/$M_{\ast}$ (and \fg) with increasing sSFR at a given redshift. The solid 
black line is the best fit to the $z \sim 2$ data, and the dashed line has the same slope,  normalised to the local sample.  
The grey horizontal lines define the region of MS galaxies. {\bf b)}  Gas to stellar mass ratios (and gas fractions) 
as a function of stellar mass for the same sample as in the left panel. The data suggest a decrease of  \mgas/$M_{\ast}$ ~with increasing $M_{\ast}$ with  \mgas/$M_{\ast}$ ~$\propto M^{-0.5}_{\ast}$.}
\label{fig:fgas} %
\end{figure*}

As a reference point, we also consider the large compilation of normal galaxies in the nearby universe ($\langle z \rangle \approx 0.05$) and local ULIRGs by da Cunha et al.\ (2010a,b). The dust mass estimates in these studies are derived based on a two-component fit for warm and cold dust, known to give results consistent with those from the DL07 models (Magrini et al.\ 2010). Therefore, a direct comparison with our sample is meaningful after applying a correction factor of $\sim$ 1.7 to the dust masses of the Da Cunha et al. (2010) sample, to account for the different $\kappa$ value adopted in their study.

Figure \ref{fig:stack} reveals a remarkable similarity in the average SEDs of $z \sim 2$ MS galaxies for different sSFR bins, indicating a very small variation of the SED shape of the galaxies within the MS. This is further manifested by the weak dependence of \lir/\md ~with sSFR/sSFR$_{\rm MS}$ (inset panel), both for the $z \sim$ 2 sample and also for the local, normal galaxies (slope of 0.11 $\pm 0.04$). We reach similar conclusions when considering the SEDs for different stellar mass bins:  the shape of the SED and the value of \lir/\md ~for both $z \sim 2$  and $z \sim 0$ MS galaxies appear to be independent of the stellar mass. The observed trends between \lir/\md ~vs.\ sSFR/sSFR$_{\rm MS}$ and \lir/\md ~vs.\ $M_{\ast}$ are in striking agreement with those derived by our analytical approach, favouring a small variation of SFE within the MS, and therefore i) the existence of a global SF-law for MS galaxies, and ii) a step-like dependence of SFE with sSFR/sSFR$_{\rm MS}$. Indeed, the scenario of continuously increasing SFE would imply a considerable increase of  \lir/\md ~(shown in the inset panel of Figure \ref{fig:stack} (top) with a dashed line), and a noticeable change of SED shape of galaxies within the MS, something that is not supported by our data.

Our data can also be used to infer the dependence of \fg ~on sSFR and $M_{\ast}$. Namely,  
from the derived \md ~ and Z values, we can use the \gdr $- Z$ relation to estimate \mgas, and subsequently \fg\ or \mgas/$M_{\ast}$. In Figure \ref{fig:fgas}, we plot the derived  \mgas/$M_{\ast}$~as a function of sSFR/sSFR$_{\rm MS}$ and $M_{\ast}$ for the normal galaxies at $z \sim 2.0 $ and $z \sim 0$. We find a clear trend of increasing \mgas/$M_{\ast}$ ~with increasing sSFR for both samples,  
with:
\begin{equation}
\begin{centering}
 {\rm \frac{M_{\rm gas}} {M_{\ast}} =(2.05\pm0.32) \times [\frac{sSFR}{sSFR_{\rm MS}}]^{0.87\pm0.15}} 
\end{centering}
\end{equation}
\noindent very close to the slope  derived in equation 17. This result further supports that it is variations of \fg, rather than variations of star formation efficiency, that are responsible for the thickness of the SFR-$M_{\ast}$ relation at any redshift. We also reveal a clear trend of decreasing \fg ~with increasing stellar mass, with:
\begin{equation}
\begin{centering}
 {\rm log(\frac{M_{\rm gas}} {M_{\ast}}) =(5.63\pm0.39) - (0.51\pm0.10) \times log(M_{\ast})} 
\end{centering}
\end{equation}
\noindent ($M_{\ast}$ in \msol) in agreement with various studies that predict a similar behavior (e.g., Daddi et al.\ 2010a, Saintonge et al.\ 2011, Dav\'e et al. 2012, Popping et al.\ 2012, Fu et al.\ 2012). Note that these trends  are due to the 
sublinear slopes of the SFR vs.\ $M_{\ast}$ and \mgas ~vs.\ SFR relations (see Daddi et al 2010a).

We recall that the \mgas ~derived with the \gdr ~method traces the total gas of a galaxy, i.e., $M_{\rm H_{\rm 2}}$ + $M_{\rm HI}$. As we have discussed above, in our analysis we have assumed that $H_{\rm 2}$ dominates the gas mass of relatively massive ($M_{\ast} > 10^{10}$~\msol) high-z galaxies. An indirect way to investigate this assumption is to compare the inferred \mgas ~and \lir ~of the $z \sim 2$ sample to the \lir-$M_{\rm H_{\rm 2}}$ ~star formation law for MS galaxies as derived by Daddi et al.\ (2010), based on the dynamical properties of the individually detected  BzK galaxies considered in our study. If $M_{\rm HI}$ contributed significantly to the total gas mass, then the galaxies would appear to show an excess in $M_{\rm H_{\rm 2}}$ with respect to \lir. This comparison, for various subsets of our $z \sim 2$ sample, is shown in Figure 13. 
The very good agreement between the estimated \mgas ~and the SF-law seems to validate our assumption (i.e., negligible contribution from HI), at least for the stellar mass range considered in this study. Indeed, our \md\ measurements are (obviously) luminosity-weighted, and thus effectively SFR-weighted. Similarly, the metallicity estimates that we adopt from the literature are luminosity/SFR-weighted as well, so it is reasonable that 
our final estimates are most sensitive to the gas connected to the ongoing SFRs, i.e. the molecular gas.

\begin{figure}[!h]
\centering
\includegraphics[scale=0.39]{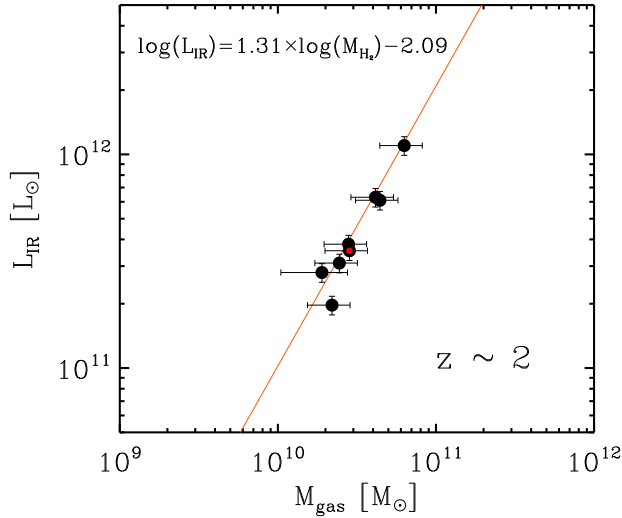}
\caption{ \lir ~vs.\ \mgas ~for various stacked subsets   in sSFR and $M_{\ast}$ bins (black circles) as well as of the total sample (black circle with red dot) of $z \sim2 $ normal galaxies. The line depicts the \lir-$M_{\rm H_{\rm 2}}$ star formation law of local and high$-z$ disk galaxies presented by Daddi et al. (2010b).}
\label{fig:lirm} %
\end{figure}
\section{THE EVOLUTION OF \lir/\md ~IN MS GALAXIES}

Aside from the similar trend of a weak increase of \lir/\md ~with sSFR for $z \sim 2$ and $z \sim 0$ normal galaxies, Figure 
\ref{fig:stack}  implies that the shape of the SED of MS galaxies at a given redshift, as traced by \lir/\md, is not expected to vary 
significantly with increasing \lir, (i.e., SFR). Moreover, normal galaxies at $z \sim 2$ appear to exhibit systematically higher 
values of \lir/\md ~than their local counterparts, suggesting a possible trend of \lir/\md ~with cosmic time.  To  explore this 
further, we enhance our data set by performing SED fitting to the mean SED of MS galaxies at $z \sim 1$ derived through the 
stacking procedure described in Section 2. The stacked data points and best fit models for the $z \sim 1$ and $z \sim 2$ MS 
galaxies are shown in Figure \ref{fig:sed4}. The final sample available for our analysis consists of $z \sim 0$ normal galaxies 
from da Cunha et al.\ 2010 and Draine et al.\ (2007), local ULIRGs from da Cunha et al.\ 2010b, individually detected MS 
galaxies at $z \sim 0.5$ and $z \sim 1.5$, mean SEDs of MS galaxies at $z \sim 1$ and $z \sim 2$, as well as SMGs at various 
redshifts. We also recall that \lir/\md ~is proportional to the mean radiation field  $\langle U \rangle$  ~in the DL07 models 
(equation 4). Using the derived  $\langle U \rangle$, \lir ~and \md ~values for our individuals sources as well as for the stacked 
samples we find an average{\footnote {There is a small scatter ($<5\%$) in the derived average scaling factor due our definition of \lir as $L_{\rm 8-1000\,\mu m}$ instead of \lir=$L_{\rm 
0-\infty}$.}  scaling factor between the two quantities of  $P_{0} \approx 125$.
\begin{figure}
\centering
\includegraphics[scale=0.35]{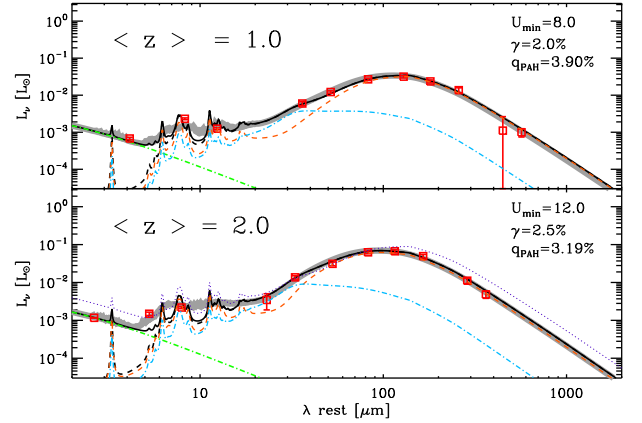}
\caption{Observed average SEDs of main sequence galaxies at 
$z \sim 1.0$ (top) and $z \sim 2.0$ (bottom) derived by stacking and overlaid with the best fit Draine \& Li 2007 models taking into account the redshift distribution of the stacked sources (grey solid line) 
and assuming the median redshift of each sample (solid black line). The black dashed line is the DL07 model without the stellar component that is depicted with a green dotted-dashed line. Orange dashed and cyan dot-dashed lines show separate contributions of starlight and emission from dust heated by $U=U_{min}$ (diffuse ISM component) and dust heated by $U_{min} < U < U_{max}$ (``PDR'' component) respectively. The fitted parameters from the best fit Draine \& Li (2007) model fits are listed within each panel. For comparison the best fit 
SED of the $z \sim1.0$, is also shown in the bottom panel with a blue-dotted line.}
\label{fig:sed4} %
\end{figure}
\begin{figure*}
\centering
\includegraphics[scale=0.39]{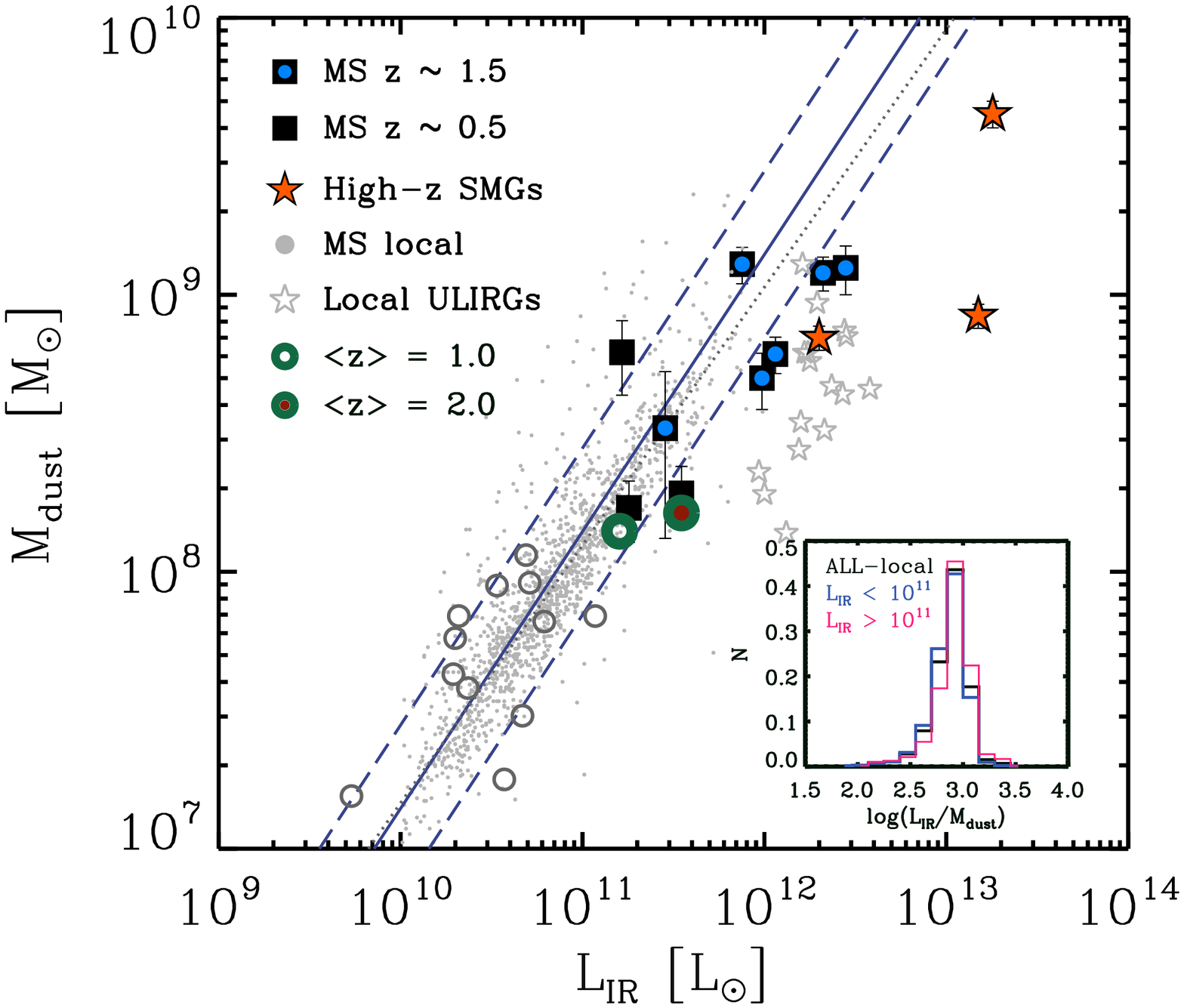}
\includegraphics[scale=0.39]{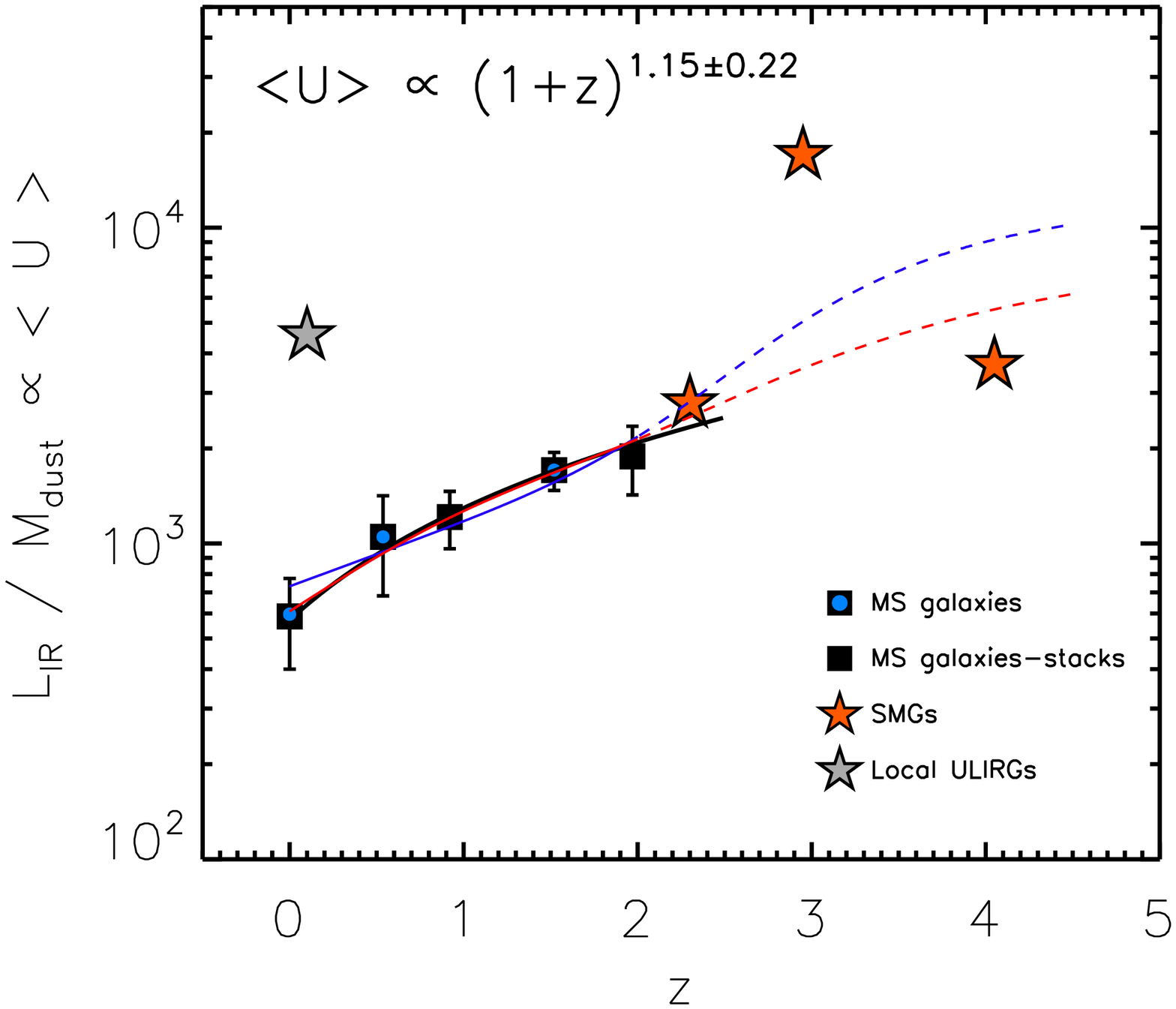}
\caption{{\bf Left:} Dust mass as a function of infrared luminosity for: local normal galaxies from the sample of da Cunha et al.\ (2010) and the SINGS sample of Draine et al.\ 2007 (grey points and circles respectively), local ULIRGs from da Cunha et al. 2010b (grey stars), individually detected MS galaxies at $z \sim 0.5$ (black squares) and $z \sim1.5$ (black squares with blue circle), stacked MS galaxies at $z \sim1.0$ (empty green circle) and $z \sim 2.0$ (filled green circle) and high$-z$ SMGs (orange stars).
The solid line corresponds to a linear relation between \md ~and \lir, consistent with the data for local normal galaxies, although the best fit yields a slope of 0.95 (grey dotted line). The dashed lines show the linear relation with a $\pm$ 0.3~dex offset. 
{\bf Right:} Evolution of the mean radiation field $\langle U \rangle \propto$ \lir/\md ~ 
as a function of redshift. Black squares with blue circles depict individually detected normal star-forming galaxies at various redshifts. At $z \sim 0$ we use the average value as derived based on the samples of Draine et al.\ (2007), da Cunha et al.\ (2010) and Dale et al.\ (2012) while at high$-z$ the median values from the individually detected sources at  $z \sim 0.5$ and at $z \sim 1.5$. Black squares 
represent the stacking results  at $z \sim 1$ and $z \sim 2$. Orange stars represent high$-z$ SMGs considered in the study, while the grey star denotes the position of the local ULIRGs based on data from da Cunha et al.\ (2010b). The black solid line is the best fit to the MS galaxies,  yielding a relation of $\langle U \rangle \propto (1+z)^{1.15}$. The blue and red solid lines show  the expected evolution of SFE/Z with redshift  for the case of $M_{\ast} = 1 \times 10^{10}$ \msol ~and $M_{\ast} = 5.0 \times 10^{10}$ \msol ~respectively, highlighting a co-evolution between SFE/Z and $\langle U \rangle$.}

\label{fig:shape} %
\end{figure*}

In Figure  \ref{fig:shape} (left) we plot \md ~as a function of \lir ~for the whole set of sources described above. Focusing at first on the local samples, we see that $z\sim0$, normal galaxies follow at tight relation with a slope close to but not equal  to unity, ($\xi_{\rm ml}$ = 0.95), possibly mirroring the \lir~$-$ \td ~relation observed both in the local as well as in the high$-z$ universe  (e.g., Hwang et al.\ 2010). However, we notice that this is a very mild increase of \lir/\md. Indeed, the \lir/\md ~of the whole sample of the local normal galaxies exhibits a Gaussian distribution with a mean value of $\langle \log($\lir/\md)$\rangle = 2.89$ and a dispersion of 0.15~dex. To further examine whether sources with higher \lir ~tend to have significantly higher \lir/\md, we split the sample into two luminosity bins, \lir ~$ > 10^{11}$ \lsol ~and \lir ~$< 10^{11}$ \lsol, and find that the two sub-samples follow almost identical distributions (Figure \ref{fig:shape} left, inset panel). In contrast, local ULIRGs are strong outliers from the \md-\lir ~relation formed by normal galaxies, with significantly higher \lir/\md ~values, indicative of their warmer ISM.  The IRAS selection for the normal galaxies considered here 
could, in principle, introduce a bias towards warmer sources, and miss cooler sources whose SEDs peak at longer wavelengths. Recent \h ~observations of local galaxies seem to confirm this, revealing a population of normal galaxies in the local universe that are 
systematically colder than the IRAS selected sample (e.g., Smith et al.\ 2012). However, while the \h-selected galaxies would increase the scatter and shift the mean  of  \lir/\md ~distribution towards a lower value,  they are also found to follow a  \lir~$-$~\md ~relation with a slope close to unity, leaving the overall picture described above unaffected. 
At $z=0$ the ULIRGs have an order of magnitude larger $L_{\rm IR}/M_{\rm dust}$ than normal (spiral) galaxies, consistent with the fact that have
a much higher SFE.

Moving to the high$-z$ samples of MS galaxies, we notice a small deviation from the local relation, exhibiting on average higher \lir/\md ~values. Having already seen that we do not expect large variations of \lir/\md ~within MS galaxies at the same redshift, this indicates a small change in the shape of the SEDs of MS galaxies towards stronger radiation fields as we move back in cosmic time. This is clearly depicted in Figure \ref{fig:shape} (right) where we plot \lir/\md, or equally  $\langle U \rangle$,  as a function of redshift, for our samples, along with that of local galaxies. Our data suggest an increase of $\langle U \rangle \propto (1+z)^{1.15}$ for MS galaxies, pointing also towards an increase in their \td ~with look-back time. This is not to be confused with the general conception that ULIRGs tend to get colder at higher redshifts. Indeed, when comparing high$-z$ MS ULIRGs to local galaxies of comparable \lir~ (i.e., (U)LIRGs), the latter have much larger  \lir/\md ~values, indicative of a much stronger starlight intensity field and a much warmer ISM, in line with recent studies that find distant ULIRGs to be colder than their local counterparts  (e.g., Symeonidis et al.\ 2009, Hwang et al.\ 2010,  Magdis et al.\ 2010, Muzzin et al.\ 2010).  We also note that the evolution of \td ~with redshift is directly demonstrated by the comparison of SEDs of the $z \sim 2$ and $z \sim 1$ samples, with the latter peaking at a longer wavelength. Since our stacks consist of galaxies with photometric redshift too, we also explored the possibility that the difference in the derived $\langle U \rangle$, arises from a systematic bias in the redshift distribution of the stacked samples. However, we find that for the two samples to peak at the same wavelength, this bias should be $\Delta$ $z \approx$ 0.35, which is unlikely given that the systematics in the derived photometric redshifts is $\Delta$ $z << 0.1$ (e.g., Grazian et al.\ 2006, Santini et al.\ 2009, Pannella et al. in prep).  

As we have seen in  equation 19, the dust-weighted luminosity, (or mean radiation field) 
is directly proportional to SFE/Z. It is easy to show how a similar evolution with redshift of  SFE/Z  can be expected based on a simple analytical approach.
Two trends would drive the evolution of SFE/Z. One is the 
evolution (decrease) of metallicity at given stellar mass ($C_{Z}(z)$) derived by the mass-metallicity relation at various redshifts (Tremonti et al.\ 2004 for z=0, Savaglio et al.\ 2005 for $z = 0.5$, Erb et al.\ (2006) for $z=2$, Sommariva et al.\ 2012 for $z = 3$).  The other is the evolution of the normalization of the sSFR$_{\rm MS}$ at fixed stellar mass ($C_{MS}(z)$), which scales as (1+z)$^{2.8}$. 
At fixed stellar mass, we have:
\begin{equation}
{\rm \log\frac{SFE}{Z}(z)  = C + (1-\xi_{\rm gas}) \times C_{MS}(z) +C_{Z}(z)}
\end{equation}
\noindent and the emerging trend of SFE with redshift for the case of $M_{\ast} = 1 \times 10^{10}$ \msol\ and 
$M_{\ast} = 5.0 \times 10^{10}$ \msol ~are shown in Figure \ref{fig:shape} (right), highlighting the striking agreement between the observed and the predicted trends. We stress that the metallicity evolution is mapped up to $z \sim 3$, so the values beyond this redshift are based on extrapolation. However, 
at very low metallicities, the SFR and \lir ~could deviate from their linear relation, with \lir\
tracing only a small fraction of the total star formation activity as dust attenuation becomes small and the directly-visible UV emission becomes significant. In that scenario, we would expect some flattening of the SFE evolution beyond $z \sim 2$. In any case, we conclude that the observed increase of $\langle U \rangle$ with redshift up to $z \sim 2.0$ is in excellent agreement with the expectations 
 based on our analytical approach. We stress that the observed increase of 
$\langle U \rangle$ with redshift is in line with both alternative scenarios, i.e, a weak and strong dependence of SFE on sSFR/sSFR$_{\rm MS}$ for galaxies within the main sequence. 

\begin{figure*}
\centering
\includegraphics[scale=0.6]{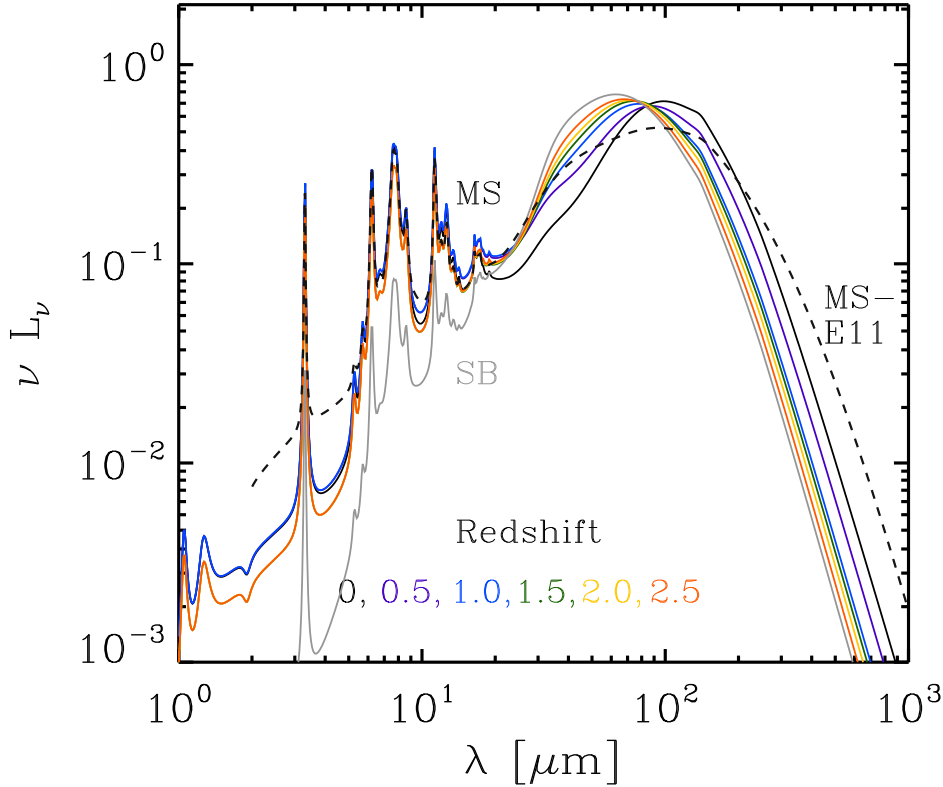}
\caption{Template SEDs of MS galaxies at various redshift built based on DL07 models and the evolution of $\langle U \rangle$ with cosmic time as (1+z)$^{1.15}$. The templates assume no evolution of the $\gamma$ parameter in the DL07, i.e, the fractional contribution of the PDRs to the total IR emission of a galaxy and a small evolution of the $q_{PAH}$ beyond $z\sim2.0$, as indicated from our data. The templates are color-coded based on their redshift and normalised to 1\lsol. For starburst galaxies we show the best fit template of GN20 (grey line). The black-dashed line shows the 
MS template SED of Elbaz et al. (2011). The templates can be found at {\it http://georgiosmagdis.pbworks.com}}
\label{fig:uz} %
\end{figure*}

\subsection{Template SEDs of MS Galaxies}
We have seen that MS galaxies at a given redshift tend to have uniform radiation fields, parametrized by \lir/\md, and an overall small variation in  shape of their SED. We also presented evidence that the mean radiation field of MS galaxies increases with redshift as $\sim (1+z)^{1.15}$.  
This indicates that we could build template SEDs of MS galaxies at different redshifts. In the DL07 models, the mean radiation field $\langle U \rangle$ is an interplay between $\gamma$, which is the fraction of dust emission arising from PDRs, and $U_{min}$, which is the radiation field heating the diffuse ISM. We stress that the same $\langle U \rangle$ could arise from a different combination of $\gamma$ and $U_{min}$ and it does not monotonically define the shape of the SED. Although our sample is not 
sufficiently large to break this degeneracy, we can use the mean $\langle U \rangle$ values at each redshift to construct template SEDs of MS galaxies, under the assumption that $\gamma$ does not change significantly with time. We adopt a constant $\gamma$ = 0.02, similar to that reported for local galaxies by Dale et al.\ (2012) and Draine et al.\ (2007). Indeed, if we were to reproduce the same $\langle U \rangle$, with higher $\gamma$ values then the shape of the SED would become unphysically flat close to the peak. Also, the galaxies in our sample have a mean $\gamma$ very close to that value,  so it seems like a valid simplification. Then, using the relation between $\langle U \rangle$ and redshift derived in the previous section, we built template SEDs of MS galaxies for a grid of redshifts up to $z = 2.5$ using the DL07 models. Beyond $z > 2.5$, we assume a flattening of the evolution of $\langle U \rangle$, similar to that observed for the sSFR (e.g., Gonzalez et al. 2010). 
For the remaining free parameter, $q_{PAH}$, we adopt the mean value of 3.19\% up to redshift $z \sim 1.5$ and  $q_{PAH} = 2.50\%$ at $z > 1.5$ as indicated by our data. The resulting MS SEDs are shown in Figure \ref{fig:uz}. 

We note that the overall philosophy is very similar to that of Elbaz et al.\ (2011),
who presented  a single template for MS galaxies, with a unique far-IR to mid-IR luminosity ratio, IR8, at all redshifts up to $z > 2$ and a single template for star-bursting systems with a significantly higher value of IR8.  Here, we extend this approach, but introduce some evolution of the SEDs of MS galaxies, following the increase of  $\langle U \rangle$ with redshift. Our templates result in IR8 values
that also mildly increase from IR8 $\sim$ 4.0 at $z = 0$ to IR8 $\sim$ 6.0 at $z > 2$, in agreement with Elbaz et al.\ (2011; and in preparation) and with Reddy et al.\ (2012).
Similar to Elbaz et al.\ (2011), for starbursts galaxies, we show here a unique template, using the 
best fit SED of GN20. Despite the simple and straightforward nature of these templates, we note that B\'ethermin et al.\ (2012b) find that they manage to reproduce recent \h\ source counts (e.g., Berta et al.\ 2011; B\'ethermin et al.\ 2012), including counts measured in redshift slices. We note that to investigate the contribution of an AGN in our templates, we also repeated the SED fitting for the stacked samples, excluding this time photometric points that are likely to be ``polluted'' by an AGN activity (i.e., $\lambda_{\rm rest} < 60\,\mu$m). The inferred $\langle U \rangle$ values are very close to the values derived when considering the whole set of photometric points, in agreement with various studies that find a negligible effect of an AGN in the far-IR colours of a galaxy (e.g., Hatziminaoglou et al. 2010, Mullaney et al. 2012). We conclude that our templates should be representative for the bulk of MS galaxies, even for those that contain an average AGN contribution in the mid-IR. The templates can be found at {\it http://georgiosmagdis.pbworks.com}.

\section{DISCUSSION}
It seems that the main physical parameter that drives the star formation rate of  a normal galaxy is the gas fraction (or \mgas/$M_{\ast}$). Indeed, according to our analysis, at fixed stellar mass and redshift,  there is a strong dependence of sSFR on
\mgas/$M_{\ast}$, suggesting that the variations within the main sequence are due to variations in the \fg ~of a galaxy. Interestingly, 
we find that within the MS, \mgas/$M_{\ast}$ varies almost linearly with sSFR/sSFR$_{\rm MS}$ (slope of $\sim$0.9). It is important to recall that for the gas fractions inferred in this study we use the total gas mass of the galaxies. However,  \lir, and subsequently SFR, is known to strongly correlate with the HCN luminosity, which is a measure of dense gas ($H_{\rm 2}$ volume densities $>$10$^{4}$ cm$^{-3}$), for spirals, local (U)LIRGs and QSOs, with an almost linear relation (Gao \& Solomon 2004; Juneau et al.\ 2009). The inferred dependence of SFR to the total gas fraction, along with the fact the we do not observe substantial changes in the star formation efficiency, is indicative of a rather constant fraction of dense gas among MS galaxies. Alternatively, it suggests that the fluctuations of galaxies above and below MS are relatively long lasting, allowing the dense gas fraction to adjust. Indeed, given that the typical timescale for dense gas consumption is 10$^{8}$~yr (and given that it is only a fraction of the total gas),  fluctuations have to last a similar or longer time. This is also the timescale required to allow integrated U$-$V colors to change (Salmi et al 2012).  Furthermore, the  fact that \fg ~decreases with $M_{\ast}$, implies that accretion does not keep up with
SF  and outflows at least at $z < 2$. 

Another important implication of our analysis is that any attempt to constrain the evolution of 
\fg ~ (or \mgas/$M_{\ast}$) with cosmic time should be focused on narrow stellar mass ranges or take into account the \fg  ~dependence on both $M_{\ast}$ and sSFR. Rescaling the  \fg ~estimates of the total stacked samples of $z\sim1$ and $z\sim2$ main sequence galaxies  based on the derived 
trends between \mgas/$M_{\ast} -M_{\ast}$ and \mgas/$M_{\ast} -sSFR/sSFR_{\rm MS}$, in Figure 17 we show  the evolution of  \fg ~with redshift for galaxies of $sSFR/sSFR_{\rm MS}$ = 1 and $M_{\ast} = 5 \times 10^{10}$ \msol.  The evolution of \fg ~is in excellent agreement with the the observationally motivated tracks of Sargent et al. (2012b in prep), that are built assuming \mgas ~$\propto$ SFR$^{0.81}$ at all redshifts and an increase of sSFR  as $(1+z)^{2.8}$ up to $z \sim 2.5$ 
followed by a flattening towards higher redshifts. We note that recently, Magdis et al. (2012a) provided observational evidence for the plateau 
in the evolution of the \fg~ at $z > 2$.
\begin{figure}
\centering
\includegraphics[scale=0.32]{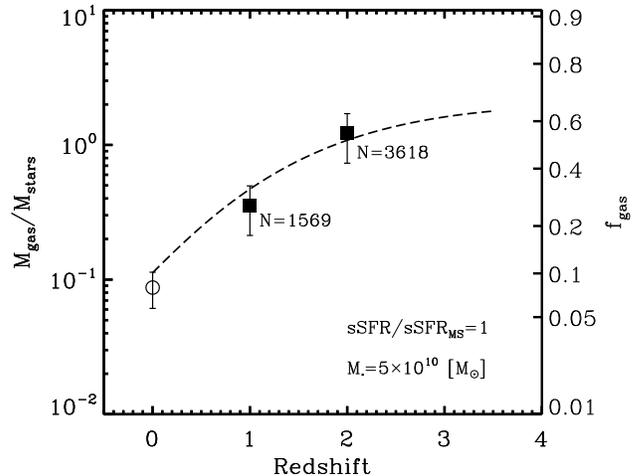}
\caption{Evolution of  \mgas/$M_{\ast}$ with redshift of galaxies with $sSFR/sSFR_{\rm MS} = 1$ and $M_{\ast} = 5\times10^{10}$ \msol. The black squares correspond to \mgas/$M_{\ast}$ of  the total stacked samples of $z \sim 1$ and $z \sim 2$ main sequence galaxies. The number next to the stacked points corresponds 
to the total number of objects that are included in the stack. The open circle at $z \sim 0$, corresponds to the mean  \fg ~of the  Leroy et al. (2008) sample.  All measurements have been corrected to refer to $M_{\ast} = 5\times10^{10}$ \msol. The black dashed line depicts the evolution of   \mgas/$M_{\ast}$ with redshift  for the case of  $M_{\ast}$ = 5.0 $\times$ 10$^{10}$ \msol, as implied by the cosmic evolution of sSFR (Sargent et al. 2012 in preparation).}
\label{fig:ffgas} %
\end{figure}

\begin{table*}
{\footnotesize
\caption{Derived properties for stacked sub-samples of $z \sim 2$   MS galaxies.}             
\label{tab:4}      
\centering                          
\begin{tabular}{l c c c c c c c c c c c c}        
\hline\hline                 
Stack$^{a}$ &$\log$ \lir & $\log$ \md& $\log <M_{\ast}> $ & $IR8^{b}$& $Z_{\rm PP04}$ & $\log$ \mgas$^{c}$ \\
& \lsol &$M_\odot$ &$M_\odot$ &&& $M_\odot$\\
\hline
sSFR1 & 11.44 $\pm$ 0.05 &8.18 $\pm$ 0.11 &10.09 &4.5$\pm$0.7 &8.44&10.28 $\pm$ 0.19\\
sSFR2 & 11.49 $\pm$ 0.05 & 8.26 $\pm$ 0.09 &10.04&4.7$\pm$0.7& 8.41&10.38 $\pm$ 0.13\\
sSFR3 & 11.58 $\pm$ 0.08 & 8.32 $\pm$ 0.09 &10.04&5.2$\pm$0.9&8.41&10.44 $\pm$ 0.16\\
sSFR4 & 11.80 $\pm$ 0.06 & 8.51 $\pm$ 0.09 &10.17&5.8$\pm$1.0&8.43&10.61 $\pm$ 0.13\\
mass1 & 11.29 $\pm$ 0.06 & 8.08 $\pm$ 0.10 &9.84&4.9 $\pm$0.7&8.35&10.34 $\pm$ 0.15\\
mass2 & 11.78 $\pm$ 0.06 & 8.52 $\pm$ 0.08 &10.33&5.0$\pm$0.8&8.48&10.64 $\pm$ 0.13\\
mass3 & 12.05 $\pm$ 0.05 & 8.83 $\pm$ 0.08 &10.80&4.9$\pm$0.7&8.65&10.80 $\pm$ 0.13\\
Total& 11.55 $\pm$ 0.06 & 8.27 $\pm$ 0.11 &10.10&5.1$\pm$0.6&8.40&10.45 $\pm$ 0.11\\
\hline                                   
Total-z1& 11.21 $\pm$ 0.09 & 8.14 $\pm$ 0.13 &10.44&4.5$\pm$0.6&8.62&10.04 $\pm$ 0.12\\
\hline
\end{tabular}\\
Notes:\\
a: The stacking sub-samples in sSFR and stellar mass bins are given in increasing order or $<$sSFR$>$/$sSFR_{\rm MS}$ and  $<M_{\ast}>$.
The last line corresponds to the total stacked sample of $z \sim 1$ MS galaxies. \\
b: Defined as \lir/$L_{\rm 8}$.\\
c: Estimated through the  \gdr$-Z$ relation based on the derived \md ~and the assumed $Z_{\rm PP04}$.}
\end{table*}

While it seems that  there is no value of \fg, stellar mass, or sSFR that could be regarded as representative for all MS galaxies at a given redshift. we find that MS galaxies appear to share very similar IR-SED shapes, with a characteristic $\langle U \rangle \propto$ \lir/\md ~at each cosmic epoch.  We find that MS galaxies tend to exhibit higher  \lir/\md ~values as we move back in time, suggesting that there is a evolution within the MS as a function of time towards warmer \td. Nevertheless, we stress that high-z MS galaxies with ULIRGs-like \lir, while warmer than local, normal galaxies, are colder than local ULIRGs, in agreement with various studies that find an evolution in the \td ~of ULIRGs (e.g., Hwang et al.\ 2010, Muzzin et al.\ 2010). On the other hand, the uniform SED shape of 
MS galaxies at a given redshift suggests that \td ~does not change substantially with \lir ~within the MS. Consequently, it is possible that the strong evolution of \td ~with \lir ~reported by several studies (e.g., Magdis et al.\ 2010, Hwang et al.\ 2010) is a result of mixing MS galaxies at various redshifts, or MS and starburst systems, the latter known to dominate the star-forming galaxy population at high luminosities (e.g., Rodighiero et al.\ 2011, Sargent et al.\ 2012).   Although the limited number of star-bursting galaxies in this study does not allow us to reach  
solid conclusions, based on Figure 15 (right)  we can speculate that a similar, but less strong, evolution of  $\langle U \rangle$, and therefore \td, with look-back time is present for this population too
(see B\'ethermin et al. 2012b for a possible implementation). A larger sample galaxies with elevated sSFR with respect to the MS is necessary to extend this investigation. 

We stress that the methodology presented in this study opens an alternative window for a systematic 
investigation of the gas properties for large samples of  high$-z$ galaxies,  for which \h ~has provided a robust  characterisation of their SEDs. While this approach does not require CO emission line measurements and is free from the uncertain nature of \aco, we stress that both techniques yield consistent results.
However, it is important to recall the limitations of this methodology. The key assumption of this study is that the \gdr $- Z$ relation observed in the local Universe does not evolve substantially with time, implying that roughly  half of the metals of high$-z$ galaxies as well are locked into dust. Evidently, if the fraction is lower at higher redshifts, then the inferred \mgas\ would be underestimated. However, as we argued before, this is unlikely to be a strong effect, the opposite might actually be more plausibly expected.  
We also stress that the \gdr $- Z$ relation is poorly constrained at the low metallicity end ($Z$ $<$ 8.0), where it might even become strongly non-linear. However, since the lowest metallicity considered in our study is  
$\sim 8.3$ (Figure 4 left, horizontal line), we do not expect to be affected from these uncertainties. Another possible caveat is also the assumed metallicities, for which we rely on empirical relations and measurements from the literature. For example, since the adopted relation to derive metallicities for the $ z \sim 2$ sample has been calibrated on UV-selected galaxies, one could imagine a possible bias towards underestimating the actual metallicity of galaxies in our sample (e.g.,  Onodera et al. 2010).  If indeed  the metallicity is underestimated, we would overestimate the true \mgas ~of the galaxies. However, we do not expect this effect to impact our main conclusions, given the flat shape of the observed stellar mass$-Z$ relation (e.g., Erb et al.\ 2006). We also note that while stacking data for thousands of sources erases interesting details on individual sources, it also serves as a tool to average out the dispersion in the $M_{\rm \ast}$ $-$ Z  and \gdr ~$-$ Z relations. However,  this method is sensitive to \md ~measurements that could, in principle, be affected by systematic biases, up to a factor of $\sim$ 3 if one allows for extreme 
variations in the properties of the dust grains. Also, we stress that using the templates presented in this study to infer the total \lir ~ of individual galaxies from a single detection in Rayleigh-Jeans tail, requires extreme caution due to the scatter in $\langle U \rangle$ which is not fully constrained at high$-z$. Future, direct observations of individual sources in the Rayleigh-Jeans tail will guide this investigation and also address the physical origin of the scatter, 
which could be linked to the thickness both of the $M_{\ast}-Z$ relation inside the main sequence as well as of the star formation law. Finally,  as indicated in Table 4, while the IR8 of the $z \sim 2$ stacked samples does not appear to vary with stellar mass (hence SFR), there is evidence that it could increase with increasing distance to the MS, in agreement with Elbaz et al. (2011). Variations in IR8, $<U>$ and possibly of other physical parameters  suggest that a single invariant SED for the whole MS should be treated as a good first order approximation.

\section{CONCLUSIONS}
We have used a sample of 9 normal galaxies at 
$z \sim 0.5$ and $z \sim1.5$ and 3 high$-z$ SMGs, all 
selected to benefit from low-J CO observations and high quality far-IR photometry, to estimate their gas content and \aco\ conversion factors. In addition, large stacked samples
of $>4000$ mass-selected main sequence galaxies at $0.5<z<2.5$ were used to investigate the evolution of their \md\ and \mgas\ content over the last 10 billions years.
The main findings and implications are summarized as follows:

\begin{itemize}

\item For high$-z$ normal galaxies we derive an average \aco ~value of $5.5 \pm 0.4$, similar 
to that of local spirals, and $\sim 5$ times higher that than of local galaxies with comparable \lir\ (i.e., (U)LIRGs). We also find that high$-z$ MS galaxies are ~$\sim 12$ times less efficient at converting molecular gas to stars when compared to local starbursts, with SFE $\sim$ 14 \lsol/\msol, corresponding to an average gas consumption time scale of 0.6 Gyr. 
\item For MS galaxies at any redshift, we find a weak dependence of the dust mass-weighted luminosity (\lir/\md), or equally of the mean radiation field  $\langle U \rangle$, on sSFR/sSFR$_{\rm MS}$ and $M_{\ast}$, complemented by a mild evolution with cosmic time as $(1+z)^{1.15}$. Also,  for \lir ~= $L_{\rm 8-1000\mu m}$, we find average an scaling factor between $\langle U \rangle$ and \lir/\md ~of $P_{\rm 0} \approx 125$.
\item We find  clear trends of  increasing \mgas/$M_{ast}$ (or equally of \fg) with increasing sSFR/sSFR$_{\rm MS}$, with a slope of $\xi_{frac1} \approx 0.9$ and decreasing \mgas/$M_{ast}$ ~with increasing $M_{\ast}$ with a slope of $\xi_{frac2} \approx -0.5$. Since at a given redshift  \fg~ varies with  stellar mass and  sSFR, 
any attempt to explore the evolution of \fg ~with redshift should take into account these emerging trends.
\item All the above, taken together, lead to the conclusion that variation in the gas fraction (\fg) is the driving physical parameter responsible for the spread in the SFR$-M_{\ast}$ correlation traced by normal galaxies at any redshift, and points towards a single, tight \lir$-$\mgas ~relation for MS galaxies.
\item We find that at a given redshift, MS galaxies seem to have uniform IR SED shapes, as parametrized by the mean radiation field $\langle U \rangle \propto$ \lir/\md ~of the DL07 models. Based on this result and on the derived  evolution of $\langle U \rangle$ with redshift, we build and provide a set of redshift-dependent template SEDs of MS galaxies at various redshifts, and show that the evolution of the SED of MS galaxies is primarily driven by the cosmic evolution of metallicity, and is only marginally  linked to the rise of average sSFR with redshift.
\end{itemize}

Together with the pilot study of Magdis et al.\ (2011), we have demonstrated for the first time the importance of studying the continuum SEDs of galaxies in order to place constraints on their gas content, through dust mass and metallicity estimates. We believe that this is a powerful technique:  until we are in position to formally estimate \aco ~ for high$-z$ galaxies using the resolving power of ALMA, it provides an alternative approach for the study of \mgas ~in galaxies throughout out cosmic time. However, from our detailed investigation we can point out two important results regarding the derivation of dust masses, that similar studies should keep in mind: 
\begin{itemize}
\item Dust mass estimates based a single temperature modified black body model are, on average, a factor of $\sim$ 2 lower compared to those derived using the physically motivated and more realistic models of Draine \& Li (2007).
\item While rest-frame submm data ($\lambda_{\rm rest} > 200\,\mu$m) reduce the uncertainties in the inferred \md ~by a factor of two, excluding mm continuum data from the fit does not have an impact on the derived \md ~estimates of MS galaxies. On the other hand, it appears that in the absence of mm continuum data, \md ~estimates of star-bursting systems can be grossly overestimated. 
\end {itemize}
\section{ACKNOWLEDGEMENTS}
GEM acknowledges support from the John Fell Oxford University Press (OUP) Research Fund and the University of Oxford and useful discussions with F. Galliano, DKX, LM and SB. E. Daddi, M. B{\'e}thermin and M. T Sargent were supported by grants ERC- StG UPGAL 240039 and ANR-08-JCJC-0008. Support for this work was also provided by NASA through an award issued by JPL/Caltech. PACS has been developed by a consortium of institutes led by MPE (Germany) and including UVIE (Austria); KU Leuven, CSL, IMEC (Belgium); CEA, LAM (France); MPIA (Germany); INAFIFSI/OAA/OAP/OAT, LENS, SISSA (Italy); IAC (Spain). This development has been supported by the funding agen- cies BMVIT (Austria), ESA-PRODEX (Belgium), CEA/CNES (France), DLR (Germany), ASI/INAF (Italy), and CICYT/MCYT (Spain). SPIRE has been de- veloped by a consortium of institutes led by Cardiff University (UK) and including Univ. Lethbridge (Canada); NAOC (China); CEA, LAM (France); IFSI, Univ. Padua (Italy); IAC (Spain); Stockholm Observatory (Sweden); Imperial College London, RAL, UCL-MSSL, UKATC, Univ. Sussex (UK); and Caltech, JPL, NHSC, Univ. Colorado (USA). This development has been supported by national funding agencies: CSA (Canada); NAOC (China); CEA, CNES, CNRS (France); ASI (Italy); MCINN (Spain); Stockholm Observatory (Sweden); STFC (UK); and NASA (USA). This paper used data from ESO program IDs 078.F-9028(A),
079.F-9500(A), 080.A-3023(A) and  081.F-9500(A).


\begin{thebibliography}{}
\bibitem[1988]{balluch}Berta, S., Magnelli, B., Nordon, R., et al.\ 2011, A\&A, 532, A49 
\bibitem[1988]{balluch}B{\'e}thermin, M., Dole, H., Beelen, A., \& Aussel, H.\ 2010, A\&A, 512, A78
\bibitem[1988]{balluch} B{\'e}thermin, M., Le Floc'h, E., Ilbert, O., et al.\ 2012a, A\&A, 542, A58
\bibitem[1988]{balluch} B{\'e}thermin, M., et al.\ 2012b, ApJ, 757, L23 
\bibitem[1988]{balluch}Bigiel, F., Leroy, A., Walter, F., et al.\ 2008, AJ, 136, 2846 
\bibitem[1988]{balluch}Blitz, L., \& Rosolowsky, E.\ 2006, ApJ, 650, 933 
\bibitem[1988]{balluch}Brinchmann, J., Charlot, S.,White, S. D. M., et al.\ 2004, MNRAS, 351, 1151
\bibitem[1988]{balluch} Bruzual, G., \& Charlot, S.\ 2003, MNRAS, 344, 1000 
\bibitem[1988]{balluch} Calura, F., Pipino, A., Matteucci, F.\  2008, A\&A, 479, 669
\bibitem[1988]{balluch} Calura, F., Pipino, A., Chiappini, C., Matteucci, F., \& Maiolino, R.\ 2009, A\&A, 504, 373 
\bibitem[1988]{balluch} Carilli, C. L., et al.\ 2010, ApJ, 714, 1407
\bibitem[1988]{balluch} Carilli, C. L., Hodge, J., Walter, F., Riechers, D., Daddi, E., Dannerbauer, H., Morrison, G. E.\ 2011, ApJ. 739, 33
\bibitem[1988]{balluch}Chabrier, G.\ 2003 PASP, 115, 763
\bibitem[1988]{balluch} Conley, A., Cooray, A., Vieira, J.~D., et al.\ 2011, ApJL, 732, L35 
\bibitem[1988]{balluch} Cowie, L.~L., Barger, A.~J., Hu, E.~M., Capak, P., \& Songaila, A.\ 2004, AJ, 127, 3137 
\bibitem[1988]{balluch}da Cunha, E., Eminian, C., Charlot, S., \& Blaizot, J.\ 2010, MNRAS, 403, 1894 
\bibitem[1988]{balluch} da Cunha, E., Charmandaris, V., D{\'{\i}}az-Santos, T., et al.\ 2010, A\&A, 523, A78 
\bibitem[1988]{balluch} Daddi, E., Cimatti, A., Renzini, A., et al.\ 2004, ApJ, 617, 746 
\bibitem[1988]{balluch} Daddi, E., Dickinson, M., Morrison, G., et al.\ 2007, ApJ, 670, 156 
\bibitem[1988]{balluch} Daddi, E., Dannerbauer, H., Elbaz, D., Dickinson, M., Morrison, G., Stern, D., Ravindranath, S.\ 2008, ApJ, 673, 21
\bibitem[1988]{balluch} Daddi, E., Dannerbauer, H., Krips, M., Walter, F., Dickinson, M., Elbaz, D., Morrison, G. E.\ 2009, ApJ, 694, 1517
\bibitem[1988]{balluch} Daddi, E., Bournaud, F., Walter, F., Dannerbauer, H., et al.\ 2010, ApJ, 713, 686
\bibitem[1988]{balluch} Daddi, E., Elbaz, D., Walter, F., et al.\ 2010b, ApJ, 714, 118
\bibitem[1988]{balluch} Dannerbauer, H.; Daddi, E., Riechers, D. A., Walter, F., Carilli, C. L., Dickinson, M., Elbaz, D., Morrison, G. E.\ 2009, ApJ, 698, 178
\bibitem[1988]{balluch}Dale, D.~A., Aniano, G., Engelbracht, C.~W., et al.\ 2012, ApJ, 745, 95 
\bibitem[1988]{balluch}Dame, T.~M., Hartmann, D., \& Thaddeus, P.\ 2001, ApJ, 547, 792 
\bibitem[1988]{balluch}Dav{\'e}, R., Finlator, K., \& Oppenheimer, B.~D.\ 2012, MNRAS, 421, 98 
\bibitem[1988]{balluch} Downes, D. \& Solomon, P. M.\ 1998, ApJ, 507, 615
\bibitem[1988]{balluch} Draine, B. T. \& Li, A.\ 2007, ApJ, 657, 810
\bibitem[1988]{balluch} Draine, B. T., et al.\ 2007, ApJ, 663, 866
\bibitem[1988]{balluch}Dunne, L., Eales, S., Edmunds, M., et al.\ 2000, MNRAS, 315, 115 
\bibitem[1988]{balluch}Dunne, L., \& Eales, S.~A.\ 2001, MNRAS, 327, 697 
bibitem[1988]{balluch}Edmunds, M.~G.\ 2001, MNRAS, 328, 223 
\bibitem[1988]{balluch}Elbaz, D., Daddi, E., Le Borgne, D., et al.\ 2007, A\&A, 468, 33
\bibitem[1988]{balluch}Elbaz, D., Hwang, H.~S., Magnelli, B., et al.\ 2010, A\&A, 518, L29
\bibitem[1988]{balluch}Elbaz, D., Dickinson, M., Hwang, H. S., et al.\ 2011, A\&A, 533, 119
\bibitem[1988]{balluch}Erb, D.~K., Shapley, A.~E., Pettini, M., et al.\ 2006, ApJ, 644, 813 
\bibitem[1988]{balluch}Feldmann, R., Gnedin, N.~Y., \& Kravtsov, A.~V.\ 2012, ApJ, 747, 124
\bibitem[1988]{balluch} Frayer, D.~T., Huynh, M.~T., Chary, R., et al.\ 2006, ApJL, 647, L9
\bibitem[1988]{balluch}Fu, J., Kauffmann, G., Li, C., \& Guo, Q., 2012, arXiv:1203.5280 
\bibitem[1988]{balluch}Galliano, F., Hony, S., Bernard, J.-P., et al.\ 2011, A\&A, 536, A88
\bibitem[1988]{balluch}Gao, Y., \& Solomon, P.~M.\ 2004, ApJS, 152, 63 
\bibitem[1988]{balluch}	Geach, James E., Smail, Ian,  Moran, Sean M., MacArthur, Lauren A., Lagos, Claudia del P., Edge, Alastair C.\ 2011, ApJ, 730, 19
\bibitem[1988]{balluch}Genzel, R., Tacconi, L. J., Gracia-Carpio, J., et al.\ 2010, MNRAS, 407, 2091
\bibitem[1988]{balluch}Genzel, R., Tacconi, L. J., Combes, F., et al.\ 2012, ApJ, 746, 69
\bibitem[1988]{balluch} Giavalisco, M., Ferguson, H.~C., Koekemoer, A.~M., et al.\ 2004, ApJ, 600, L93 
\bibitem[1988]{balluch}Grazian, A., Fontana, A., de Santis, C., et al.\ 2006, A\&A, 449, 951 
\bibitem[1988]{balluch}Gonz{\'a}lez, V., Labb{\'e}, I., Bouwens, R.~J., et al.\ 2010, ApJ, 713, 115
\bibitem[1988]{balluch} Hatziminaoglou, E., Omont, A., Stevens, J.~A., et al.\ 2010, A\&A, 518, L33 
\bibitem[1988]{balluch} Hodge, J., et al.\ 2012, ApJ in press, arxiv:1209.2418
\bibitem[1988]{balluch}Hwang, H.~S., Elbaz, D., Magdis, G., et al.\ 2010, MNRAS, 409, 75 
\bibitem[1988]{balluch} Inoue, A.~K.\ 2003, PASJ, 55, 901 
\bibitem[1988]{balluch}Israel, F.P.\ 1997 A\&A 328, 471
\bibitem[1988]{balluch} Ivison, R.~J., Swinbank, A.~M., Swinyard, B., et al.\ 2010, A\&A, 518, L35 
\bibitem[1988]{balluch}Ivison, R.~J., Papadopoulos, P.~P., Smail, I., et al., 2011\ MNRAS, 412, 1913 
\bibitem[1988]{balluch}Jauzac, M., Dole, H., Le Floc'h, E., et al.\ 2011, A\&A, 525, A52
\bibitem[1988]{balluch}Juneau, S., Narayanan, D.~T., Moustakas, J., et al.\ 2009, ApJ, 707, 1217 
\bibitem[1988]{balluch}Karim, A., Schinnerer, E., Mart{\'{\i}}nez-Sansigre, A., et al.\ 2011, ApJ 730, 61
\bibitem[1988]{balluch}Kennicutt, R.C.\ 1998a, ARA\&A 36, 189
\bibitem[1988]{balluch}Keres, D., Yun, M.~S., \& Young, J.~S.\ 2003, ApJ, 582, 659 
\bibitem[1988]{balluch}Kewley, L. J. \& Ellison, S. L. 2008, ApJ, 681, 1183
\bibitem[1988]{balluch}Lara-L{\'o}pez, M.~A., Cepa, J., Bongiovanni, A., et al.\ 2010, A\&A, 521, L53
\bibitem[1988]{balluch} Le Borgne, D.; Elbaz, D.; Ocvirk, P.; Pichon, C.\ 2009, A\&A, 504, 727
\bibitem[1988]{balluch}Leroy, A. K., Walter, F., Brinks, E., Bigiel, F., de Blok, W. J. G., Madore, B., Thornley, M. D. 2008\ AJ, 136, 2782
\bibitem[1988]{balluch}Leroy, Adam K., Bolatto, Alberto, Gordon, Karl, et al.\ 2011, ApJ, 737, 12 
\bibitem[1988]{balluch}Li, A. \& Draine, B. T.\ 2001, ApJ, 554, 778
\bibitem[1988]{balluch} Magdis, G. E., Rigopoulou, D., Huang, J.-S., Fazio, G. G.\ 2010a, MNRAS, 401, 1521
\bibitem[1988]{balluch} Magdis, G. E., Elbaz, D., Daddi, E., Morrison, G. E., Dickinson, M., Rigopoulou, D., Gobat, R., Hwang, H. S.\ 2010b, ApJ, 714, 1740
\bibitem[1988]{balluch}Magdis, G. E., Daddi, E., Elbaz, D., et al.\ 2011, ApJ, 740, 15
\bibitem[1988]{balluch} Magdis, G.~E., Daddi, E., Sargent, M., et al.\ 2012, ApJ, 758, 9
\bibitem[1988]{balluch}Magnelli, B., Lutz, D., Santini, P., et al.\ 2012, A\&A, 539, A155 
\bibitem[1988]{balluch}Magrini, L., Bianchi, S., Corbelli, E., et al.\ 2011, A\&A, 535, A13
\bibitem[1988]{balluch}Mannucci, F., Cresci, G., Maiolino, R., Marconi, A., Gnerucci, A.\ 2010, MNRAS, 408, 2115
\bibitem[1988]{balluch}Morrison, G. E., Owen, F. N., Dickinson, M., Ivison, R. J., Ibar, E.\ 2010, ApJS, 188, 178
\bibitem[1988]{balluch}Mu{\~n}oz-Mateos, J.~C., Gil de Paz, A., Boissier, S., et al.\ 2009, ApJ, 701, 1965 
\bibitem[1988]{balluch}Mullaney, J.~R., Pannella, M., Daddi, E., et al.\ 2012, MNRAS, 419, 95 
\bibitem[1988]{balluch}Muzzin, A., van Dokkum, P., Kriek, M., et al.\ 2010, ApJ, 725, 742 
\bibitem[1988]{balluch}Narayanan, D., Krumholz, M., Ostriker, E. C., Hernquist, L.\ 2011, MNRAS, 418, 664
\bibitem[1988]{balluch} Narayanan, D., Krumholz, M.~R., Ostriker, E.~C., \& Hernquist, L.\ 2012, MNRAS, 421, 3127 
\bibitem[1988]{balluch}Noeske, K. G., Weiner, B. J., Faber, S. M., et al.\ 2007, ApJ, 660, 43
\bibitem[1988]{balluch}Obreschkow, D., Croton, D., De Lucia, G., Khochfar, S., \& Rawlings, S.\ 2009, ApJ, 698, 1467 
\bibitem[1988]{balluch}Onodera, M., Arimoto, N., Daddi, E., et al.\ 2010, ApJ, 715, 385 
\bibitem[1988]{balluch} Oliver, S.~J., Bock, J., Altieri, B., et al.\ 2012, MNRAS, 3269 
\bibitem[1988]{balluch}Pannella, M., Carilli, C. L., Daddi, E., et al.\ 2009, ApJ, 698, 116
\bibitem[1988]{balluch}Papadopoulos, Padelis P., van der Werf, Paul, Xilouris, E., Isaak, Kate G., Gao, Yu\ 2012, ApJ, 751, 10
\bibitem[1988]{balluch}Peng, C. Y., Ho, L. C., Impey, C. D., Rix, H-W.\ 2002, AJ, 124, 266
\bibitem[1988]{balluch} Peng, Y.-j., Lilly, S.~J., Kova{\v c}, K., et al.\ 2010, ApJ, 721, 193  
\bibitem[1988]{balluch} Perera, T.~A., Chapin, E.~L., Austermann, J.~E., et al.\  2008, MNRAS, 391, 1227 
\bibitem[1988]{balluch} Pettini, M., \& Pagel, B.~E.~J.\ 2004, MNRAS, 348, L59 
\bibitem[1988]{balluch}Pilbratt, G. L.,  Riedinger, J. R., Passvogel, T., et al.\  2010, A\&A, 518, 1
\bibitem[1988]{balluch}Pope, A., et al.\ 2006, MNRAS, 370, 1185
\bibitem[1988]{balluch}Poglitsch, A., Waelkens, C., Geis, N. et al.\ 2010, A\&A, 518, L2
\bibitem[1988]{balluch}Popping, G., Caputi, K.~I., Somerville, R.~S., \& Trager, S.~C., 2012, arXiv:1201.3826 
\bibitem[1988]{balluch}Reddy, N., Dickinson, M., Elbaz, D., et al.\ 2012, ApJ, 744, 154 
\bibitem[1988]{balluch} Riechers, D.~A., Cooray, A., Omont, A., et al.\ 2011, ApJL, 733, L12 
\bibitem[1988]{balluch}Riechers, D.~A., Hodge, J., Walter, F., Carilli, C.~L., \& Bertoldi, F.\ 2011, ApJL, 739, L31 
\bibitem[1988]{balluch}	Rodighiero, G., Daddi, E., Baronchelli, I., et al.\ 2011, ApJ, 739, 40
\bibitem[1988]{balluch}Rodighiero, G., Cimatti, A., Gruppioni, C., et al.\  2010A\&A, 518, 25
\bibitem[1988]{balluch}Rodr{\'{\i}}guez Zaur{\'{\i}}n, J., Tadhunter, C.~N., \& Gonz{\'a}lez Delgado, R.~M.\ 2010, MNRAS, 403, 1317 
\bibitem[1988]{balluch}Saintonge, A., Kauffmann, G., Kramer, C., et al.\ 2011, MNRAS, 415, 32
\bibitem[1988]{balluch} Salmi, F., Daddi, E.,  Elbaz, D., et al.\ 2012, arXiv:1206.1704 
\bibitem[1988]{balluch}  Sanders, D.~B., \& Mirabel, I.~F.\ 1996, ARA\&A, 34, 749 
\bibitem[1988]{balluch}  Santini, P., Fontana, A., Grazian, A., et al.\ 2009, A\&A, 504, 751 
\bibitem[1988]{balluch} Sargent, M.~T., B{\'e}thermin, M., Daddi, E., \& Elbaz, D.\ 2012, ApJL, 747, L31
\bibitem[1988]{balluch}Savaglio, S., Glazebrook, K., Le Borgne, D., et al.\ 2005, ApJ, 635, 260
\bibitem[1988]{balluch} Schruba, A., Leroy, A.~K., Walter, F., et al.\ 2012, AJ, 143, 138 
\bibitem[1988]{balluch} Smail, I., Swinbank, A.~M., Ivison, R.~J., \& Ibar, E.\ 2011, MNRAS, 414, L95 
\bibitem[1988]{balluch} Smith. et al.\ 2012, arxiv:1208.3079
\bibitem[1988]{balluch}Solomon, P. M., Downes, D., Radford, S. J. E., Barrett, J. W.\ 1997, ApJ, 478, 144
\bibitem[1988]{balluch}Solomon, P.~M., \& Vanden Bout, P.~A.\ 2005, ARA\&A, 43, 677 
\bibitem[1988]{balluch}Sommariva, V., Mannucci, F., Cresci, G., et al.\ 2012, A\&A, 539, A136 
\bibitem[1988]{balluch} Stark, D.~P., Ellis, R.~S., Chiu, K., Ouchi, M., \& Bunker, A.\ 2010, MNRAS, 408, 1628
\bibitem[1988]{balluch} Strong, A.~W., \& Mattox, J.~R.\ 1996, A\&A, 308, L21 
\bibitem[1988]{balluch} Symeonidis, M., Page, M.~J., Seymour, N., et al.\ 2009, MNRAS, 397, 1728
\bibitem[1988]{balluch} Swinbank, A.~M., Smail, I., Longmore, S., et al.\ 2010, Nature, 464, 733
\bibitem[1988]{balluch}Tacconi, L. J., Genzel, R., Smail, I., et al.\ 2008, ApJ 680, 246
\bibitem[1988]{balluch}Tacconi, L. J., Genzel, R., Neri, R., et al.\ 2010, Nature 463, 781
\bibitem[1988]{balluch}Takagi, T., Ono, Y., Shimasaku, K., \& Hanami, H.\ 2008, MNRAS, 389, 775 
\bibitem[1988]{balluch} Teplitz, H.~I., Chary, R., Elbaz, D., et al.\ 2011, AJ, 141, 1 
\bibitem[1988]{balluch}Tremonti, C.~A., Heckman, T.~M., Kauffmann, G., et al.\ 2004, ApJ, 613, 898
\bibitem[1988]{balluch}Weingartner, J.~C., \& Draine, B.~T.\ 2001, ApJ, 563, 842 
\bibitem[1988]{balluch}Wei{\ss}, A., Kov{\'a}cs, A., Coppin, K., et al.\ 2009, ApJ, 707, 1201
\bibitem[1988]{balluch}Wilson, C.~D.\ 1995, ApJL, 448, L97 
\end{thebibliography}
\end{document}